\documentclass{aa}

\usepackage{graphicx}
\usepackage{enumitem}
\usepackage[justification=raggedright,singlelinecheck=false]{caption}
\usepackage{booktabs}
\usepackage{subfigure}
\usepackage{tabularx}
\renewcommand{\arraystretch}{1.29}
\usepackage{makecell}
\usepackage{multicol}
\usepackage{stfloats}
\usepackage{amsmath}
\usepackage{txfonts}
\usepackage[
  colorlinks=true,   
  linkcolor=blue,    
  citecolor=blue,    
  urlcolor=blue,     
  pdfborder={0 0 1}
]{hyperref}
\usepackage{cleveref}
\usepackage{orcidlink}

\newcommand{\Msun}{\mathrm{M}_{\odot}}
\newcolumntype{Y}{>{\raggedright\arraybackslash}X}
\newcolumntype{Z}{>{\centering\arraybackslash}X}

\begin{document}

   \title{Infrared emission from $z\sim6.5$ quasar host galaxies: a direct estimate of dust physical properties}

   \author{M. Costa \inst{1,2}, R. Decarli \inst{2}, F. Pozzi \inst{1}, P. Cox \inst{3}, R. Meyer \inst{4}, A. Pensabene \inst{5}, B. Venemans \inst{6}, F. Walter, F. Xu
           }

   \author{
        M. Costa\inst{1,2}\orcidlink{0009-0009-1622-8636}
        \and R. Decarli\inst{2}\orcidlink{0000-0002-2662-8803}
        \and F. Pozzi\inst{1}\orcidlink{0000-0002-7412-647X}
        \and P. Cox\inst{3}\orcidlink{0000-0003-2027-8221}
        \and R. A. Meyer\inst{4}\orcidlink{0000-0001-5492-4522}
        \and A. Pensabene\inst{5,6,7}\orcidlink{0000-0001-9815-4953}
        \and B. P. Venemans\inst{8}\orcidlink{0000-0001-9024-8322}
        \and F. Walter\inst{9}\orcidlink{0000-0003-4793-7880}
        \and F. Xu\inst{10,11,2}\orcidlink{0000-0003-0754-9795}
        }
        
        \institute{
        Department of Physics and Astronomy, University of Bologna, via Gobetti 93/2 40129, Bologna, Italy
        \and
        INAF -- Istituto Nazionale di Astrofisica, via Gobetti 93/3, I-40129, Bologna, Italy.
        \and
        Sorbonne University, UPMC Paris 6 and CNRS, UMR 7095, Institut d’Astrophysique de Paris, 98b bd. Arago, 75014 Paris, France
        \and
        Department of Astronomy, University of Geneva, Chemin Pegasi 51, 1290 Versoix, Switzerland
        \and
        Cosmic Dawn Center (DAWN)
        \and
        DTU Space, Technical University of Denmark, Elektrovej 327, DK2800 Kgs. Lyngby, Denmark
        \and
        Department of Physics “G. Occhialini”, University of Milano-Bicocca, Piazza della Scienza 3, I-20126, Milan, Italy
        \and
        Leiden Observatory, Leiden University, Niels Bohrweg 2, NL-2333 CA Leiden, The Netherlands
        \and
        Max-Planck Institute for Astronomy, Königstuhl 17, 69118 Heidelberg, Germany
        \and
        Department of Astronomy, School of Physics, Peking University, Beijing 100871, People's Republic of China
        \and
        Kavli Institute for Astronomy and Astrophysics, Peking University, Beijing 100871, China
        }

  \date{Submitted to A\&A on August 11, 2025; accepted for publication on November 27, 2025}
  \abstract{
  Quasars at the dawn of Cosmic Time ($z>6$) are fundamental probes to investigate the early co-evolution of supermassive black holes and their host galaxy. Nevertheless, their infrared spectral energy distribution remains at the present time poorly constrained, due to the limited photometric coverage probing the far-infrared wavelength range where the dust modified black-body is expected to peak ($\sim80$ $\mathrm{\mu m}$).
  Here we present a study of the high-frequency dust emission via a dedicated ALMA Band 8 ($\sim$400 GHz) campaign targeting 11 quasar host galaxies at $6<z<7$. Combined with archival observations in other ALMA bands, this program enables a detailed characterization of their infrared emission, allowing for the derivation of dust masses ($M_{d}$), dust emissivity indexes ($\beta$), dust temperatures ($T_{d}$), infrared luminosities ($L_{IR}$), and associated star formation rates (SFRs).\\
  Our analysis confirms that dust temperature is on average higher in this sample (34-65 K) if compared to local main-sequence galaxies' values, and that this finding can be linked to the increased star formation efficiency we derive in our work, as also suggested by the [CII]$_{158\mu m}$ deficit. Most remarkably, we note that the average value of $T_d$ of this sample doesn't differ from the one that is observed in luminous, ultra-luminous and hyper-luminous infrared galaxies at different redshifts that show no signs of hosting a quasar. 
  Finally, our findings suggest that the presence of a bright AGN does not significantly bias the derived infrared properties, although further high-frequency, high-spatial resolution observations might reveal more subtle impacts on sub-kiloparsec scales.
  }


   \maketitle

\section{Introduction}  
    Quasar (QSO) host galaxies at $z>6$ are among the most extreme systems present in the Universe. Here large reservoirs of gas \citep[with molecular gas masses $M_{H_2}>10^{10}$ $\Msun$,][]{2022A&A...662A..60D, 2024A&A...684A..33K} fuel both rapid AGN accretion \citep[with mass accretion rates $\dot{M}_{BH}>10$ $\mathrm{\Msun/yr}$, see e.g.][]{2022ApJ...941..106F, 2022A&A...665A.107T} and intense bursts of star formation \citep[with star formation rates SFR $>100$ $\mathrm{M_\odot/yr}$, see e.g.][]{2018ApJ...866..159V}. This means that, according to the downsizing scenario \citep[see e.g.][]{2006A&A...453L..29C}, these sources witness at $z\sim6$ the peak of their SFR, making them plausible candidates for being the progenitors of the most massive elliptical galaxies that we observe in the local universe \cite[see e.g.][]{2006ARA&A..44..141R}.
    One of the most popular scenarios suggests gas-rich major mergers to play a primary role: the rapid inflow of gas that these events are able to trigger fuels at the same time nuclear star formation episodes and feeds gas onto the central supermassive black hole (SMBH), thus activating the AGN. This marks the beginning of a dust-obscured accretion phase, that lasts until dust is removed by either stellar or AGN feedback, and the black hole becomes visible as a traditional QSO in the optical band \citep[][]{2008ApJS..175..356H}. Eventually, the winds driven by the accretion and star-formation processes will be able to disperse the residual gas, thus completely quenching the two processes \citep[][]{2015ARA&A..53...51S}, finally leading to the formation of a massive elliptical galaxy with a quiescent SMBH \citep[see e.g.][]{2023Natur.619..716C}.\medskip \\
    \indent The study of the stellar continuum in (type 1) QSO host galaxies can be extremely challenging due to the unfavorable contrast of their rest-frame optical starlight against the bright central AGN. As a result, obtaining measurements of key stellar population-related properties becomes difficult, unless we achieve the high angular resolutions necessary to separate the host galaxy’s emission from the AGN contribution \citep[see e.g.][]{2022ApJ...939L..28D, 2023ApJ...955...15T}. This means that, at large cosmological distances, the best method to investigate QSO host galaxies today is through observations in the submillimeter and millimeter wavelengths,
    particularly using facilities such as the \emph{Atacama Large Millimeter/Submillimeter Array} (ALMA), the \emph{Northern Extended Millimeter Array} (NOEMA) or the \emph{Jansky Very Large Array} (JVLA) in the radio band. Thanks to these observations we are now able to exploit multiple interstellar medium (ISM) tracers to gauge the gas mass and density both in the molecular and ionized phase \citep[see e.g.][]{2019ApJ...881...63N, 2021A&A...652A..66P}, to study the kinematics of the system \citep[see e.g.][]{2021ApJ...911..141N, 2024ApJ...968....9W}, and to search for outflows \citep[see e.g.][]{2023ApJ...944..134B, 2025ApJ...982...72S}.
    At the same time, sampling the rest-frame far-infrared (FIR) continuum emission originating from interstellar dust heated either by young stars or by AGN radiation is fundamental to investigate QSO host galaxies' key properties, such as dust masses ($M_d$), dust emissivity indexes ($\beta$) and, most importantly, dust temperatures ($T_d$). Reliable estimates of these parameters will result in a robust constrain of the infrared (IR) spectral energy distribution (SED) of the source, thus allowing us to pin down its infrared luminosity ($L_{IR}$) and the associated star formation rate (SFR). To date, given the scarce amount of data sampling the rest-frame wavelength of $\sim80$ $\mathrm{\mu m}$, these properties remain poorly constrained and the study of high-$z$ QSO hosts relies on the assumption of templates \citep[see e.g.][]{2018ApJ...854...97D, 2022A&A...666A..17G, 2022MNRAS.515.3126I}.
    To provide estimates of the dust temperature specifically is of crucial importance, due to its strong impact on the energy budget, as $L_{IR} \propto T_d^4$  or $L_{IR} \propto T_d^{4+\beta}$ assuming optically thick or thin conditions, respectively (\citeauthor{2016MNRAS.461.1328E} \citeyear{2016MNRAS.461.1328E}). \medskip \\
    \indent Moreover, in the last few years, several works have been finding evidence of an evolution of $T_d$ across redshift in UV-selected galaxies \citep[][]{2018A&A...609A..30S, 2020ApJ...902..112B, 2020MNRAS.498.4192F}, the origin of which they attribute to a higher SFR surface density ($\Sigma_{SFR}$) induced by the higher star formation efficiency (SFE = SFR/$M_{H_2}$) or specific SFR (sSFR =  SFR/$M_*$, with  $M_*$ being the stellar mass content) predicted by models at high-$z$ \citep[][]{2019MNRAS.489.1397L, 2019MNRAS.487.1844M, 2022MNRAS.513.3122S}. The behavior of this trend at $z>6$, however, is still quite uncertain: recent works tried to extend the $T_d-z$ relation into the Epoch of Reionization, although relying on a somewhat limited photometric coverage \citep[][]{2022MNRAS.513.3122S, 2024ApJ...971..161M}.
    Dust temperature may also play a role in defining the negative correlation between the [CII]$_{158\mu m}$ line luminosity ($L_{[CII]})$ to $L_{IR}$ ratio and $L_{IR}$ itself (known in the literature as [CII]$_{158\mu m}$ deficit). This phenomenon can be ascribed to an increase of the amount of dust heated to high temperatures in the ionized region of dust-bounded
    star-forming regions \citep[][]{2013ApJ...774...68D, 2017ApJ...846...32D}, which is consistent with the picture of a larger $\Sigma_{SFR}$ in high-$z$ galaxies.
    At $z>6$, QSO hosts provide favorable cases for estimating dust temperatures, as their brightness and well-sampled SEDs yield multiple data points for reliable analysis. Here, however, the unresolved emission coming from a hot dust component heated by the AGN can bias the derived results and mimic the effect of an increased $\Sigma_{SFR}$ by enhancing both the derived $L_{IR}$ and $T_d$ \citep[][]{2020ApJ...904..130V, 2024A&A...689A.220T}. Whether the amount of dust heated by the AGN is sufficient to account for a non-negligible fraction of the observed flux at the rest-frame wavelength of $\sim 80$ $\mu$m is still a matter of debate \citep[][]{2021MNRAS.503.3992S}. To pin down this contribution, high-resolution ALMA observations may be needed, as they are able to probe the emission coming from the central region of the source, which might be powered by the AGN \citep[see e.g.][]{2023MNRAS.523.4654T, 2025A&A...695L..18M}.\medskip \\
    \indent In this paper we present a systematic survey of the 400 GHz emission in $z>6$ QSOs, a new set of Band 8 ALMA ACA observations targeting 11 $6<z<7$ QSOs. The combination of these data with flux measurements that sample the Rayleigh-Jeans tail of the dust emission allows us to pin down the IR SED of these sources without the need of the assumption of a given $T_d$, resulting in a direct estimate of this parameter instead. In section \ref{sec2} we present the sample, the observations and the process of data reduction. In section \ref{sec3} we describe the adopted methods, the data analysis and the algorithms. Finally, in section \ref{sec4}, we discuss the obtained results, putting them in a broader scientific context and linking our findings on $T_d$ with the results of various works, discussing how QSOs, star-forming and starburst galaxies compare to each other when it comes to dust temperature. 
   In this work, we adopt the $\Lambda$CDM cosmology from \citeauthor{2020A&A...641A...6P} \citeyear{2020A&A...641A...6P}: $H_0=67.4$ km$\cdot$s$^{-1}\cdot$ Mpc$^{-1}$, $\Omega_m=0.315$ and $\Omega_\Lambda=0.685$.

\section{Sample, observations and data reduction} \label{sec2}

\begin{table*}[h!]
    \centering
    \caption{List of the sources analyzed in this work.}
    \begin{tabularx}{\textwidth}{ZZZZZ}
         \midrule
         \midrule
         ID & RA & Dec & z &  Ref\\
         \midrule
         AJ025-33 & 01:42:43.727 & -33:27:45.470 & 6.3373 & \makebox[0pt][c]{\cite{2020ApJ...904..130V}}\\ 
         J0305-3150 & 03:05:16.916 & -31:50:55.900 & 6.6139 & \makebox[0pt][c]{\cite{2020ApJ...904..130V}} \\
         J0439+1634* & 04:39:47.110 & +16:34:15.820 & 6.5188 &  \makebox[0pt][c]{\cite{2019ApJ...880..153Y}}\\
         J2318-3029 & 23:18:33.100 & -30:29:33.370 & 6.1456 & \makebox[0pt][c]{\cite{2020ApJ...904..130V}}\\
         J2348–3054 & 23:48:33.334 & -30:54:10.240 & 6.9007 & \makebox[0pt][c]{\cite{2020ApJ...904..130V}}\\
         PJ007+04 & 00:28:06.560 & +04:57:25.680 & 6.0015 &  \makebox[0pt][c]{\cite{2020ApJ...904..130V}}\\
         PJ009–10 & 00:38:56.522 & -10:25:53.900 & 6.0040 &  \makebox[0pt][c]{\cite{2020ApJ...904..130V}}\\
         PJ036+03 & 02:26:01.876 & +03:02:59.390 & 6.5405 &  \makebox[0pt][c]{\cite{2020ApJ...904..130V}}\\
         PJ083+11 & 05:35:20.900 & +11:50:53.600 & 6.3401 & \makebox[0pt][c]{\cite{2020ApJ...903...34A}}\\
         PJ158-14 & 10:34:46.509 & -14:25:15.855 & 6.0681 & \makebox[0pt][c]{\cite{2020ApJ...900...37E}} \\
         PJ231–20 & 15:26:37.841 & -20:50:00.660 & 6.5865 & \makebox[0pt][c]{\cite{2021A&A...652A..66P}}\\
         \midrule
    \end{tabularx}
    \\[9pt]
    \parbox{\linewidth}{\small Notes -- columns: QSO name; radial ascension; declination; redshift; reference for the redshift. *this this source is strongly lensed.}
    \label{tab:Sample}
\end{table*}

\subsection{ALMA Band 8 observations}
The sample of this study consists of 11 type 1 radio quiet QSOs at $6<z<7$ observed with ALMA ACA Band 8 (ALMA project IDs: 2019.2.00053.S and  2021.2.00064.S, PI: Roberto Decarli). The sources for this observational campaign are selected so that their flux at the rest-frame wavelength of 158 $\mu$m is greater than 1.5 mJy and had no public flux measurement performed by \emph{Herschel/SPIRE}: they are the QSOs AJ025-33, J0305-3150, J0439+1634, J2318-3029, J2348–3054, PJ007+04, PJ009–10, PJ036+03, PJ083+11, PJ158-14 and PJ231–20. The list of all the targets can be found in Table \ref{tab:Sample}. These observations were carried out between 19 July 2021 and 25 September 2022 and, within the frequency span of Band 8, they have been tuned to 407 GHz, where the atmospheric transparency is most favorable. The calibrated visibilities have been obtained by reducing the raw data using the default pipeline scripts executed in the Common
Astronomy Software Applications \citep[CASA,][]{2022PASP..134k4501C}. After checking for the presence of emission lines, the continuum emission has then been imaged running the TCLEAN task in \emph{multi-frequency-synthesis} mode, adopting a 3 $\sigma$ cleaning threshold (with $\sigma$ being the rms of the noise) and a natural weighting scheme, in order to minimize the noise.
We exploit the continuum maps obtained this way to derive a flux measurement by fitting a 2D Gaussian to the sources via CASA task IMFIT. A summary of the characteristics of the observations, as well as the flux of the continuum emission can be found in Table \ref{tab:Band8} in the appendix. In Figure \ref{fig:EsB8} we report the continuum map of the QSO J0305-3150 as example, while the complete set of continuum maps can be found in Figure \ref{fig:B8im} in the appendix.
\begin{figure}[h!]
    \centering
    \includegraphics[width=0.7995\linewidth]{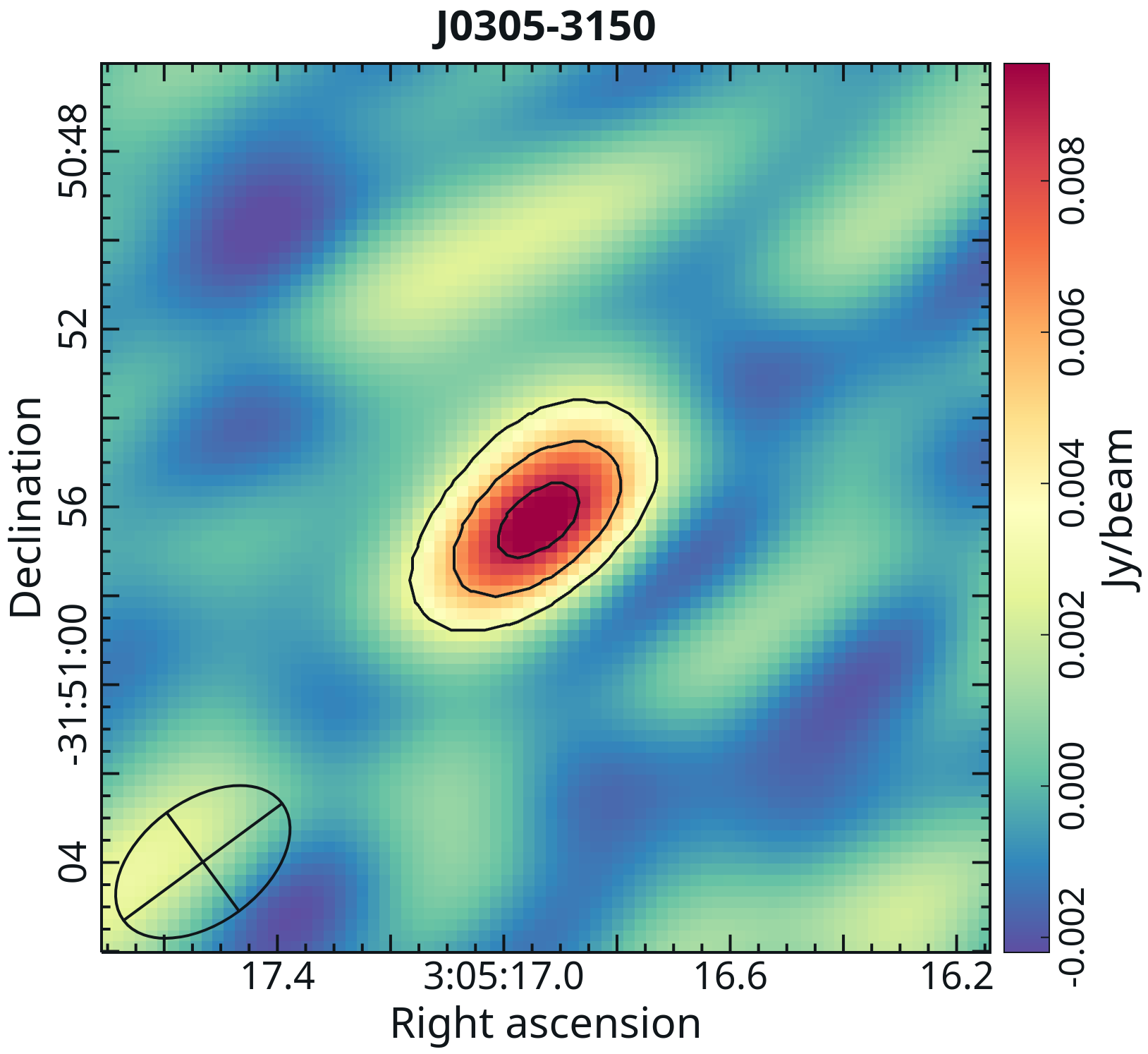}
    \caption{407 GHz dust continuum map of the QSO
    J0305-3150 (contours at 3, 6 and 9 $\sigma$).}
    \label{fig:EsB8}
\end{figure}
\subsection{Multi-wavelength data and size measurements}
To perform the modified black-body (MBB) fit, we collect additional dust continuum flux measurements at various frequencies available for the sources in our sample. These data are gathered from both the specific literature and the observations available in the ALMA archive, which we reduce and analyze using the same procedure described in the previous subsection. 
The complete list of flux measurements used in the SED fitting procedure is reported in Table \ref{tab:gigatab}, in the appendix. Few corrections to both our Band 8 and the published values have been necessary. J0439+1634 is a strongly lensed QSO; we therefore scaled down the fitted flux value by the galaxy-integrated average magnification factor $\mu=4.5$ \citep[][]{2021ApJ...917...99Y}. PJ231-20, instead, was found to have a companion galaxy located $\sim 2$ arcsec south to the quasar \citep[][]{2017Natur.545..457D}. Having a beam size of $4.34\times2.92$ arcsec, our observations lack the angular resolution necessary to disentangle the flux from the two galaxies. In this paper, we used as reference for the other flux measurements the data from \citep[][]{2021A&A...652A..66P} where, however, the two sources are resolved. For this reason, we reduced and analyzed very high resolution (with a beam size of $0.14\times0.13$ arcsec) Band 8 observations of this source from the ALMA project 2022.1.00321.S (PI: Roberto Decarli), in which is possible to disentangle the quasar and the companion galaxy. Here we measured for the quasar a flux of 8.4 $\pm$ 0.4 mJy at the observed frequency of 398 GHz. In this map, the combined flux of the companion and QSO host agrees with the flux measured in the lower-resolution observation. \medskip \\
\indent The appendix features also a size measurement of the emitting region derived from spatially resolved observations (see Table \ref{tab:sizes}), as this parameter is necessary to fit a MBB to the data. The angular extent of the emitting region $\Omega$ is computed as:
\begin{align}
    \Omega=\pi \cdot \dfrac{9\cdot a\cdot b}{8\cdot \ln 2}
\end{align}
where $a$ and $b$ are respectively the full-width half-maximum of the major and minor axis of the 2D Gaussian fitted to the image plane in the case of spatially resolved observations. This equation corresponds to the area of an ellipse that subtends $\sim98.9 \%$ of the total volume (flux) of that Gaussian. We will discuss this assumption and its impact on the fit in the following section, in light of the obtained results. For the QSO's J0439+1439 size, it was necessary to convert the effective radius ($r_e$) reported in \citeauthor{2021ApJ...917...99Y} \citeyear{2021ApJ...917...99Y} to the FWHM of a 2D Gaussian according to \citeauthor{2010MNRAS.404..458V} \citeyear{2010MNRAS.404..458V}:
\begin{align}
    FWHM = 2\cdot r_e \cdot \Bigg(\dfrac{\ln{2}}{k}\Bigg)^n
\end{align}
where $n = 1.58$ is the Sersic index and $k = 1.9992n-0.3271$. In the case of PJ083+11 instead, the angular size was computed from the physical size reported in \citeauthor{2020ApJ...903...34A} \citeyear{2020ApJ...903...34A}.

\subsection{NOEMA Band 2 complementary data}
In addition, we use a dedicated NOEMA program (ID: S24CH, P.I.s: Michele Costa, Leindert Boogaard) to secure dust continuum photometry of the Rayleigh-Jeans tail of the emission in PJ007+04 and PJ009-10. These observations sample the dust continuum at the observed frequency of 158 GHz, as well as the emission lines CO(9-8) at 1036.912 GHz, CO(10-9) at 1151.985 GHz, and H$2$O($3_{2,1}$-$3_{1,2}$) at 1162.912 GHz rest-frame. A summary of the observations, as well as the flux of the continuum emission measured at the observed frequency of 158 GHz can be found in Table \ref{tab:Noema}. All of them were carried out in array configuration C. The calibrated visibilities were obtained by reducing the raw data using the Grenoble Image and Line Data Analysis Software (GILDAS\footnote{\url{https://www.iram.fr/IRAMFR/GILDAS}}). We imaged the cubes adopting a 3 $\sigma$ cleaning threshold, natural weighting, and a channel width of 50 km/s. To search for emission lines, we extracted the spectrum from a region centered on the brightest pixel of the continuum image, with the region size matching the clean beam. These spectra are reported in Figure \ref{fig:NOEMAims}, along with the images of the dust emission obtained combining together the line-free channels from both the spectral windows. A continuum flux measurement was obtained from these maps with the same procedure adopted for ALMA Band 8 observations. No lines were found in the spectrum of PJ007+04, while for PJ009-10 we report a 3.4 $\sigma$ detection of the CO(9-8) line and a 2.4 $\sigma$ detection of the CO(10-9) line, obtained by fitting a Gaussian profile to the data centered at the expected frequency of the transition. We find a flux of $353^{+27}_{-26}$ mJy $\cdot$ Km/s for the CO(9-8) line and of $323^{+27}_{-26}$ mJy $\cdot$ Km/s for the CO(10-9) line. \\

\section{Methods and data analysis} \label{sec3}
\begin{figure}[h!]
    \centering
    \includegraphics[width=\linewidth]{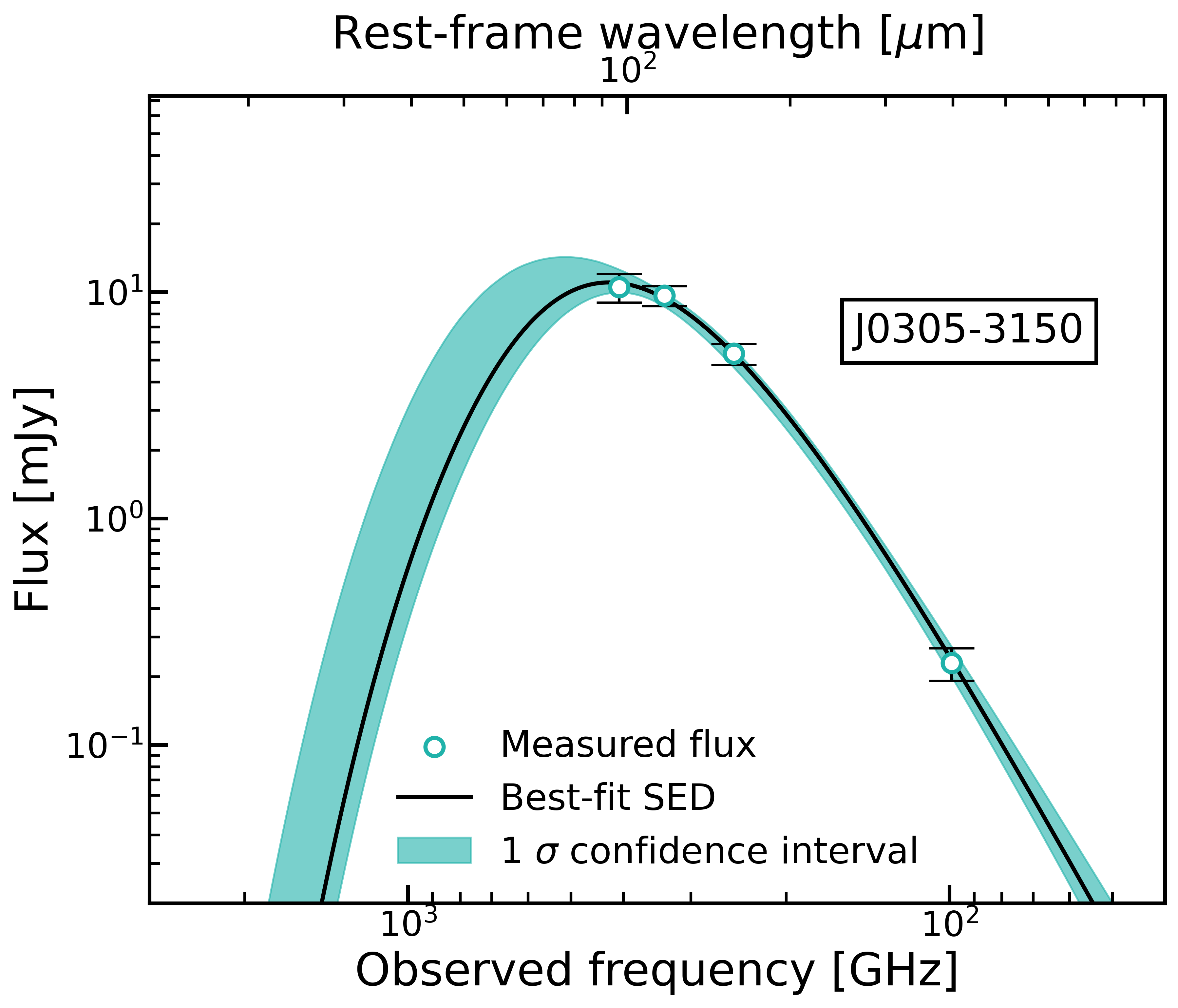}
    \vspace{-0.3cm}\\
    \includegraphics[width=0.925\linewidth]{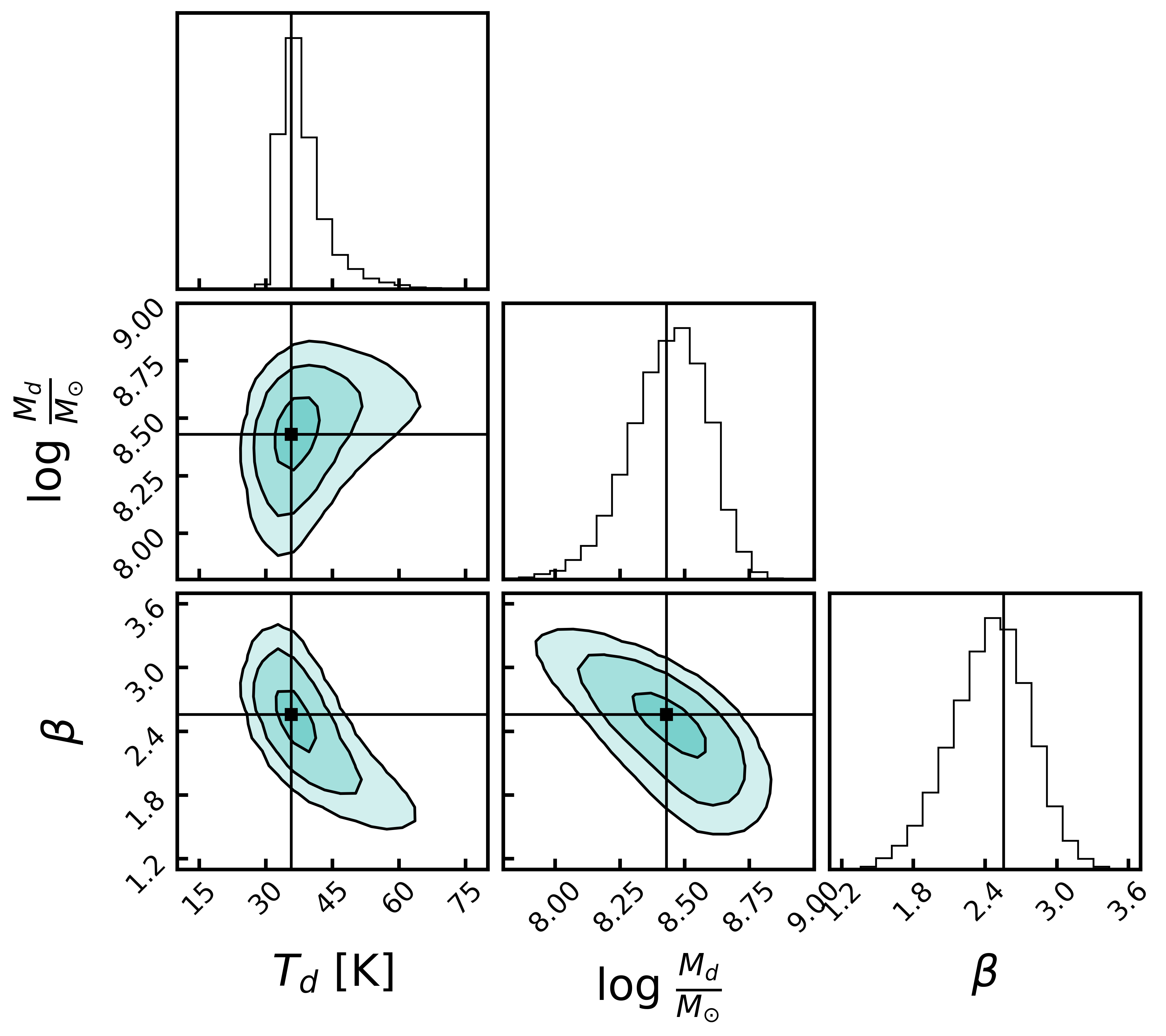}
    \caption{Top panel: fitted dust SED of the QSO J0305-3150. Bottom panel: corner plot of the posterior probability distribution of the fit parameters.}
    \label{fig:EsAJ}
\end{figure}

The FIR emission of dusty astronomical objects is often modeled using multiple MBB components with different dust temperatures. In normal galaxies \citep[see, e.g.][]{2012MNRAS.425..763G}, the heating of a cold dust component ($T_d\sim20$ K) may be ascribed to the global starlight radiation field, including the contribution of evolved stars, while a warmer dust component ($T_d\sim60$ K) is often associated to the emission from star-forming regions. However, in high-$z$ QSOs, the high frequency regime where this last component is expected to contribute most significantly is hardly accessible by current facilities. For this reason, the FIR SED of these sources is often fitted using only a single dust component, even though the contribution from dust heated to higher temperatures near the SED peak may not be negligible or could even be the primary source of emission. Furthermore, in luminous QSOs, an additional hot dust component ($T_d > 60$ K), potentially heated directly by the AGN, may dominate at rest-frame mid-infrared (MIR) wavelengths \citep[see, e.g.][]{2006ApJ...642..694B}.
In this paper we carry out a spatially unresolved analysis, assuming that the emission arises from a single dust component given by:
\begin{align}
    S(\nu_{obs})= \dfrac{\Omega}{(1+z)^3}[B(T_d(z),\nu)-B(T_{CMB}(z),\nu)](1-e^{-\tau(\nu)})
    \label{eq:mbb}
\end{align}
where $\nu_{obs}$ is the emitted frequency $\nu$ redshifted to the observer rest-frame, $\Omega$ the angular size of the emitting region, $B(T_d(z),\nu)$ the spectrum of a black-body with temperature $T_d(z)$, $B(T_{CMB}(z),\nu)$ the CMB spectrum at a given redshift and $\tau(\nu)$ the optical depth. The optical depth is defined as:
\begin{align}
    \tau(\nu)=k(\nu)\dfrac{M_d}{A_g}
    \label{eq:od}
\end{align}
where $k(\nu)$ is the opacity, $M_d$ is the dust mass, and $A_g$ is the physical size of the emitting region ($A_g=(\Omega\cdot D_L^2)/(1+z)^4$, with $D_L$ being the luminosity distance). Following \citeauthor{2006ApJ...642..694B} \citeyear{2006ApJ...642..694B} we parametrize the opacity as a power law:
\begin{align}
    k(\nu)=k_0\bigg(\dfrac{\nu}{\nu_0}\bigg)^\beta
    \label{eq:op}
\end{align}
where $k_0=0.4$ cm$^2$/gr, $\nu_0=250$ GHz, and $\beta$ is called emissivity index.
To account for the effect of the interaction of dust with the CMB at high-redshift we follow \citeauthor{2013ApJ...766...13D} \citeyear{2013ApJ...766...13D}:
\begin{align}
    T_d(z)=\bigg(T_d^{4+\beta}+T_0^{4+\beta}\Big[(1+z)^{4+\beta}-1\Big]\bigg)^{1/(4+\beta)}
    \label{eq:tcorr}
\end{align}
where $T_0 = 2.73$ K is the CMB temperature at $z = 0$, $T_d(z)$ the temperature of the dust at a given redshift, and $T_d$ the intrinsic temperature of the dust at the net effect of the heating provided by the CMB. This step allows for the comparison of the dust temperature in objects at different redshifts.
The set of equations \ref{eq:mbb}, \ref{eq:od}, \ref{eq:op} and \ref{eq:tcorr} results in a full model for a source that emits as a MBB at a single temperature. We leave as free parameter the dust temperature, the dust mass, and the emissivity index, and explore the 3D parameter space using a Markov chain Monte Carlo (MCMC) algorithm implemented in the EMCEE package \citep[][]{2013PASP..125..306F}. We assume uniform priors probability distribution for $T_d$ and $M_d$ in the ranges $T_d$ $\in$ [1, 150] K and
$\log{M_d/M_\odot}$ $\in$ [6, 10], while for $\beta$ we model the prior probability distribution as a Gaussian with mean value $\mu_G=2.2$ and standard deviation $\sigma_G = 0.6$ \citep[see, e.g.][]{2025MNRAS.540.1560B}. We run the algorithm using 45 chains, 5000 steps and a burn-in phase of 1000 steps. In Figure \ref{fig:EsAJ}, we show the fitted SED of the QSO J0305-3150 as an example, together with the corresponding corner plots. The complete set can be found in Figure \ref{sedimages} in the appendix. The SED of the QSO PJ158-14 is not shown, as the fitting algorithm did not converge due to the availability of only two flux measurements, one in ALMA Band 8 and one in ALMA Band 6.

\noindent For each source we then compute the IR luminosity as:
\begin{equation}
    L_{IR} = 4\pi D_L \int_{8\mu m}^{1000\mu m}S(\lambda)d\lambda
\end{equation}
where $D_L$ is the luminosity distance and $S(\lambda)$ the rest-frame specific flux emitted by the source, while for the SFR we adopt the conversion factor from \citeauthor{2011ApJ...737...67M} \citeyear{2011ApJ...737...67M}:
\begin{equation}
    \dfrac{SFR}{M_\odot \cdot yr^{-1}} = 1.496 \cdot 10^{-10} \dfrac{L_{IR}}{L_\odot}
\end{equation}
which relies on the assumption of a Kroupa initial mass function \citep[see][]{2003ApJ...598.1076K}. Finally, we derive the optical depths at the rest-frame frequency of 1900 GHz ($\tau_{1900}$) exploiting equations \ref{eq:od} and \ref{eq:op} and the values of $M_d$ and $\beta$ estimated in the fit.

\section{Results} \label{sec4}

\subsection{SED fitting results}

\begin{table*}[b!]
    \centering
    \caption{Infrared physical properties derived from SED fitting of the sources.}
    \begin{tabularx}{\textwidth}{ZZZZZZZ}
         \midrule
         \midrule
         ID & \makecell[c]{$\mathrm{T_d}$  (K)} & $\mathrm{\log (M_d/M_\odot)}$ & $\mathrm{\beta}$ & \makebox[0pt][c]{$\mathrm{L_{IR}}$ ($\mathrm{10^{12}L_\odot}$)} & \makebox[0pt][c]{SFR ($\mathrm{M_\odot \cdot yr^{-1}}$) } & $\mathrm{\tau_{1900}}$ \\
         \midrule
         \makebox[0pt][c]{AJ025-33} & $43^{\smash{+12}}_{\smash{-6}}$ & $8.0^{\smash{+0.1}}_{\smash{ -0.1}}$ & $2.3^{\smash{+0.3}}_{\smash{ -0.3}}$ & $6.3^{\smash{+6.7}}_{\smash{ -2.2}}$ & $935^{\smash{+999}}_{\smash{ -335}}$ & $0.12^{\smash{+0.07}}_{\smash{ -0.06}}$\\ 
         \makebox[0pt][c]{J0305-3150} & $37^{\smash{ +6}}_{\smash{ -3}}$ & $8.4^{\smash{ +0.1}}_{\smash{ -0.2}}$ & $2.5^{\smash{+0.3}}_{\smash{ -0.3}}$ & $7.3^{\smash{+3.7}}_{\smash{ -1.7}}$ & $1091^{\smash{+548}}_{\smash{ -261}}$ & $0.25^{\smash{+0.10}}_{\smash{ -0.09}}$\\
         \makebox[0pt][c]{J0439+1634*} & $103^{\smash{ +30}}_{\smash{ -25}}$ & $8.59^{\smash{ +0.08}}_{\smash{ -0.08}}$ & $1.0^{\smash{+0.2}}_{\smash{ -0.1}}$ & $50.8^{\smash{+60.9}}_{\smash{ -29.0}}$ & $7597^{\smash{+9104}}_{\smash{ -4332}}$ & $0.17^{\smash{+0.11}}_{\smash{ -0.06}}$\\
         \makebox[0pt][c]{J2318-3029} & $52^{\smash{ +20}}_{\smash{ -8}}$ & $8.2^{\smash{ +0.1}}_{\smash{ -0.1}}$ & $2.0^{\smash{+0.4}}_{\smash{ -0.4}}$ & $7.4^{\smash{+14.2}}_{\smash{ -3.2}}$ & $1114^{\smash{+2123}}_{\smash{ -474}}$ & $0.3^{\smash{+0.2}}_{\smash{ -0.1}}$\\
         \makebox[0pt][c]{J2348–3054} & $65^{\smash{ +5}}_{\smash{ -4}}$ & $8.2^{\smash{ +0.1}}_{\smash{ -0.1}}$ & $1.8^{\smash{+0.2}}_{\smash{ -0.2}}$ & $9.6^{\smash{+2.5}}_{\smash{ -1.9}}$ & $1441^{\smash{+381}}_{\smash{ -280}}$ & $0.34^{\smash{+0.09}}_{\smash{ -0.08}}$ \\
         \makebox[0pt][c]{PJ007+04} & $48^{\smash{ +20}}_{\smash{ -7}}$ & $8.4^{\smash{ +0.2}}_{\smash{ -0.3}}$ & $1.7^{\smash{+0.5}}_{\smash{ -0.5}}$ & $4.3^{\smash{+6.6}}_{\smash{ -1.7}}$ & $645^{\smash{+987}}_{\smash{ -249}}$ & $0.3^{\smash{+0.2}}_{\smash{ -0.2}}$\\
         \makebox[0pt][c]{PJ009–10} & $50^{\smash{ +29}}_{\smash{ -13}}$ & $8.4^{\smash{ +0.1}}_{\smash{ -0.1}}$ & $1.9^{\smash{+0.5}}_{\smash{ -0.5}}$ & $13.6^{\smash{+30.6}}_{\smash{ -7.1}}$ & $2039^{\smash{+4590}}_{\smash{ -1063}}$ & $0.02^{\smash{+0.02}}_{\smash{ -0.01}}$\\
         \makebox[0pt][c]{PJ036+03} & $44^{\smash{ +2}}_{\smash{ -2}}$ & $8.2^{\smash{ +0.1}}_{\smash{ -0.2}}$ & $2.2^{\smash{+0.2}}_{\smash{ -0.2}}$ & $5.3^{\smash{+0.7}}_{\smash{ -0.6}}$ & $787^{\smash{+98}}_{\smash{ -91}}$ & $0.23^{\smash{+0.04}}_{\smash{ -0.03}}$ \\
         \makebox[0pt][c]{PJ083+11} & $34^{\smash{ +6}}_{\smash{ -3}}$  & $8.7^{\smash{ +0.2}}_{\smash{ -0.2}}$ & $2.3^{\smash{+0.4}}_{\smash{ -0.4}}$ & $6.1^{\smash{+2.8}}_{\smash{ -1.4}}$ & $907^{\smash{+410}}_{\smash{ -208}}$ & $0.22^{\smash{+0.10}}_{\smash{ -0.09}}$\\
         \makebox[0pt][c]{PJ158–14} & --- &  --- & --- & --- & --- & --- \\
         \makebox[0pt][c]{PJ231–20} & $63^{\smash{ +7}}_{\smash{ -5}}$ &  $8.51^{\smash{ +0.08}}_{\smash{ -0.09}}$ & $1.7^{\smash{+0.2}}_{\smash{ -0.2}}$ & $11.0^{\smash{+5.5}}_{\smash{ -2.8}}$ & $1643^{\smash{+822}}_{\smash{ -419}}$ & $0.6^{\smash{+0.2}}_{\smash{ -0.2}}$\\
         \midrule
    \end{tabularx}
    \\[9pt]
    \parbox{\linewidth}{\small
     Notes -- we report as confidence result the median value of the posterior probability distribution, while the upper and lower uncertainties correspond to the 16th and 84th percentile respectively. *this source is strongly lensed.}
    \label{tab:Sampleprop}
\end{table*}

The physical parameters of the dust retrieved by the (converged) fits, as well as the
QSOs' IR luminosities, star formation rates, and optical depths at 1900 GHz, are summarized in Table \ref{tab:Sampleprop}.\\

We find a mean temperature of $48 \pm 3$ K, which is higher than the typical dust temperatures of $\sim20$ K observed for main-sequence galaxies in the local universe. This result will be
explored in detail in the following subsection, also in view of the dust
temperature evolution scenario \citep[see, e.g.][]{2022MNRAS.513.3122S}. Interestingly, the most conspicuous outlier is the QSO J0439+1634 
($T_d = 103^{+30}_{-25}\ \mathrm{K}$), 
for which the confidence value we report for the dust temperature is restricted by the upper limit of 150 K that we set as a prior in the fit (see the corresponding corner plot in Figure \ref{sedimages}, in the appendix). However, this source is affected by strong lensing, and the magnification factor for the central QSO derived from the Hubble Space Telescope (HST) imaging in \citeauthor{2019ApJ...870L..11F} \citeyear{2019ApJ...870L..11F} has been shown to be much higher with respect to the total magnification factor for the extended host galaxy. It is thus tempting to argue that the global temperature measured in our analysis might be enhanced due to the stronger magnification of a plausible hotter component associated with the inner regions of the QSO host.
This also reflects in an extreme $L_{IR}$ and SFR. Therefore, J0439+1634 is not included in the plots, nor in general discussion of the dust physical properties of the sample or in the calculation of their average values.\\

\begin{figure}[h!]
    \centering
    \includegraphics[width=\linewidth]{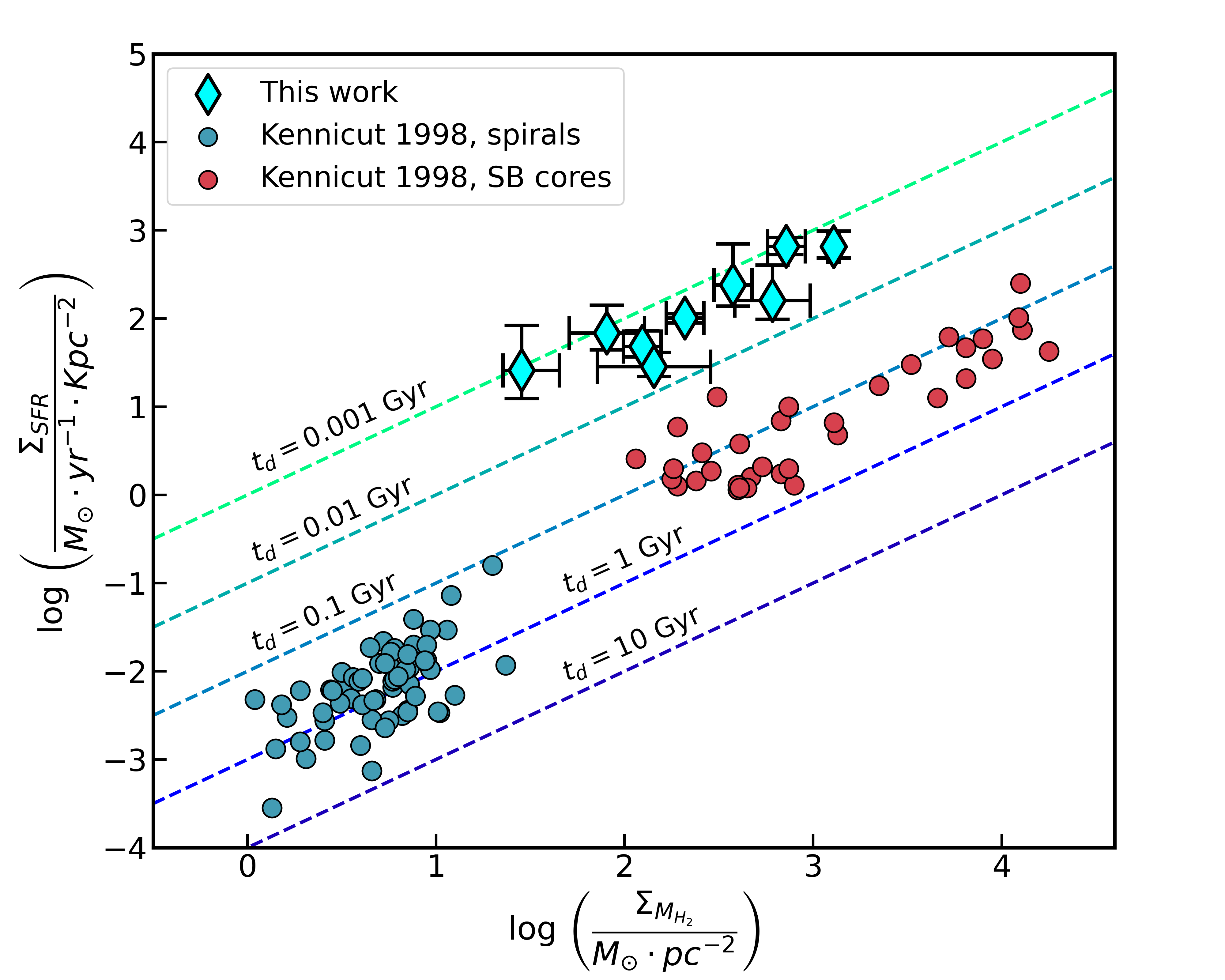}
    \caption{Molecular gas mass surface density obtained assuming a $\mathrm{GDR}=100$ vs SFR surface density. The dashed lines correspond to constant values of gas depletion time, ranging from 1 Myr to 10 Gyr.}
    \label{fig:dustmass}
\end{figure}

As for dust masses, we find a mean value for $\log M_d/M_\odot $ of $8.3 \pm 0.1$, which is quite in good agreement with the values found in \citeauthor{2024A&A...689A.220T} \citeyear{2024A&A...689A.220T} for similar sources.
It is of particular interest to study the relation between the gas mass surface density $\Sigma_{M_d}$ and the SFR surface density $\Sigma_{SFR}$: in Figure \ref{fig:dustmass} we compare these two physical quantities, assuming a gas-to-dust mass ratio (GDR) of 100 \citep[see, e.g.][]{2022A&A...662A..60D, 2023ApJ...954L..10F}. We also plot for comparison a sample of local sources from \citeauthor{1998ApJ...498..541K} \citeyear{1998ApJ...498..541K}. The strong correlation we observe (Pearson index $ R=0.84$, p-value $p=5\cdot 10^{-3}$) finds its explanation in the Kennicut-Schmidt law, while the starbursty nature of these objects seems to be linked to the low depletion time ($\tau_d<0.01$ Gyr) rather than to the enhanced gas mass content, as also assessed in \citeauthor{2024A&A...684A..33K} \citeyear{2024A&A...684A..33K}. \\

Regarding dust emissivity indexes, we obtain a mean value of $2.1\pm 0.1$. This parameter is directly related to the physical properties of dust grains and determines
the behavior of opacity as a function of frequency \citep[see, e.g.][]{2003ARA&A..41..241D}. The values found in this work seem to be consistent with the ones obtained in other studies involving similar objects \citep[see, e.g.][]{2023MNRAS.523.3119W}. For example, this last work found a mean value for $\beta$ of $1.8\pm 0.3$, with no significant evolution across cosmic time, suggesting that the effective dust properties do not change dramatically with redshift. Also, the Gaussian prior implemented in the fit has little influence on the derivation of this parameter, as the distribution of the derived emissivity indices exhibits a standard deviation much smaller than that of the adopted prior (0.3 and 0.6, respectively). If the derivation of $\beta$ were dominated by the prior, comparable values would be expected. Its main role in the fitting procedure is to keep the uncertainties under control by discouraging the chains from exploring nonphysical values. We note that similar values for $\beta$ are obtained also in \citeauthor{2023MNRAS.522.2995B} \citeyear{2023MNRAS.522.2995B} or in \citeauthor{2024ApJ...961..226L} \citeyear{2024ApJ...961..226L}. \\

Concerning infrared luminosities and star formation rates, we derive a mean value of $7.9\pm0.9 \cdot 10^{12}$ $\mathrm{L_\odot}$ and $1178 \pm 140$ $\mathrm{M_\odot \cdot yr^{-1}}$  respectively. Such high values are justified by the construction of the sample (which selects the continuum brightest QSOs at the rest-frame frequency of 1900 GHz) and by the large values derived for the dust temperatures. Given that dust emission is a thermal process (i.e. the IR radiation comes from a MBB that is in thermal equilibrium with the incident UV radiation field), the SFRs we derive here probe timescales of the order of 100 Myr. In contrast, estimates coming from optical or IR lines (such as H$\alpha$ or [OIII]$_{88\mu m}$) are often referred as instantaneous SFR tracers, as they are sensitive to timescales of around 10 Myr \citep[see, e.g.][]{2023A&A...673A.157D}. \\

\begin{figure}[h!]
    \centering
    \includegraphics[width=\linewidth]{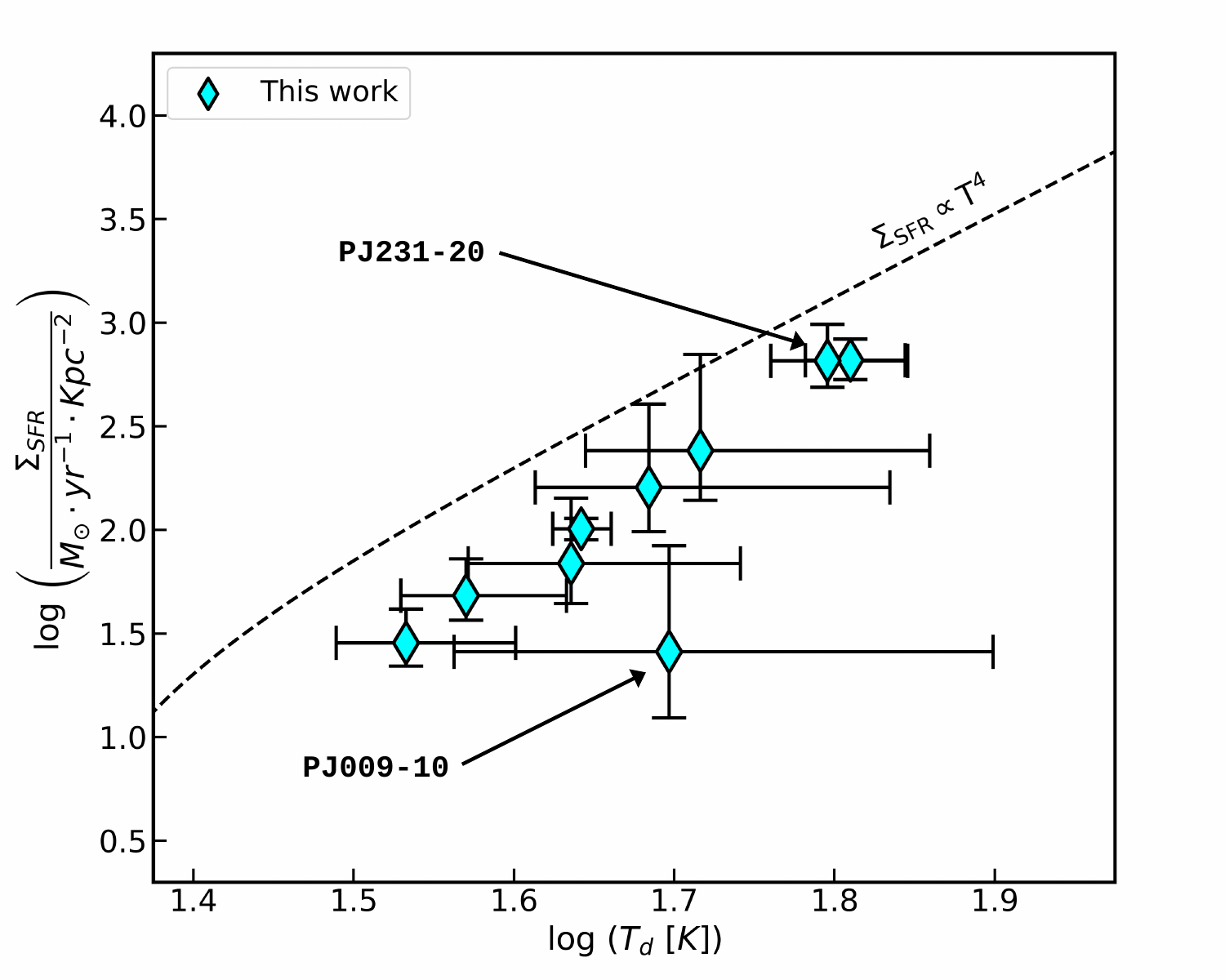}
    \caption{Dust temperature vs SFR surface density. The dashed black line corresponds to the relation obtained by analytically integrating equation \ref{eq:mbb} under optically thick conditions and assuming a $z=6.5$ background CMB radiation field. The two sources indicated by the arrows are PJ009-10 and PJ231-20, which show the lowest and largest value of $\tau_{1900}$, respectively.}
    \label{fig:dustTemp}
\end{figure}

Finally, the average optical depth at the rest frame frequency of 1900 GHz for the
sources in our sample is $0.26 \pm 0.04$. Noticeable outliers are the QSOs PJ231-20 and PJ009-10 ($\tau_{1900} = 0.6^{\smash{+0.2}}_{\smash{ -0.2}}$ and $\tau_{1900} = 0.03^{\smash{+0.02}}_{\smash{ -0.02}}$ respectively): the value of their optical depths is explained by the fact that these sources are either very compact, in the case of PJ231-20 or very extended, in the case of PJ009-10.
At the same time, our analysis seems to point out that the widespread optically thin approximation for dust emission has to be handled with care: the large values of the emissivity index found in this sample, combined with the non negligible optical depth at 1900 GHz, result in a rapid increase of $\tau(\nu)$ with frequency, which can often exceed unity at the frequency probed by our Band 8 observations. To investigate more on this topic, in Figure \ref{fig:dustTemp} we plot the logarithm of the dust temperature versus the logarithm of the SFR surface density $\Sigma_{SFR}$. We compare then our data to the theoretical relation obtained by integrating equation \ref{eq:mbb} in the case of optically thick emission ($e^{-\tau(\nu)} \sim 0$), and assuming a CMB temperature $T_{CMB}(z=6.5)=20.5$ K. This relation represents a theoretical upper limit for the sources, as it traces the $\Sigma_{SFR}$ that we would obtain if these objects emitted as perfect black-bodies. The low discrepancy and the similar slope between our data and this theoretical relation suggests that the large majority of the energy emitted by the dust is radiated at frequencies where the system is optically thick, thus making the widely assumed approximation of optically thin emission questionable for this kind of sources. An interesting point is that the QSO PJ009-10 is the most distant object from this theoretical relation, because its relatively extended shape causes a lower optical depth compared to the other sources. In contrast, the closest one is PJ231-20, which has the most compact sizes and therefore the highest value of $\tau_{1900}$. \\

\begin{figure}[h!]
    \includegraphics[width=\linewidth]{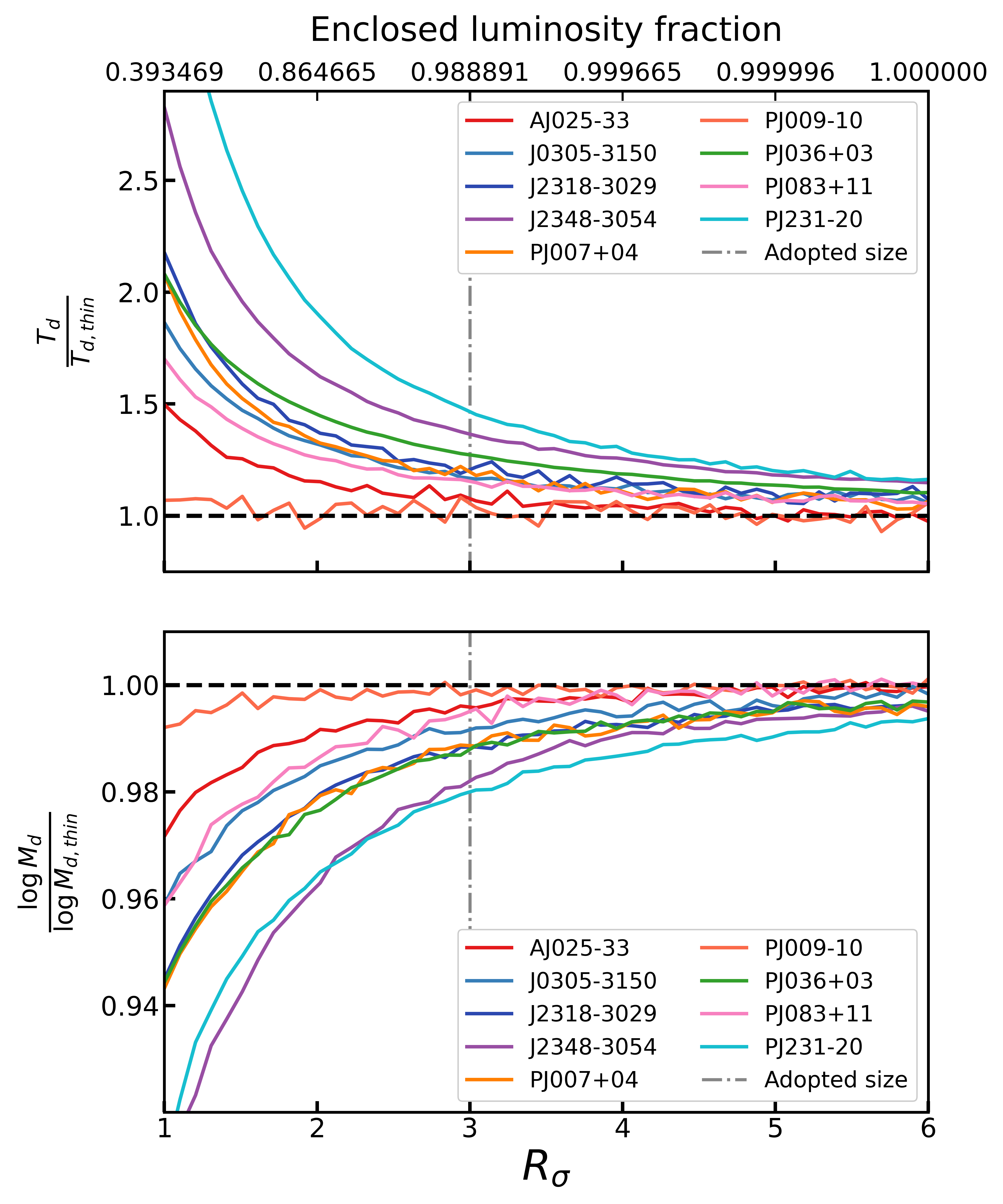}
    \caption{Median values of the $T_d$ and $M_d$ posterior probability distribution as functions of the assumed size, normalized to the optically thin results. The dashed line marks the asymptotic limit of 1 expected for low optical depth. The upper x-axis shows the fractional luminosity enclosed within the corresponding $\sigma$ distance from the Gaussian peak.}
    \label{fig:approx}
\end{figure}

Given that the compact size of the objects appears to be the primary driver of their optical thickness, and considering that our fits rely entirely on a single size measurement from ALMA Band 6, we decided to investigate how changing the definition of the angular size of the source $\Omega$ impacts our results. In Figure \ref{fig:approx}, we plot the ratios $T_d / T_{d,\hspace{0.75mm}thin}$ and $\log M_{d} / \log M_{d, \hspace{0.75mm} thin}$ as a function of the assumed source size expressed in units of $\sigma$ of the fitted Gaussian ($R_\sigma$). The subscript $thin$ refers to the quantities derived by running again the fit under the optically thin MBB emission approximation of equation \ref{eq:mbb}:
\begin{align}
        S(\nu_{obs})_{thin}= \dfrac{(1+z)}{D_L^2}[B(T_d(z),\nu)-B(T_{CMB}(z),\nu)]\cdot k(\nu)M_d
    \label{eq:mbbt}
\end{align}
Under this assumption, this equation loses the dependency on the solid angle $\Omega$, so at high-z it is often used when spatially resolved observations are not available.
We remind that in our general analysis we defined the source size as that of an ellipse encompassing 98.9\% (3$\sigma$) of the total flux (volume) of the Gaussian. These plots show that, as expected, increasing the assumed size causes the quantities derived in the general case to asymptotically approach those obtained under the optically thin approximation. This is because the progressively lower optical depth increasingly justifies the optically thin assumption. This is also noticeable in the case of PJ009-10, for which the optically thin case seems to yield reliable results. Additionally, we find that this approximation systematically underestimates the dust temperature and overestimates the dust mass, both by approximately 40\%. In contrast, the emissivity index, although it is still converging to $\beta_{thin}$ at large values of $R_{\sigma}$, does not display a clear trend as the assumed size increases, therefore we do not not include it in Figure \ref{fig:approx}.

\subsection{On the dust temperature in QSOs and galaxies across redshift}

\begin{figure*}[!b]  
\centering
\begin{minipage}{0.75\textwidth}
    \includegraphics[width=\linewidth]{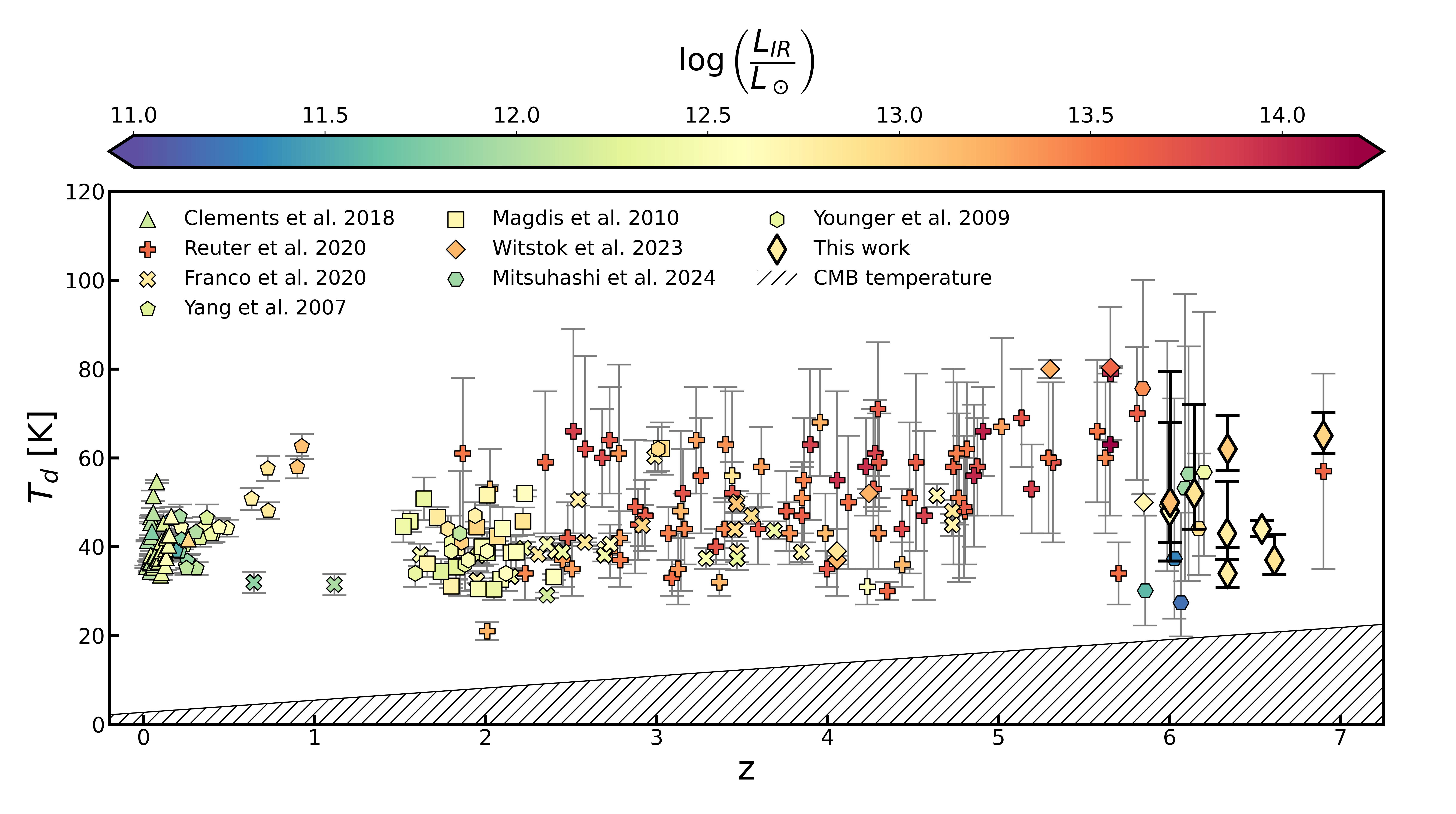}
\end{minipage}%
\hfill
\begin{minipage}{0.25\textwidth}
    \captionof{figure}{Dust temperature versus redshift in a sample of LIRGs, ULIRGs and HLIRGs, and in this work's QSOs. The dashed area corresponds to the CMB temperature. Data from 
    \citeauthor{2007ApJ...660.1198Y} \citeyear{2007ApJ...660.1198Y},
    \citeauthor{2009MNRAS.394.1685Y} \citeyear{2009MNRAS.394.1685Y},
    \citeauthor{2010MNRAS.409...22M} \citeyear{2010MNRAS.409...22M},
    \citeauthor{2018MNRAS.475.2097C} \citeyear{2018MNRAS.475.2097C},
    \citeauthor{2020A&A...643A..30F} \citeyear{2020A&A...643A..30F},
    \citeauthor{2020ApJ...902...78R} \citeyear{2020ApJ...902...78R},
    \citeauthor{2023MNRAS.523.3119W} \citeyear{2023MNRAS.523.3119W},
    \citeauthor{2024ApJ...971..161M} \citeyear{2024ApJ...971..161M}.}
    \label{fig:dtevol}
\end{minipage}
\end{figure*}

As mentioned in the previous section, we find that the temperature of the dust in our sample ranges from 34 K to 65 K, values that are higher than the ones derived for local, main sequence galaxies. 
As a matter of fact, in the last few years several works have suggested an evolution with redshift of this quantity in UV selected star-forming galaxies \citep[see, e.g.][]{2018A&A...609A..30S, 2020ApJ...902..112B, 2020MNRAS.498.4192F, 2023A&A...678A..27I}. These studies however, while agreeing on the existence of the trend, appear to be quite discordant in describing its behavior at redshift larger than 4. For example, in \citeauthor{2020ApJ...902..112B} \citeyear{2020ApJ...902..112B} the Authors find that $z$ and $T_d$ can be linked by a linear relation up to $z \simeq 7.5$, while in \citeauthor{2020MNRAS.498.4192F} \citeyear{2020MNRAS.498.4192F} they find that the relation seems to flatten at $z > 4$.
While both of the behaviors have not been fully explained by a complete physical model yet, theoretical works suggested that the increase of dust temperature with redshift could be related to the higher values of the sSFR or of the SFE in galaxies at high-z \citep[see, e.g.][]{2022MNRAS.510.5560S, 2022MNRAS.513.3122S}, which may translate in a dust distribution that is more concentrated in star-forming regions \citep[see, e.g.][]{2019MNRAS.489.1397L}. \\

\indent However, if we only take into consideration luminous, ultraluminous, and hyperluminous infrared galaxies (LIRGs, ULIRGs, and HLIRGs, respectively) that have comparable infrared luminosities to the quasar hosts in our sample, the picture is different. In Figure \ref{fig:dtevol}, we plot dust temperature versus redshift for both the sources analyzed in this work and a sample of IR-luminous, starburst galaxies selected to have $L_{\mathrm{IR}} > 10^{11}\ L_{\odot}$. For this galaxy sample, we find that the dust temperature appears to remain constant across all redshifts, while also being comparable to the values we derive for the QSO hosts in this study.
This may seem like a trivial result, as we selected these galaxies because of their high IR luminosities, and we have already discussed how $T_{\mathrm{d}}$ is an important driver of $L_{\mathrm{IR}}$. However, the evolution of dust temperature still lacks survey-like studies capable of accounting for observational biases that favor the selection of UV-brighter systems, which may exhibit hotter dust due to increased UV absorption and grain reprocessing, or even colder dust, if these objects appear UV bright because of the fact that radiation in these systems is less attenuated (and thus less reprocessed) by dust grains. \\

Despite this limitation, in \citeauthor{2022MNRAS.517.5930S}\citeyear{2022MNRAS.517.5930S}, \citeyear{2022MNRAS.513.3122S} the Authors attempt to extend the study of dust temperature across cosmic time by including single-band $T_{\mathrm{d}}$ estimates that exploit the [CII]$_{158\mu m}$ line luminosity and the underlying continuum flux for both ALPINE and REBELS galaxies, obtaining an average value of $T_d=48\pm2$ K and $T_d=47\pm2$ K respectively. The objects targeted by these programs are mostly main-sequence galaxies, which nevertheless show higher $T_d$ compared to their local counterparts. This effect is likely linked to the evolution of the $M_{\ast}$–SFR main sequence itself, as also discussed in \citeauthor{2018A&A...609A..30S} \citeyear{2018A&A...609A..30S}, which makes these galaxies progressively more similar to those we plot in Figure \ref{fig:dtevol} as redshift increases.
It is of particular interest that the $T_d$ they derive is not different to the one of $6<z<7$ QSO hosts, suggesting that the processes that heat the dust are similar in all of these objects.

\subsection{The [CII]$_{158\mu m}$ deficit}
When it comes to sources with such a high IR luminosity as the one presented in  this paper, several works \citep[see, e.g,][]{2013ApJ...774...68D, 2017ApJ...846...32D, 2018ApJ...854...97D} have found that multiple FIR fine-structure lines, such as the [CII]$_{158\mu m}$, [NII]$_{122\mu m}$ or the [OI]$_{63\mu m}$ show a decrease, or "deficit", in the ratio between their luminosity and the IR  luminosity of the source as this last quantity increases.
While the physical motivation of this phenomenon can not be ascribed to a single factor, one possible explanation was suggested to be a different spatial distribution of dust, with it being more concentrated inside the star-forming regions in higher luminosity systems \citep[see, e.g.][]{2013ApJ...774...68D}. In this scenario, a larger fraction of UV photons emitted by stars is absorbed by dust within the HII region, efficiently heating the dust grains while simultaneously reducing the UV photon flux that reaches the photodissociation region (PDR). This leads to a deficit in photoelectrons produced by the ionization of polycyclic aromatic hydrocarbons (PAHs) and dust grains, thus affecting the emission of lines such as [CII]$_{158,\mu m}$. This combination of effects can both enhance the infrared luminosity emitted by the dust, and suppress the [CII]$_{158\mu m}$ line emission . The former of the two processes would also explain why the observed line deficits concern lines that trace different ISM phases, or which emission integrated along the line of sight includes contributions from more than one of these phases \citep[see, e.g.][]{2017ApJ...845...96C, 2023A&A...679A.131R}. \\

\begin{figure}[h!]
    \centering
    \includegraphics[width=\linewidth]{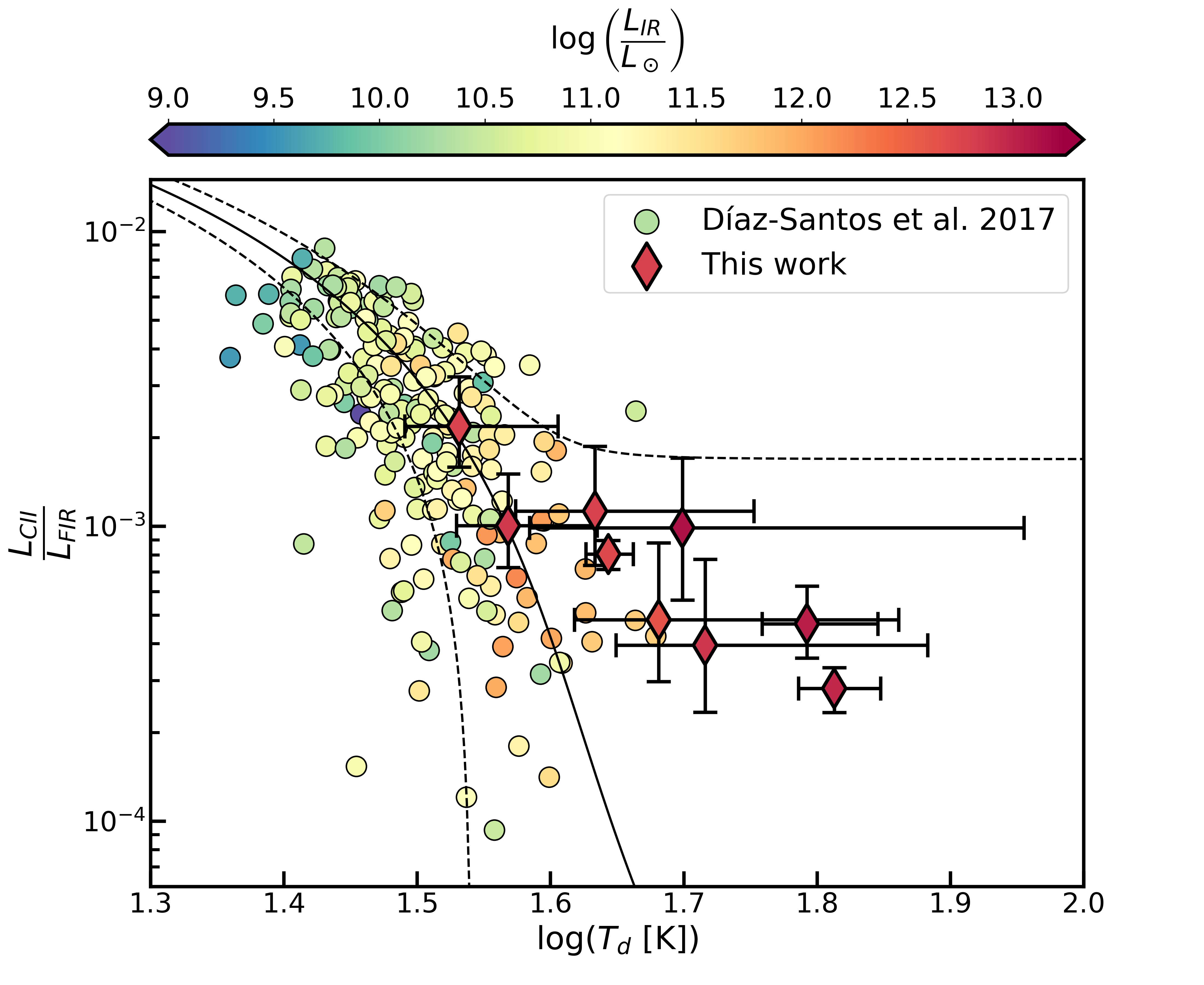}
    \caption{[CII]$_{158\mu\mathrm{m}}$ line luminosity deficit as function of dust temperature for the sample of \citeauthor{2017ApJ...846...32D} \citeyear{2017ApJ...846...32D} and for the QSOs of this work. In the former sample dust temperature is derived by the color index S$_{63\mu\mathrm{m}}$/S$_{158\mu\mathrm{m}}$, assuming MBB optically thin emission with $\beta=1.8$.  The black dashed line is the best-fit from \citeauthor{2017ApJ...846...32D} \citeyear{2017ApJ...846...32D}, while the two dashed lines represent the 1 $\sigma$ scatter from that relation. }
    \label{fig:deficit}
\end{figure}

In Figure \ref{fig:deficit}, we plot the ratio of [CII]$_{158\mu m}$ to FIR luminosity against dust temperature for a sample of local IR galaxies  from \citeauthor{2017ApJ...846...32D} \citeyear{2017ApJ...846...32D} and the QSOs in this work. This provides a direct link between the previously described scenario and the physical origin of the [CII]$_{158\mu m}$ line deficit, with high dust temperature tracing the compact, dust-bounded nature of the star-forming regions responsible for the enhanced IR luminosity. While it is true that, for fixed values of $M_d$, GDR, and physical properties of the [CII]]$_{158\mu m}$-emitting gas, a higher $T_d$ leads to a more pronounced [CII]]$_{158\mu m}$ deficit by definition, it is important to point out that the physical conditions of the gas are not independent of $T_d$. Several studies have shown that dust temperature correlates with the kinetic temperature of the [CII] colliders (typically electrons in ionized regions and neutral atoms or H$_2$ molecules in denser phases, see e.g. \citeauthor{2024A&A...687A.207Z} \citeyear{2024A&A...687A.207Z} for an example), hence altering the excitation temperature of [CII]. This coupling can enhance the line emissivity and thereby partly compensate for the [CII]]$_{158\mu m}$ deficit.
The behavior of this work's QSOs seems to agree with the empirical trend fitted in \citeauthor{2017ApJ...846...32D} \citeyear{2017ApJ...846...32D}, especially for lower values of $T_d$, while for larger dust temperatures these source effectively probe a regime of that was previously not accessed.

\subsection{Contribution of the central AGN to surrounding dust heating}
It is possible to argue that the presence of a very bright active galactic nucleus may heat the innermost-located dust of the host galaxy, which emission could contaminate our unresolved Band 8 flux measurements. This raises the concern that the values obtained from the SED fitting for this quantity may not accurately represent the one at which the dust grains in the host are at the thermodynamical equilibrium, leaving open the possibility that the temperature estimates in our sources could be biased. In fact, the presence of a temperature gradient peaking in the galaxy center has been hinted by the results obtained for the lensed QSO J0439-1634. However, the contribution of this gradient to the rest of the sample remains unconstrained. \\

Nevertheless, in Figure \ref{fig:dtevol} we can notice that the QSOs of this work appear to follow the trend in dust temperature against redshift that is observed in LIRGs, ULIRGs and HLIRGs, which is not what would be expected in the case of AGN contamination. Additionally, given that 407 GHz observations seem to not probe frequencies higher than the emission peak, and that a single temperature MBB seems to fit well the observed SED in most of the sources, the measured contribution of the AGN-heated dust component may be negligible \citep[see also][]{2025MNRAS.540.3693S}. However, to fully pin down the role that this factor plays in the dust properties derived here, high-resolution ALMA Band 9 or Band 10 observations may be needed: these would be able to sample the SED in the rest-frame mid-infrared (MIR) band, where an eventual central hot dust ($T_d > 60 K$) is expected to peak, and to disentangle it from the more diffused emission of the galaxy. Unfortunately, up to date these kind of data are available for very few sources.
For example, in \citeauthor{2023MNRAS.523.4654T} \citeyear{2023MNRAS.523.4654T}, the Authors are able to fit a MBB law to the observed emission of a QSO at $z=4.4$ by decomposing it into a galaxy-scale component and a component originating from the innermost beam of resolution (720 pc of effective radius).
They find a temperature of the cold dust component (associated to the galaxy-scale emission) of $53^{\smash{+10}}_{\smash{-11}}$ K, and of $87^{\smash{+34}}_{\smash{-18}}$ K for the hot one (which is associated to AGN heated dust, with possible contribution from the dusty torus). In \citeauthor{2025A&A...695L..18M} \citeyear{2025A&A...695L..18M} the source J2348-3054 is analyzed thanks to new spatially resolved observation, which reveal a steep gradient in $T_d$ peaking at $88^{\smash{+2}}_{\smash{-2}}$ K in the innermost region of 216 pc radius. However, when they include a torus model, this value is slightly reduced to $72^{\smash{+2}}_{\smash{-1}}$ K, suggesting the presence of a component of hot dust heated by AGN radiation that is physically distinct from the torus. Dust temperature gradients at high redshift have also been observed in galaxies that do not host an AGN. In \citeauthor{2022ApJ...934...64A} \citeyear{2022ApJ...934...64A}, for example, the authors fit the FIR emission of A1689-zD1, a strongly lensed ULIRG at $z = 7.13$ with a magnification factor of $\mu = 9.3$, which remains nearly constant across the entire spatial extent of the source ($\sim 7$ kpc). They find a galaxy-averaged dust temperature of $T_d = 41^{\smash{+17}}_{\smash{-14}}$ K, while in the central 900 pc the temperature reaches approximately 50 K. Both values are consistent, within the uncertainties, with the average dust temperatures derived in this work for QSO host galaxies. In this context, the temperature gradient in A1689-zD1 likely reflects a higher fraction of dust heated by intense central star formation.

\section{Conclusions}
In this paper we presented a sample of 11 QSO host galaxies for which we had new ALMA ACA Band 8 observations. For 10 of them we have been able to constrain their interstellar dust physical properties, as well as infrared luminosities and star formation rates. The main results can be summarized as following:
\begin{itemize}[itemsep=0.5em]
    \item The SED fitting of all the IR emission of the sources confirmed the extreme starbursty nature of QSOs host galaxies, which is likely driven by the short gas depletion times.
    \item The high value of their optical depths at 1900 GHz, as well as the behavior of their $\Sigma_{SFR}$ as function of $T_d$, revealed that the large majority of their energy is radiated at frequencies where the system is optically thick. This discourages the usage of the commonly assumed optically thin approximation when fitting their IR emission.
    \item The values of the dust temperature here derived align quite well with the one found for LIRGs, ULIRGs and HLIRGs across different cosmic times. The physical explanation of this is likely to be found in the short depletion times (large SFEs) that these sources were already confirmed to have. This scenario is consistent with a larger fraction of dust being situated in SF regions with respect to local star-forming galaxies.
    \item The dusty nature of SF region seems also to be confirmed by the link between the [CII]$_{158\mu m}$ deficit and the dust temperature, with this work's QSOs sampling a regime that was previously not probed.
    \item We find no strong evidence for the AGN to impact the results obtained by unresolved SED fitting, however to fully assess the role it plays in the energetic of the dust emission further high-frequency, high-resolution ALMA observations are needed.
\end{itemize}

{\tiny
\paragraph{\textit{\tiny Acknowledgments}:}
M.C., R.D., and F.X. acknowledge support from the INAF GO grant 2022 ''The birth of the giants: JWST sheds light on the build-up of quasars at cosmic dawn'', INAF Minigrant 2024 ''The interstellar medium at high redshift'', and by the PRIN MUR ''2022935STW'', RFF M4.C2.1.1, CUP J53D23001570006 and C53D23000950006. R.A.M. acknowledges support from the Swiss National Science Foundation (SNSF) through project grant 200020\_207349. A.P. acknowledges the support from the Fondazione Cariplo grant ''2020-0902''. This paper makes use of the following ALMA data: ADS/JAO.ALMA\#2019.2.00053.S, ADS/JAO.ALMA\#2019.1.00147.S, ADS/JAO.ALMA\#2021.2.00064.S, ADS/JAO.ALMA\#2021.2.00151.S, ADS/JAO.ALMA\#2022.1.00321.S, ADS/JAO.ALMA\#2023.1.01450.S, and ADS/JAO.ALMA\#2024.1.00071.S. ALMA is a partnership of ESO (representing its member states), NSF (USA) and NINS (Japan), together with NRC (Canada), NSTC and ASIAA (Taiwan), and KASI (Republic of Korea), in cooperation with the Republic of Chile. The Joint ALMA Observatory is operated by ESO, AUI/NRAO and NAOJ. This work is based on observations carried out under project number S24CH with the IRAM NOEMA Interferometer. IRAM is supported by INSU/CNRS (France), MPG (Germany) and IGN (Spain).}

\bibliographystyle{aa} 
\bibliography{bibliography} 

\onecolumn
\appendix
\renewcommand{\arraystretch}{1.125}
\section{ALMA Band 8 observations} \label{B8obs}

\begin{table*}[h]
    \centering
    \caption{Characteristics of the ALMA Band 8 observations analyzed in this work.}
    \begin{tabularx}{\textwidth}{ZZZZZ}
         \midrule
         \midrule
         ID & N. ant. & \makecell[c]{Beam size \\ (arcsec)} & \makecell[c]{Rms \\ (mJy/beam)} & \makecell[c]{S(407 GHz) \\ (mJy)} \\
         \midrule
         AJ025-33 & 11-13 & 1.96 $\times$ 1.62 & 0.16 & 6.8 $\pm$ 0.7  \\ 
         J0305-3150 & 8 & 4.44 $\times$ 2.75 & 0.95 & 10.4 $\pm$ 1.5\\
         J0439+1634* & 14 & 2.53 $\times$ 1.38 & 1.0 & 45.4 $\pm$ 4.7\\
         J2318-3029 & 10 & 3.96 $\times$ 2.62 & 0.65  & 6.9 $\pm$ 1.0\\
         J2348–3054 & 8 & 4.31 $\times$ 2.59 & 0.44 & 5.5 $\pm$ 0.7 \\
         PJ007+04 & 15 & 1.91 $\times$ 1.51 & 0.19 & 4.8 $\pm$ 0.6 \\
         PJ009–10 & 8 & 4.14 $\times$ 2.65 & 0.80 & 8.6 $\pm$ 1.2 \\
         PJ036+03 & 14 & 1.92 $\times$ 1.46 & 0.24 &  6.2 $\pm$ 0.7\\
         PJ083+11 & 13 & 2.91 $\times$ 1.35 & 0.58 & 10.5 $\pm$ 1.5\\
         PJ158-14 & 12 & 1.80 $\times$ 1.49 & 0.45 & 9.6 $\pm$ 1.2 \\
         PJ231–20** & 10 & 4.34 $\times$ 2.92 & 0.92 & 9.6 $\pm$ 1.6\\
         \midrule
    \end{tabularx}
    \\[9pt]
    \parbox{\linewidth}{\small Notes -- columns: QSO name; number of antennas; clean beam size; root mean squared of the noise in the clean continuum map; measured flux at the observed frequency of 407 GHz. A 10\% systematic flux calibration uncertainty was added in quadrature to the nominal uncertainty. *this source is strongly lensed, **this source has an unresolved companion.}
    \label{tab:Band8}
\end{table*}

\vspace{-0.375cm}

\begin{figure*}[h!]

    \centering
    \begin{subfigure}[]{\includegraphics[width=5.8cm]{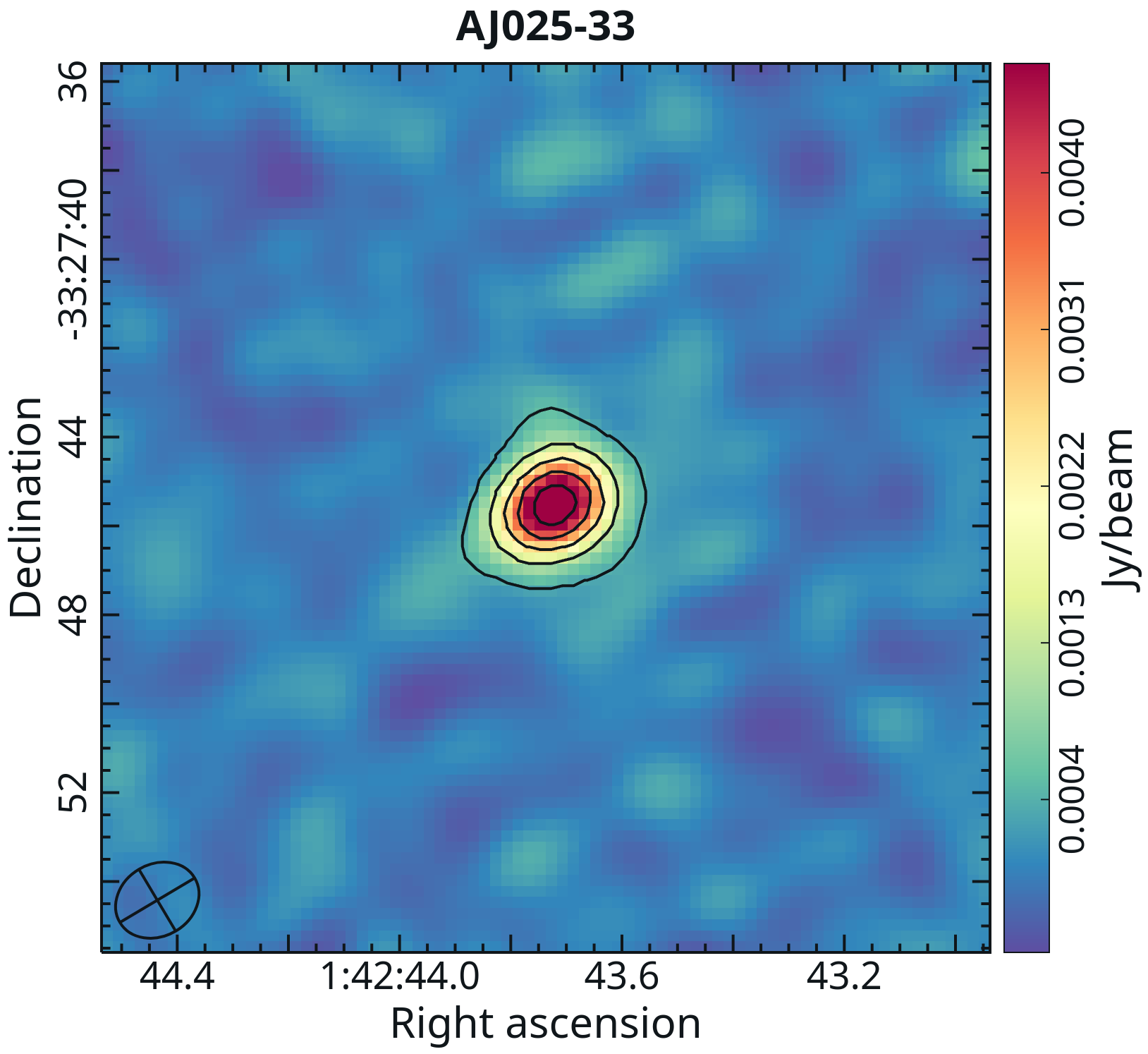}}
    \end{subfigure}
    \hspace{0.1cm}
    \begin{subfigure}[]{\includegraphics[width=5.8cm]{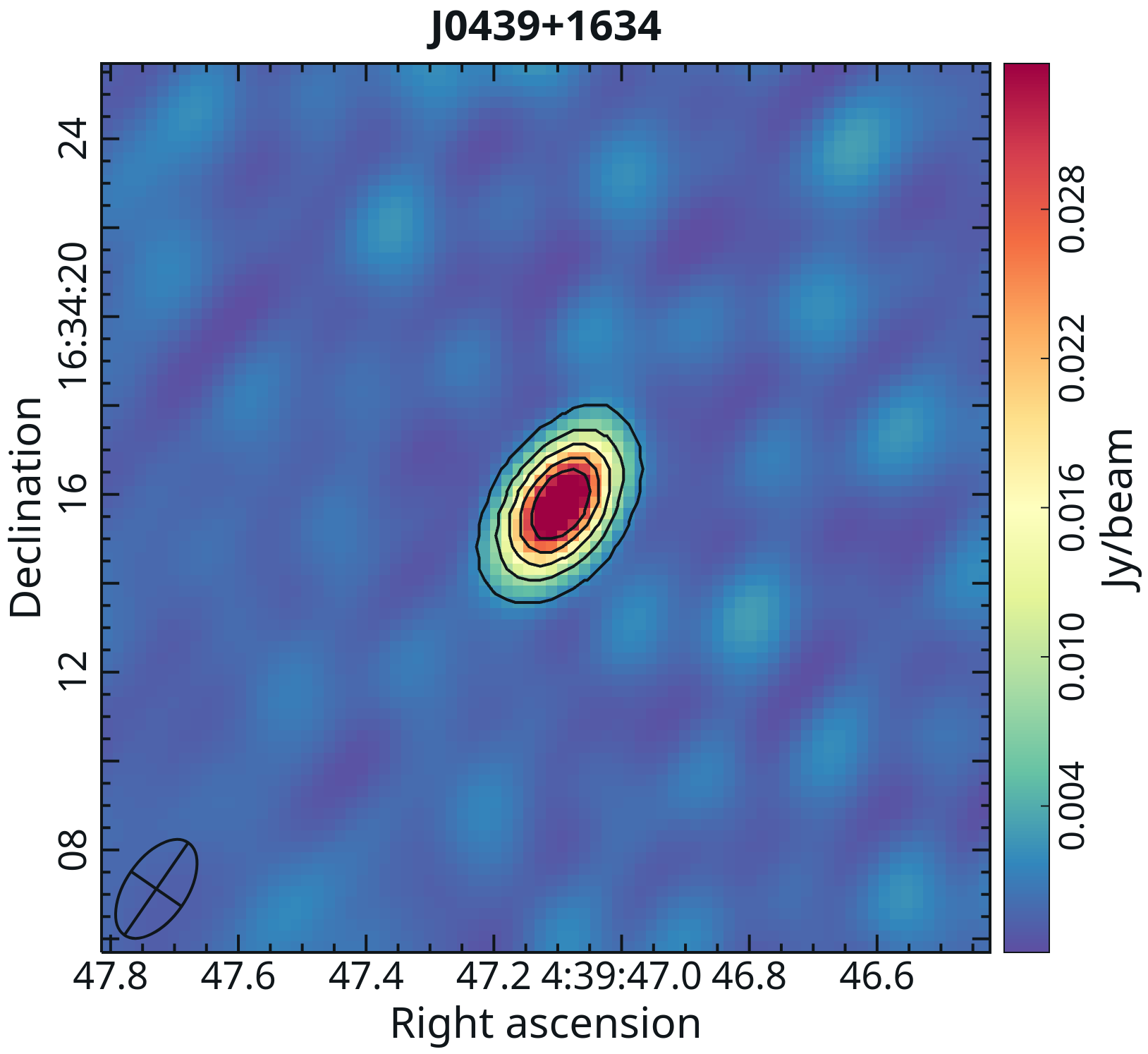}}
    \end{subfigure}
    \hspace{0.1cm}
    \begin{subfigure}[]{\includegraphics[width=5.8cm]{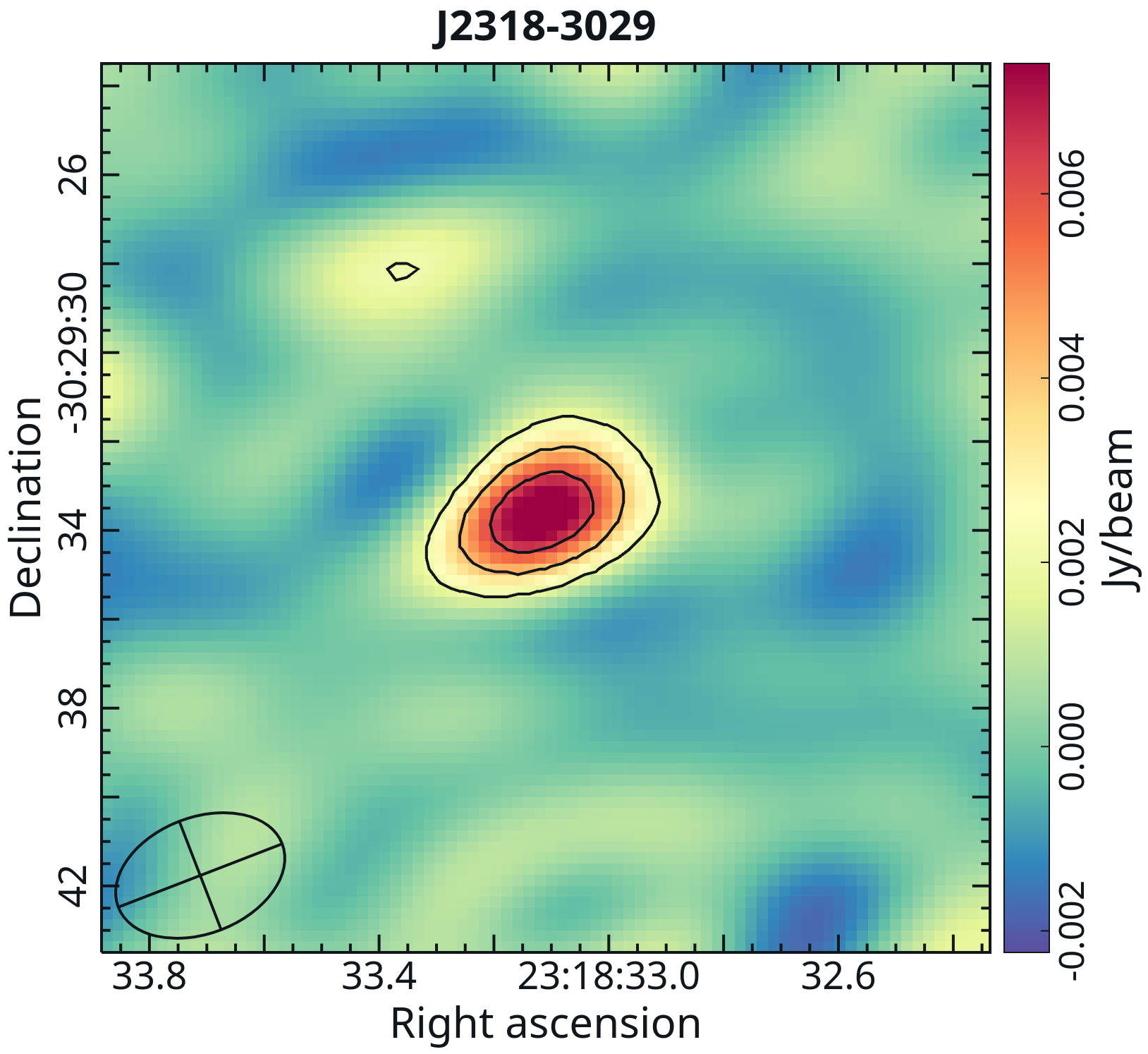}}
    \end{subfigure}
    \\
    \begin{subfigure}[]{\includegraphics[width=5.8cm]{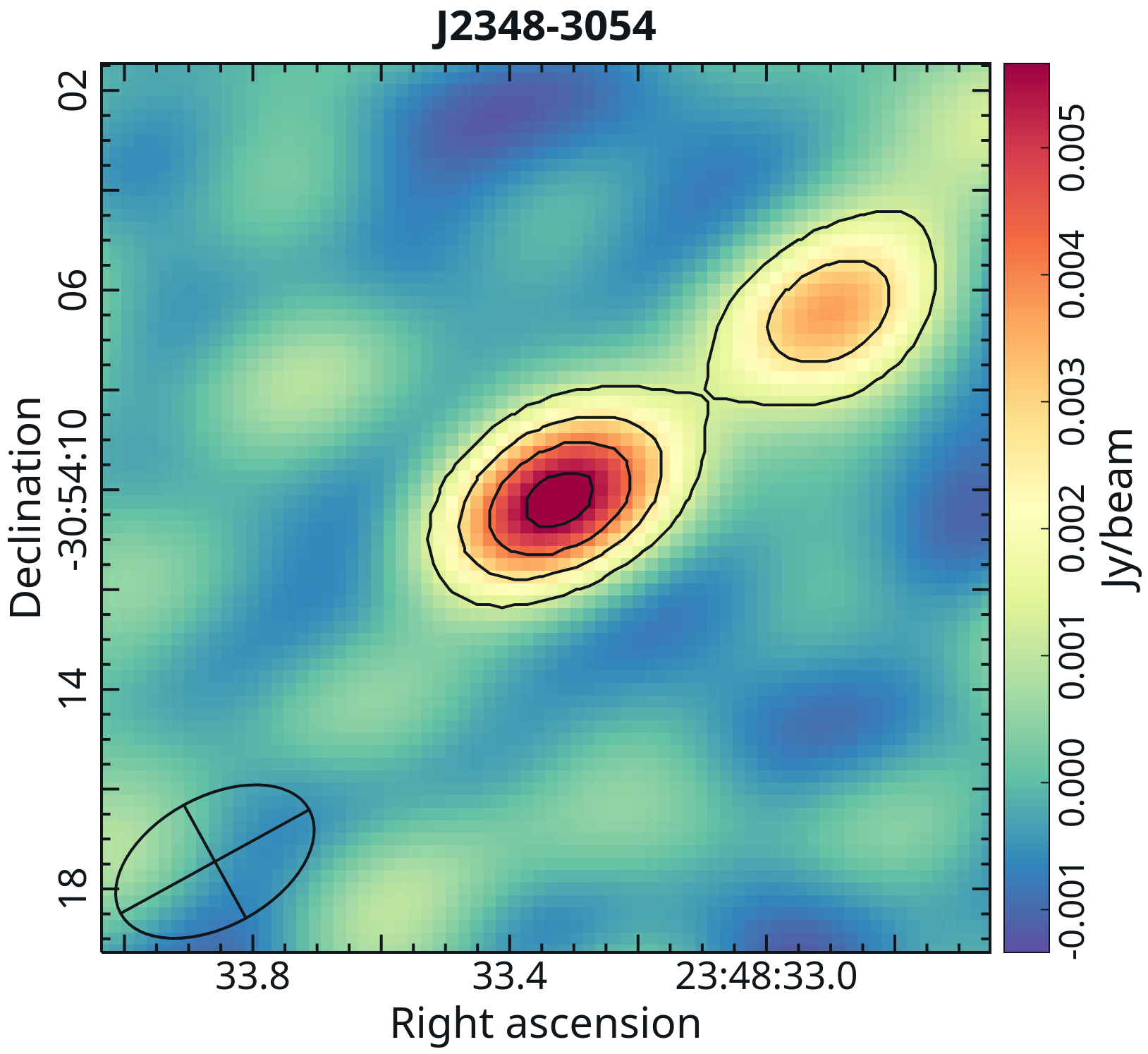}}
    \end{subfigure}
    \hspace{0.1cm}
    \begin{subfigure}[]{\includegraphics[width=5.8cm]{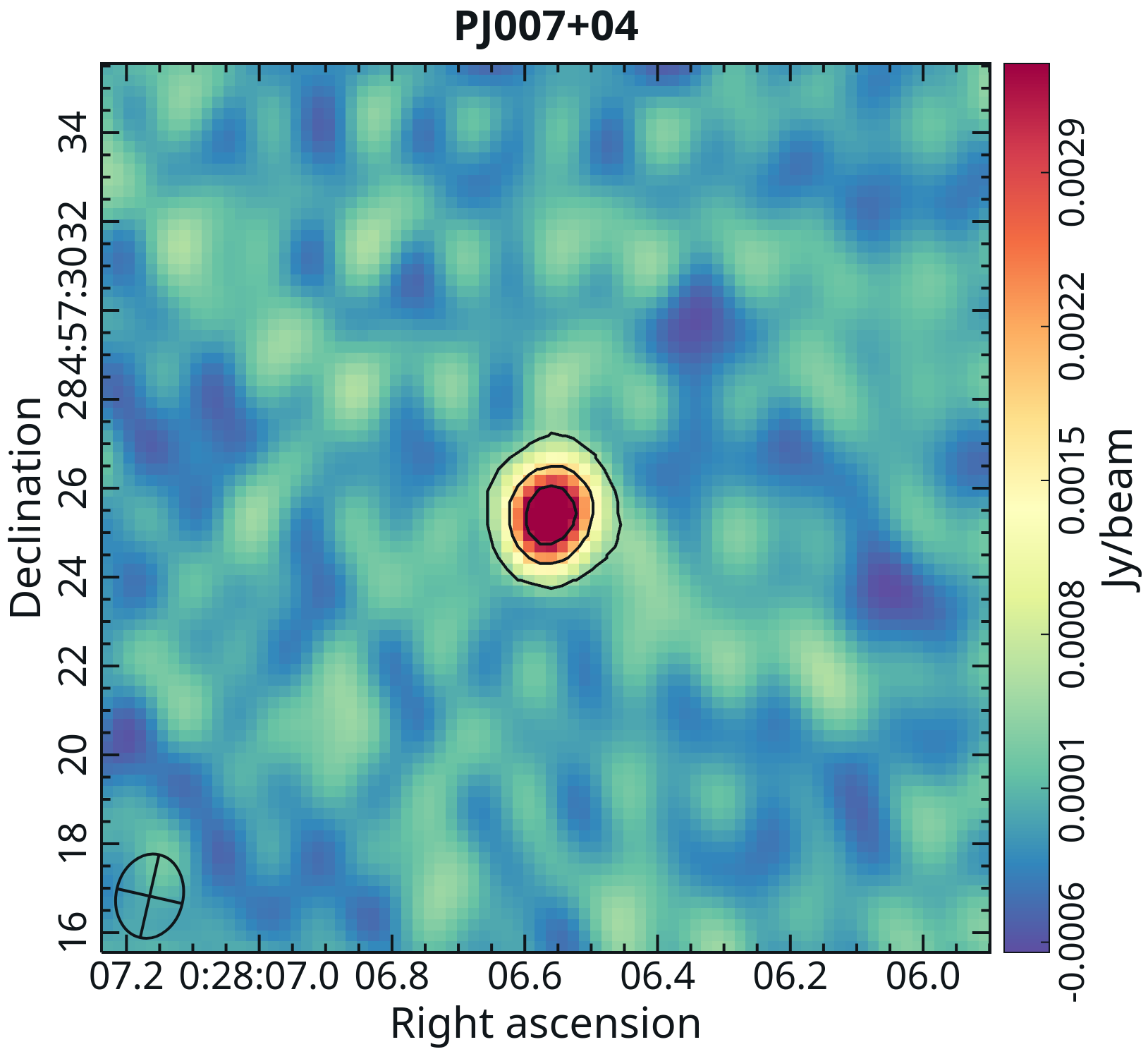}}
    \end{subfigure}
    \hspace{0.1cm}
    \begin{subfigure}[]{\includegraphics[width=5.8cm]{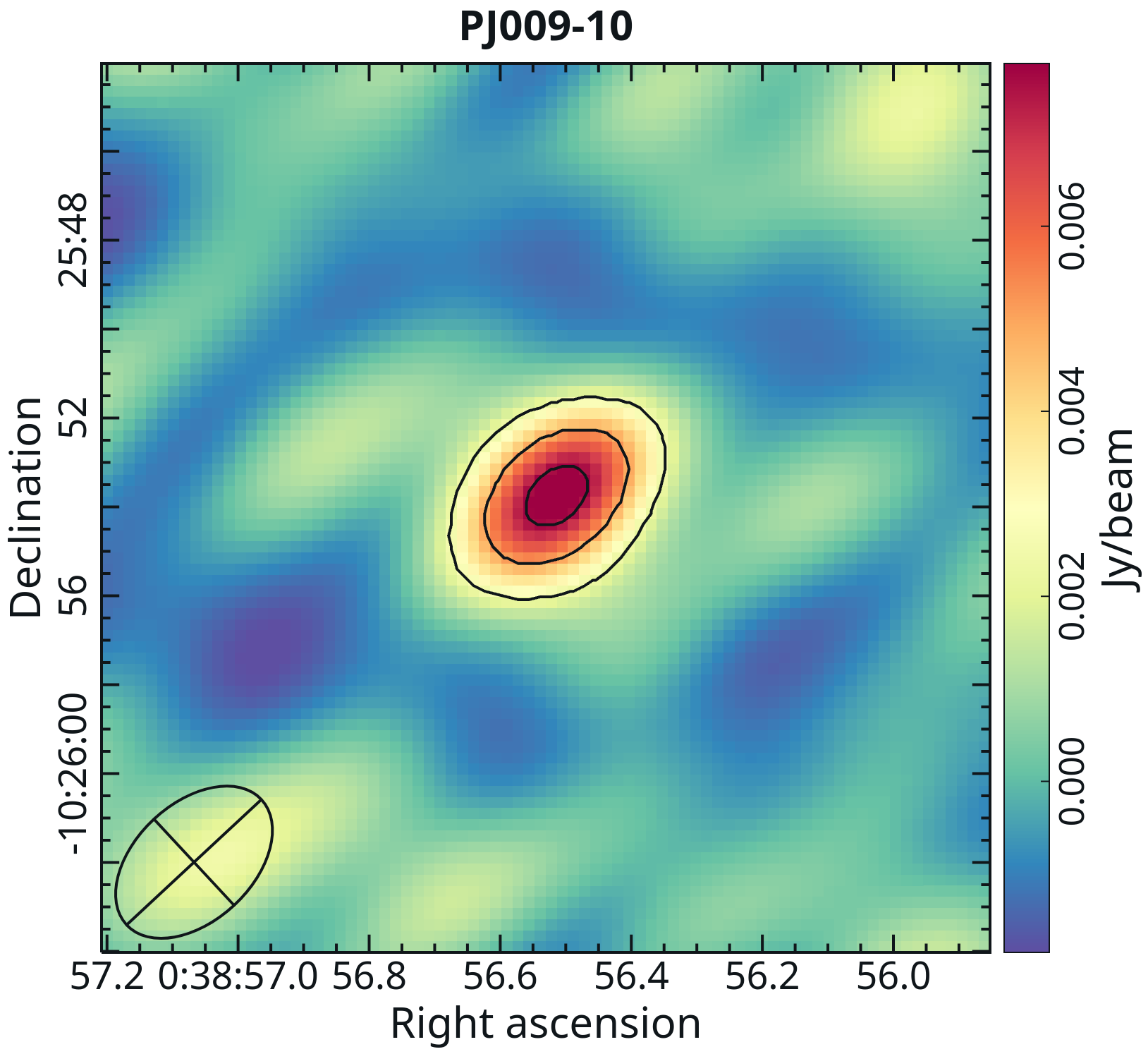}}
    \end{subfigure}
    \caption{407 GHz dust continuum maps for the sources analyzed in this work. The clean beam is reported in the bottom left corner of each panel. Panel (a): QSO AJ025-33 (contours at 3, 9, 15, 21 and 27 $\sigma$). Panel (b): QSO
    J0439+1634 (contours at 3, 9, 15, 21 and
    27 $\sigma$). Panel (c): QSO
    J2318-3029 (contours at 3, 6 and 9 $\sigma$). Panel (d): QSO
    J2348-3054 (contours at 3, 6, 9 and 12 $\sigma$). Panel (e): QSO
    PJ007+04 (contours at 3, 9 and 15 $\sigma$). Panel (f): QSO
    PJ009-10 (contours at 3, 6 and 9 $\sigma$). Panel (g): QSO
    PJ036+03 (contours at 3, 9 and 15 $\sigma$). Panel (h): QSO
    PJ083+11 (contours at 3, 6, 9 and 12 $\sigma$). Panel (i): QSO
    PJ158-14 (contours at 3, 9 and 15 $\sigma$). Panel (j): QSO
    PJ231-20 (contours at 3, 6 and 9 $\sigma$).}
    \label{fig:B8im}
\end{figure*}

\setcounter{figure}{0}

\begin{figure*}[h!]
\setcounter{subfigure}{6}
    \centering
    \begin{subfigure}[]{\includegraphics[width=5.8cm]{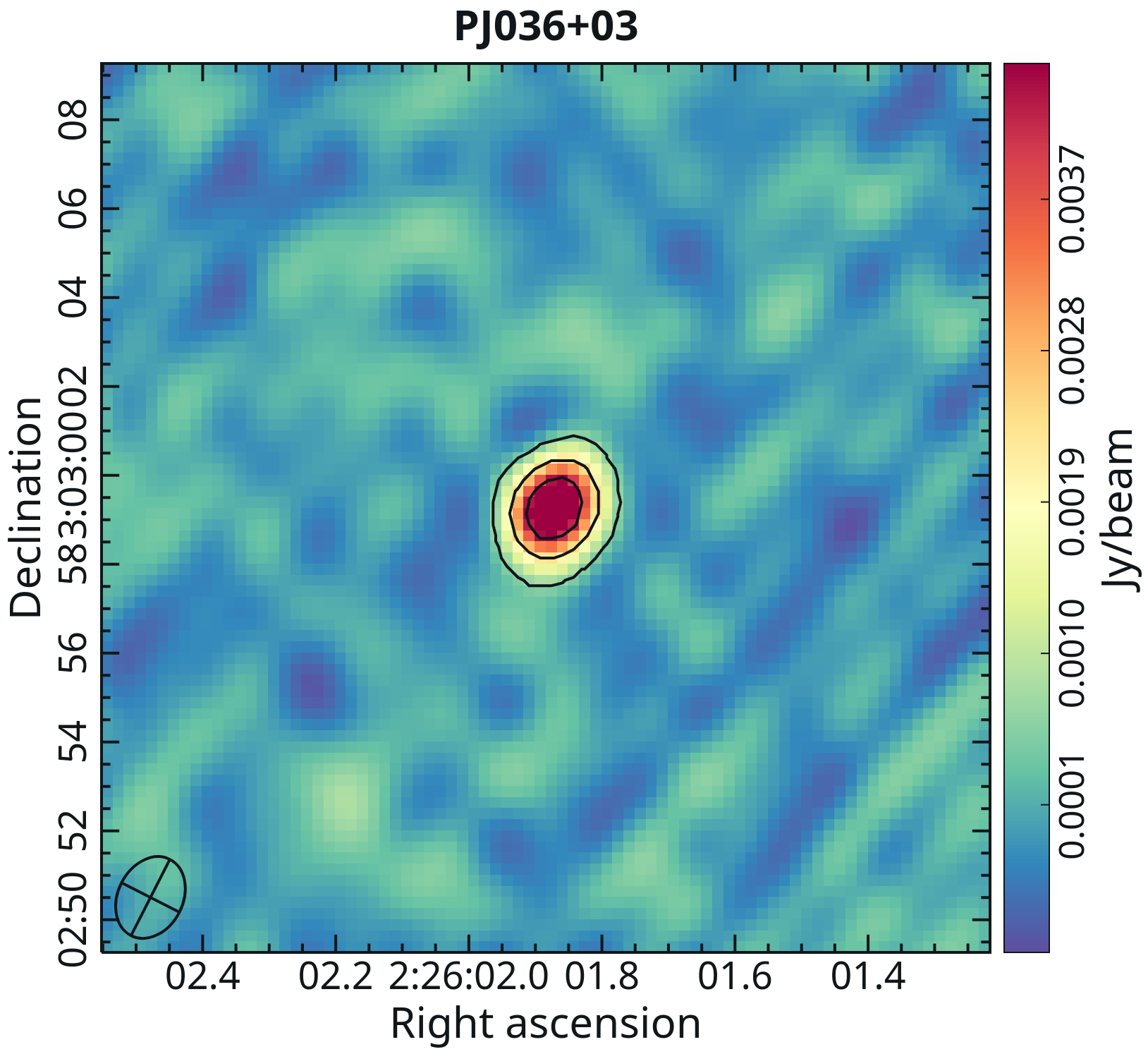}}
    \end{subfigure}
    \hspace{0.1cm}
    \begin{subfigure}[]{\includegraphics[width=5.8cm]{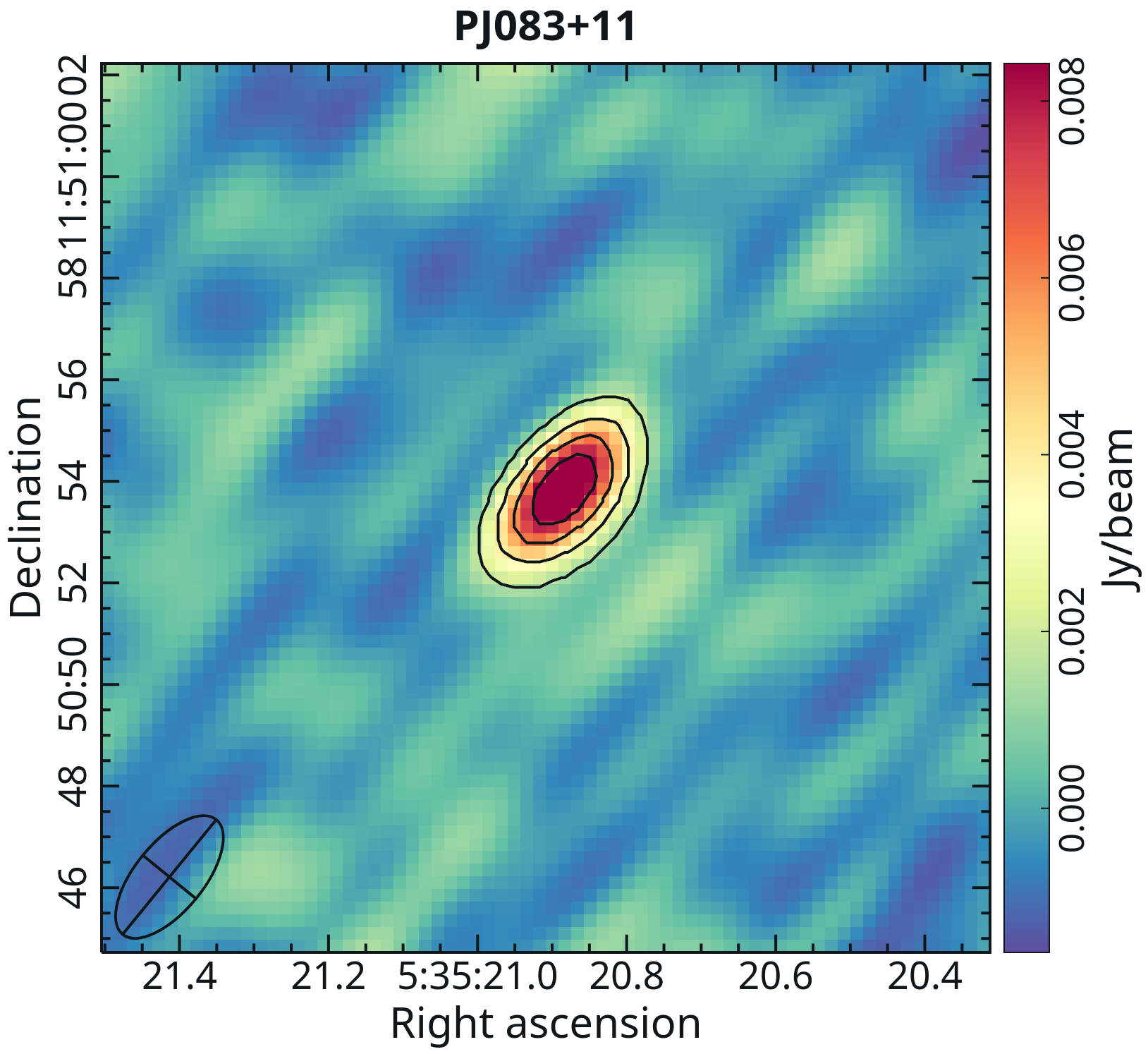}}
    \end{subfigure}
    \hspace{0.1cm}
    \begin{subfigure}[]{\includegraphics[width=5.8cm]{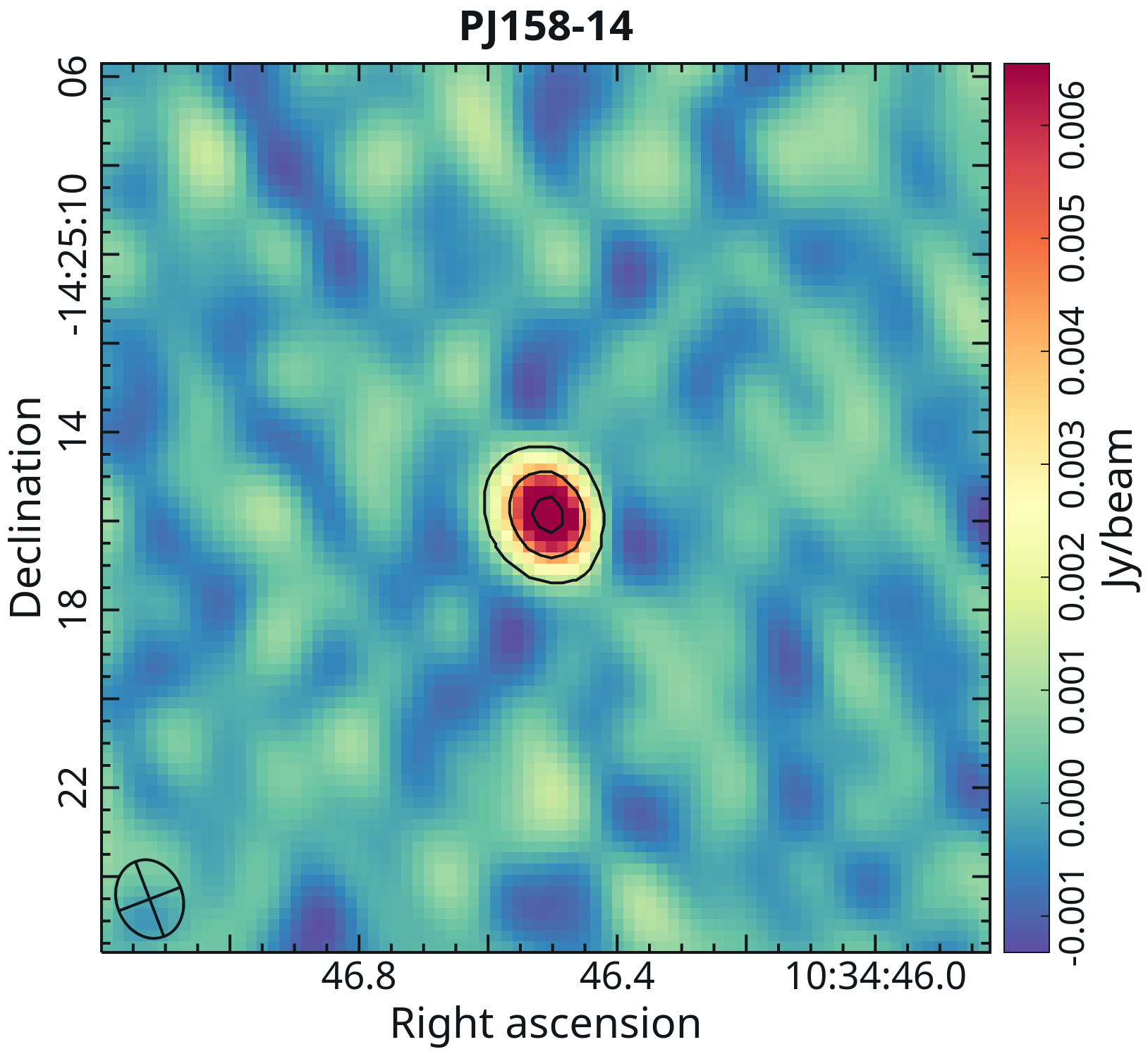}}
    \end{subfigure}
    \\
    \begin{subfigure}[]{\includegraphics[width=5.8cm]{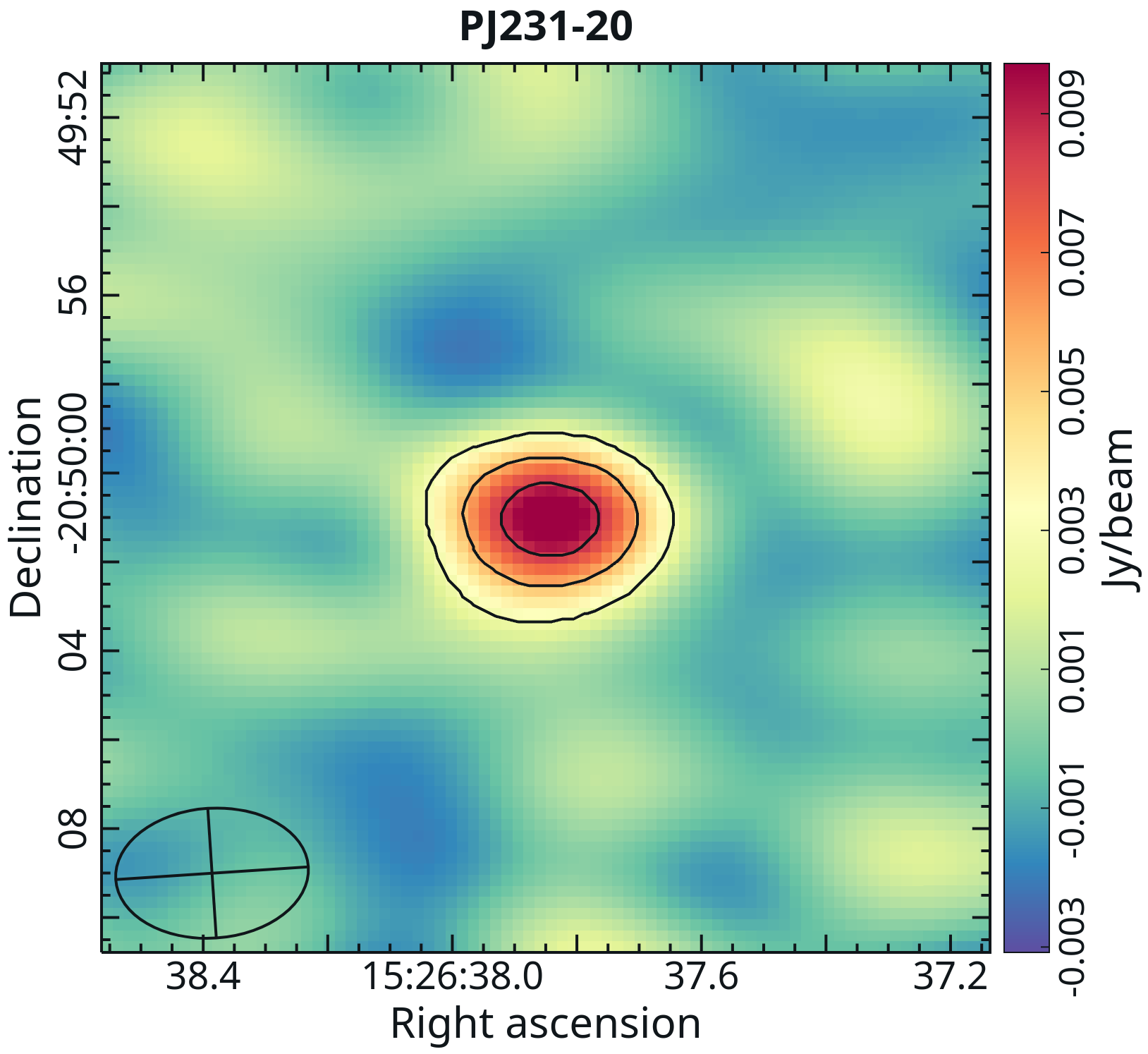}}
    \end{subfigure}
    \hspace{0.1cm}
    \phantom{
    \begin{subfigure}[]
        {\includegraphics[width=5.55cm]{Band8/PJ158-14.image-image-2024-11-28-18-17-15.png}}
    \end{subfigure}
    }
    \hspace{0.1cm}
    \phantom{
    \begin{subfigure}[]
        {\includegraphics[width=5.55cm]{Band8/PJ158-14.image-image-2024-11-28-18-17-15.png}}
    \end{subfigure}
    }
    \label{fig:B8im2}
    \caption{continued.}
\end{figure*}

\clearpage

\section{NOEMA Band 2 observations} 

\begin{table*}[h]
    \centering
    \caption{Characteristics of the NOEMA Band 2 observations analyzed in this work.}
    \begin{tabularx}{\textwidth}{ZZZZZ}
         \midrule
         \midrule
         ID & N. ant. & \makecell[c]{Beam size \\ (arcsec)} & \makecell[c]{Rms \\ (mJy/beam)} & \makecell[c]{S(158 GHz) \\ (mJy)} \\
         \midrule
         PJ007+04 & 9 & 2.91 $\times$ 2.33 & 0.030 & 0.57 $\pm$ 0.09  \\
         PJ009-10 & 10 & 3.46 $\times$ 2.27 & 0.016 & 0.67 $\pm$ 0.07  \\
         \midrule
    \end{tabularx}
    \\[9pt]
    \parbox{\linewidth}{\small Notes -- columns: QSO name; number of antennas; clean beam size; root mean squared of the noise in the clean continuum map; measured flux at the observed frequency of 158 GHz. A 10\% systematic flux calibration uncertainty was added in quadrature to the nominal uncertainty.}
    \label{tab:Noema}
\end{table*}

\begin{figure*}[h!]
    
    \centering
    \begin{subfigure}[]{\includegraphics[width=11.2cm]{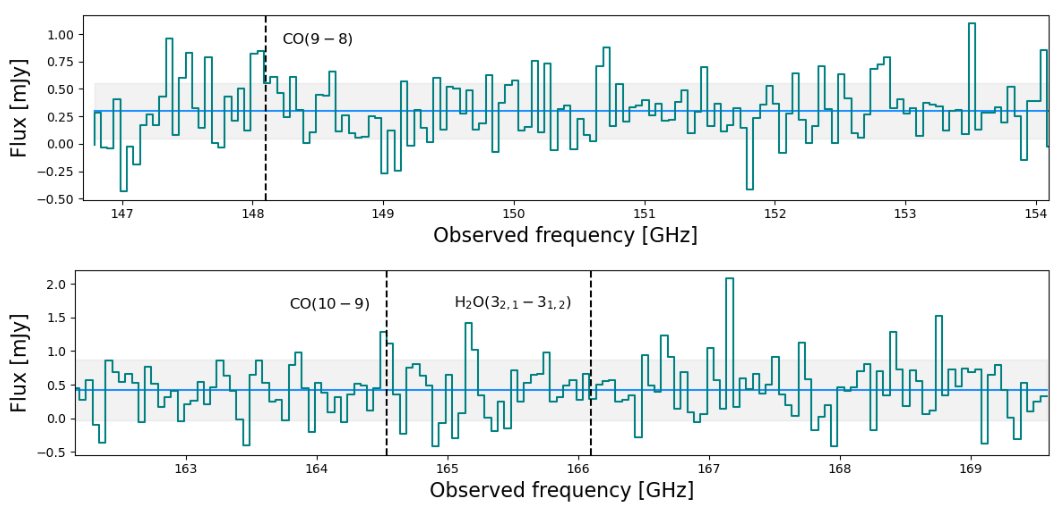}}
    \end{subfigure}
    \hfill
    \begin{subfigure}[]{\includegraphics[width=6cm]{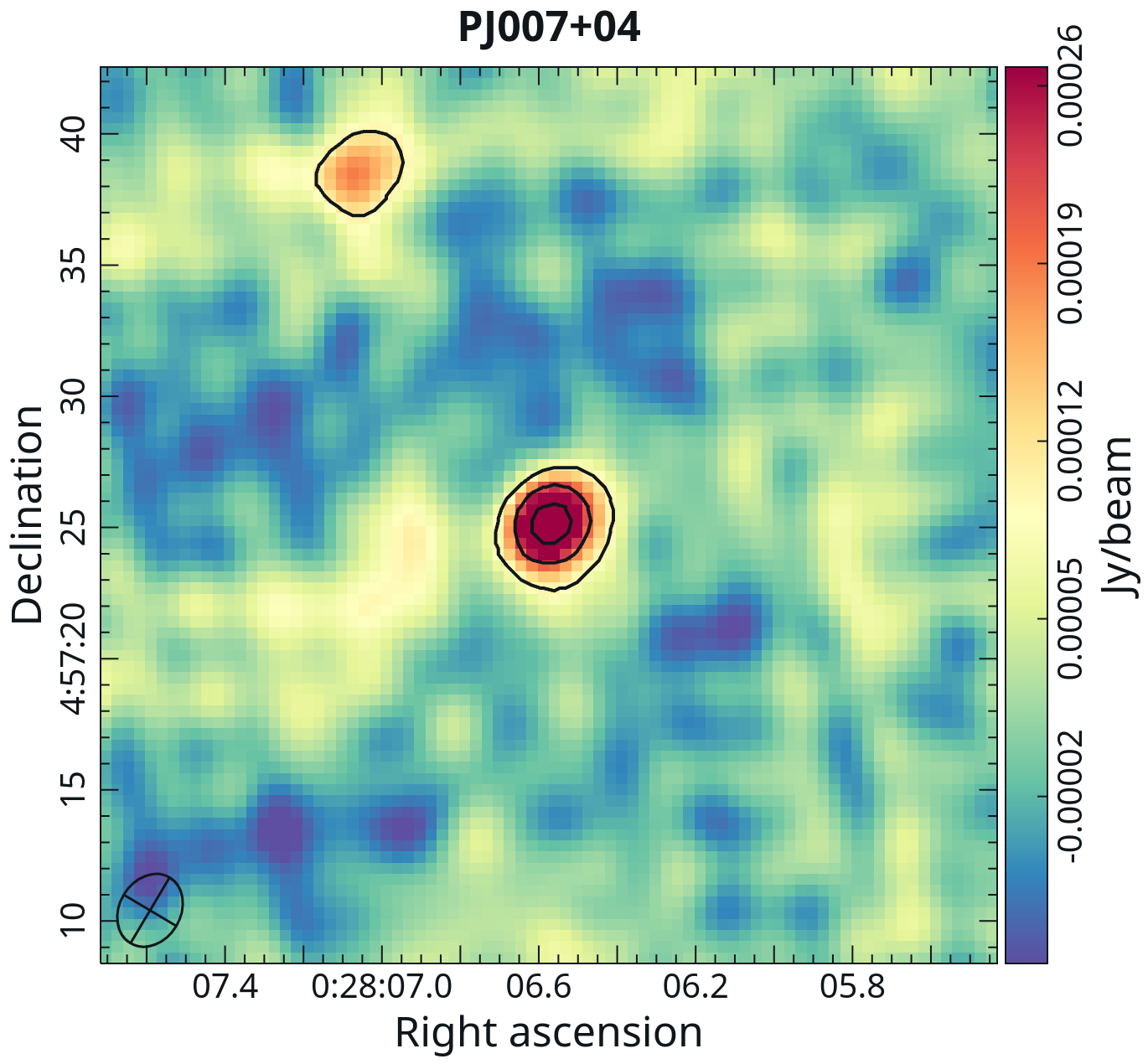}}
    \end{subfigure}

    \begin{subfigure}[]{\includegraphics[width=11.2cm]{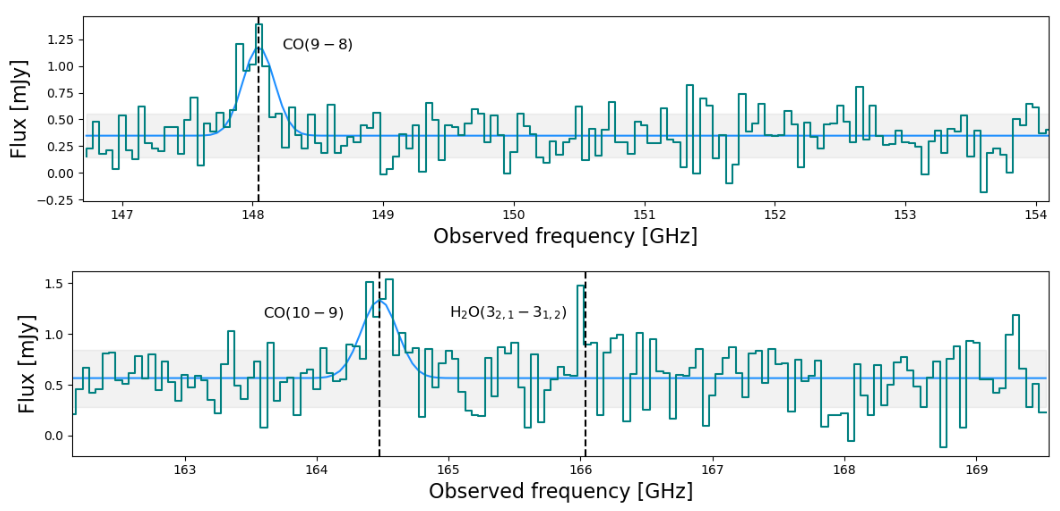}}
    \end{subfigure}
    \hfill
    \begin{subfigure}[]{\includegraphics[width=6cm]{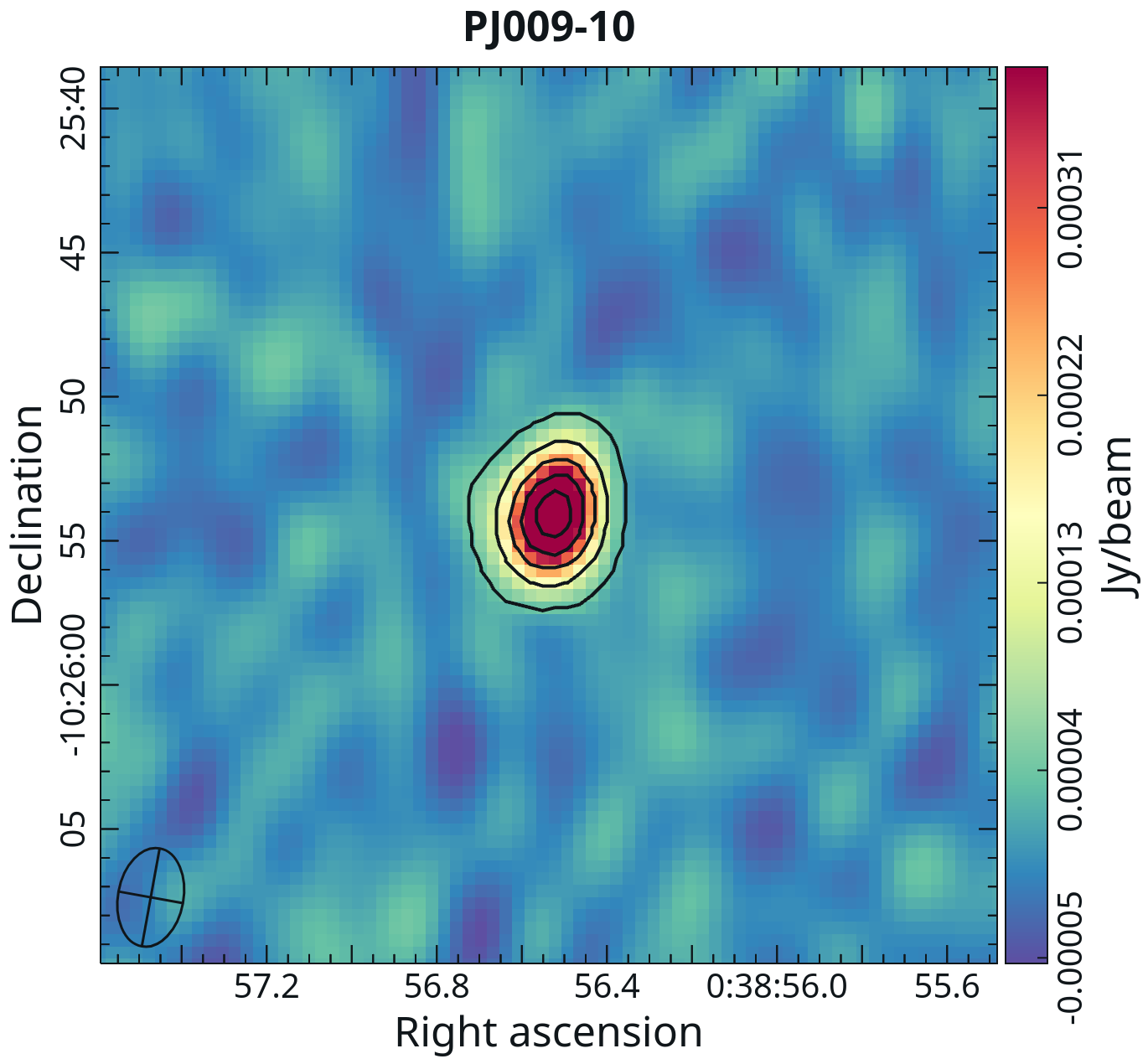}}
    \end{subfigure}

    \caption{NOEMA Band 2 spectra and dust continuum maps for the QSOs PJ007+04 and PJ009-10. The clean beam is reported in the bottom left corner of each continuum map. Panel (a): spectrum of the QSO PJ007+04 extracted over a region corresponding to the clean beam, with channel widths of 50 km/s. The light blue curve represents the fitted model, consisting of a constant continuuum and, in case of detection, some Gaussian line components. Panel (b): 158 GHz dust continuum map of the QSO
    PJ007+04 (contours at 3, 6 and 9 $\sigma$). Panel (c): same as panel (a) but for the QSO PJ009-10. Panel (d): same as panel (b) but for the QSO PJ009-10 (contours at 3, 9, 15, 21 and 27 $\sigma$).}
    \label{fig:NOEMAims}
    
\end{figure*}

\clearpage

\section{SED fitting}\label{SEDfitting}
\begin{table}[h]
    \centering
    \caption{Data used to fit the SEDs.}
    \begin{tabularx}{\columnwidth}{ZZZZ}
    \midrule \midrule
    \makecell[c]{ID} &\makecell[c]{Obs. frequency \\ (GHz)} & \makecell[c]{Flux \\ (mJy)} & \makecell[c]{Ref.}  \\ 
    \midrule
    
     AJ025-33 & 407 & 6.8 $\pm$ 0.7 & This work (a)\\
     & 342 & 4.7 $\pm$ 0.5 & This work (b)\\
     & 259 & 2.5 $\pm$ 0.3 & (1)  \\
     & 192 & 1.1 $\pm$ 0.1 & This work (b)\\
     & 103 & 0.10 $\pm$ 0.02 & This work (c)  \\
    \midrule

    J0305-3150 & 407 & 10.5 $\pm$ 1.5 & This work (a)\\
    &336 & 9.7 $\pm$ 1.0 & (2)  \\
    &250 & 5.3 $\pm$ 0.6 & (1)  \\
    &99 & 0.23 $\pm$ 0.04 & (3) \\
    \midrule
    
    J0439+1634* & 407 & 45.5 $\pm$ 4.7 & This work (a)\\
    & 353 & 26.2 $\pm$ 3.1 & (4)  \\
    & 271 & 16.9 $\pm$ 1.7 & (4)  \\
    & 255 & 15.5 $\pm$ 1.6 & (4)  \\
    & 245 & 16.0 $\pm$ 1.6 & (5) \\
    & 239 & 14.0 $\pm$ 1.4 & (4)  \\
    & 155 & 3.50 $\pm$ 0.4 & (4)  \\
    & 139 & 2.70 $\pm$ 0.3 & (4)  \\
    & 109 & 1.60 $\pm$ 0.2 & (4) \\
    & 93 & 1.30 $\pm$ 0.1 & (4) \\
    \midrule

    J2318-3029 & 407 & 6.9 $\pm$ 1.0 & This work (a)\\
    & 266 & 3.1 $\pm$ 0.3 & (1)  \\
    & 106 & 0.26 $\pm$ 0.04 & This work (c)  \\
    \midrule

    J2348-3054 & 670 & 9.1 $\pm$ 1.4 & (6) \\
    & 407 & 5.5 $\pm$ 0.7 & This work (a) \\
    & 240 & 2.28 $\pm$ 0.2 & (1)  \\
    & 95 & 0.12 $\pm$ 0.02 & (3)  \\
    \midrule
    
    PJ007+04 & 407 & 4.7 $\pm$ 0.6 & This work (a) \\
    & 271 & 2.33 $\pm$ 0.2 & (1) \\
    & 158 & 0.57 $\pm$ 0.09 & This work (d) \\
    \midrule

    PJ009-10 & 407 & 8.6 $\pm$ 1.2 & This work (a) \\
    & 346 & 5.9 $\pm$ 0.6 & This work (b) \\
    & 296 & 4.4 $\pm$ 0.5 & This work (e)  \\
    & 271 & 3.7 $\pm$ 0.5 & (1) \\
    & 202 & 1.5 $\pm$ 0.2 & This work (b) \\
    & 158 & 0.67 $\pm$ 0.07 & This work (d)\\
    \midrule

    \end{tabularx}
    \\[9pt]
    \parbox{\linewidth}{\small Notes -- (a): ALMA projects 2019.2.00053.S and 2021.2.00064.S, (b): ALMA project 2024.1.00071.S, (c): ALMA project 2019.1.00147.S, (d): NOEMA project S24CH, (e): ALMA project 2023.1.01450.S, (f): ALMA project 2021.2.00151.S, (g): ALMA project 2022.1.00321.S, (1): \cite{2020ApJ...904..130V}, (2): \cite{2025ApJ...982...72S}, (3): \cite{2017ApJ...845..154V}, (4): \cite{2019ApJ...880..153Y}, (5): \cite{2021ApJ...917...99Y}, (6): \cite{2025A&A...695L..18M}, (7): \cite{2024A&A...689A.220T}, (8): \cite{2023ApJ...944..134B}, (9): \cite{2022A&A...662A..60D}, (10): \cite{2020ApJ...903...34A}, (11): \cite{2022MNRAS.510.4976A}, (12): \cite{2021A&A...652A..66P}. A 10\% systematic flux calibration uncertainty was added in quadrature to the nominal uncertainty when it was not already included in the reference paper. *in the fit all the fluxes have been corrected for the magnification factor reported in (5).}
    \label{tab:gigatab}
\end{table}

\clearpage
\setcounter{table}{0}
\makebox[0pt][c]{}

\begin{table}[h!]
    \centering
    \caption{continued.}
    \begin{tabularx}{\columnwidth}{ZZZZ}
        
        \midrule \midrule
        \makecell[c]{ID} &\makecell[c]{Obs. frequency \\ (GHz)} & \makecell[c]{Flux \\ (mJy)} & \makecell[c]{Ref.}  \\ 
        \midrule

        PJ036+03 & 671 & 5.6 $\pm$ 0.9 & (7) \\
        & 407 & 6.2 $\pm$ 0.7 & This work (a) \\
        & 334 & 5.0 $\pm$ 0.5 & (8) \\
        & 252 & 2.55 $\pm$ 0.3 & (1) \\
        & 100 & 0.13 $\pm$ 0.02 & (9) \\
        \midrule

        PJ083+11 & 407 & 10.5 $\pm$ 1.5 & This work (a) \\
        & 348 & 8.8 $\pm$ 0.9 & (2) \\
        & 258 & 5.5 $\pm$ 0.6 & (10) \\
        & 244 & 5.10 $\pm$ 0.5 & (10) \\
        & 145 & 0.9 $\pm$ 0.2 & This work (f) \\
        \midrule

        PJ158-14 & 407 & 9.6 $\pm$ 1.2 & This work (a)\\
        & 261 & 3.2 $\pm$ 0.3 & (11) \\
        \midrule

        PJ231-20 & 398 & 8.4 $\pm$ 0.9 & This work (g)\\
        & 337 & 6.8 $\pm$ 0.7 & (2) \\
        & 250 & 3.9 $\pm$ 0.4 & (12) \\
        & 242 & 3.5 $\pm$ 0.4 & (12) \\
        & 234 & 3.3 $\pm$ 0.3 & (12) \\
        & 227 & 2.9 $\pm$ 0.3 & (12) \\
        & 205 & 2.2 $\pm$ 0.2 & (12) \\
        & 193 & 1.9 $\pm$ 0.2 & (12) \\
        & 153 & 0.86 $\pm$ 0.9 & (12) \\
        & 140 & 0.72 $\pm$ 0.07 & (12) \\
        & 106 & 0.31 $\pm$ 0.04 & (12) \\
        & 94 & 0.22 $\pm$ 0.02 & (12) \\
        \midrule
        
    \end{tabularx}
\end{table}
\vspace{0.2cm}
\begin{table}[h!]
    \centering
    \caption{Spatial extent of the sources analyzed in this work.}
    \begin{tabularx}{\textwidth}{ZZZ}
        \midrule
        \midrule
        ID & \makecell[c]{Size \\ (arcsec)} & Ref. \\
        \midrule
        AJ025-33 & 0.27 $\times$ 0.19  & (1) \\
        J0305-3150 & 0.32 $\times$ 0.28 & (1) \\
        J0439-1634* & 0.10 $\times$ 0.10 &  (2)  \\
        J2318-3029 & 0.14 $\times$ 0.12 & (1) \\
        J2348-3054 & 0.13 $\times$ 0.07 & (1) \\
        PJ007+04 & 0.13 $\times$ 0.11 & (1) \\
        PJ009-10 & 0.80 $\times$ 0.35 & (1) \\
        PJ036+03 & 0.19 $\times$ 0.16 & (1) \\
        PJ083+11** & 0.40 $\times$ 0.30 & (3) \\
        PJ158-14 & 0.20 $\times$ 0.14 & (4)  \\
        PJ231-20 & 0.11 $\times$ 0.09 & (1) \\
        \midrule
    \end{tabularx}
    \\[9pt]
    \parbox{\linewidth}{\small Notes -- the sizes are reported as FWHMs of the 2D Gaussian fitted to the images from spatially-resolved data. (1): \cite{2020ApJ...904..130V}, (2): \cite{2021ApJ...917...99Y}, (3): \cite{2020ApJ...903...34A}, (4): \cite{2022MNRAS.510.4976A}. *computed from the effective radius, **computed from the physical size.}
    \label{tab:sizes}
\end{table}

\begin{figure*}[h!]

    \centering
    \begin{subfigure}{\includegraphics[width=8.5cm]{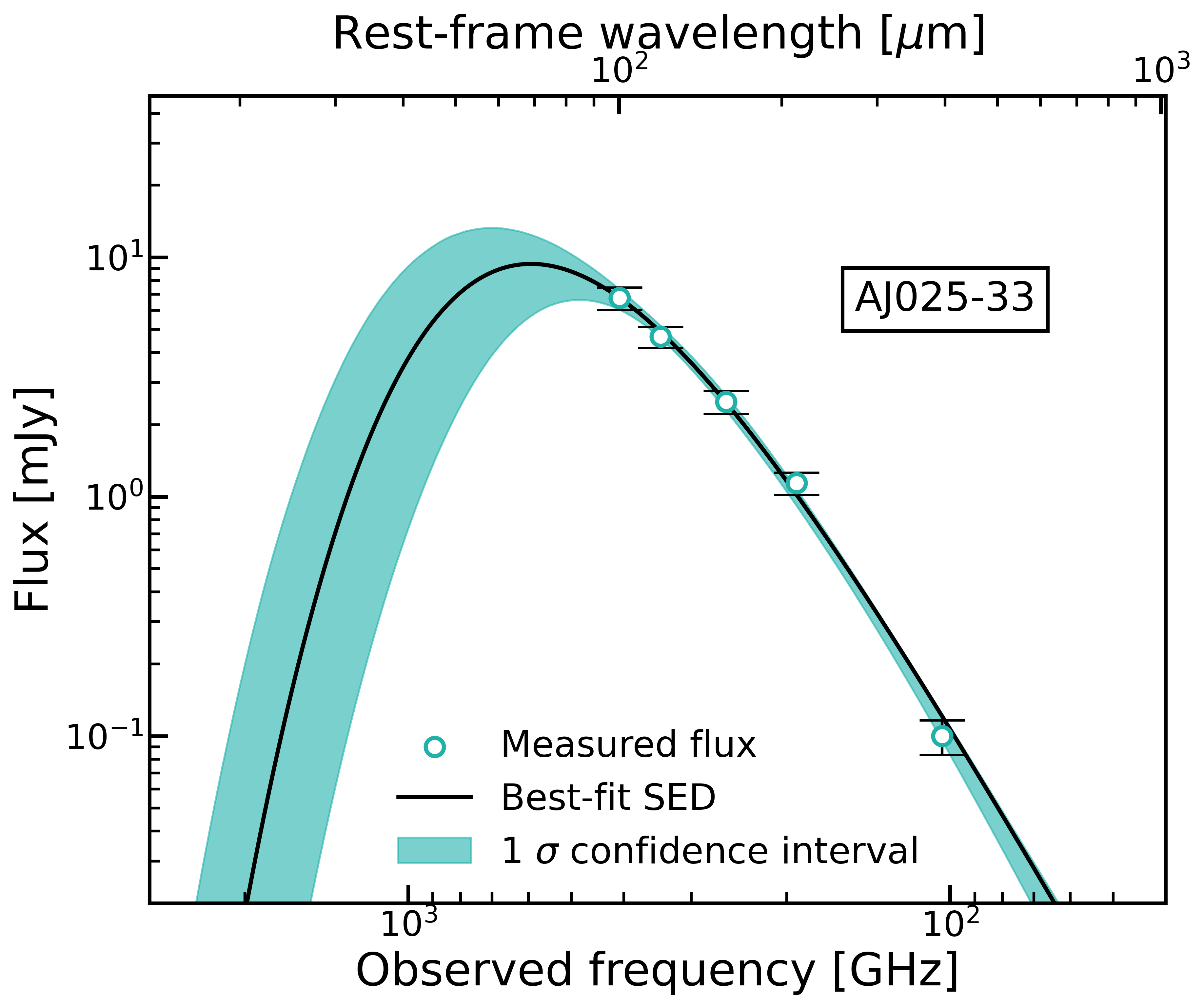}}
    \end{subfigure}
    \hspace{1.cm}
    \begin{subfigure}{\includegraphics[width=7.4cm]{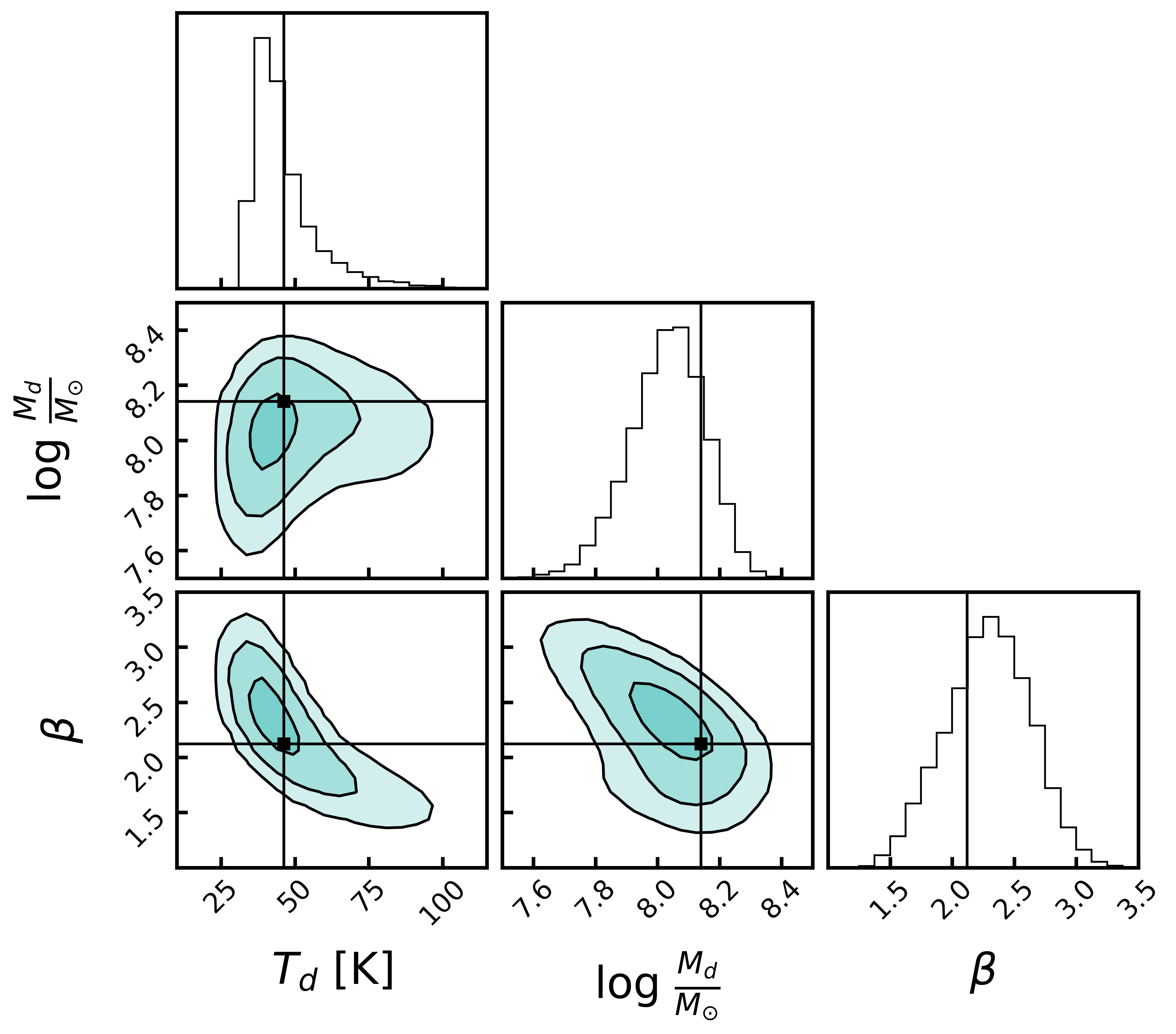}}
    \end{subfigure}

    \centering
    \begin{subfigure}{\includegraphics[width=8.5cm]{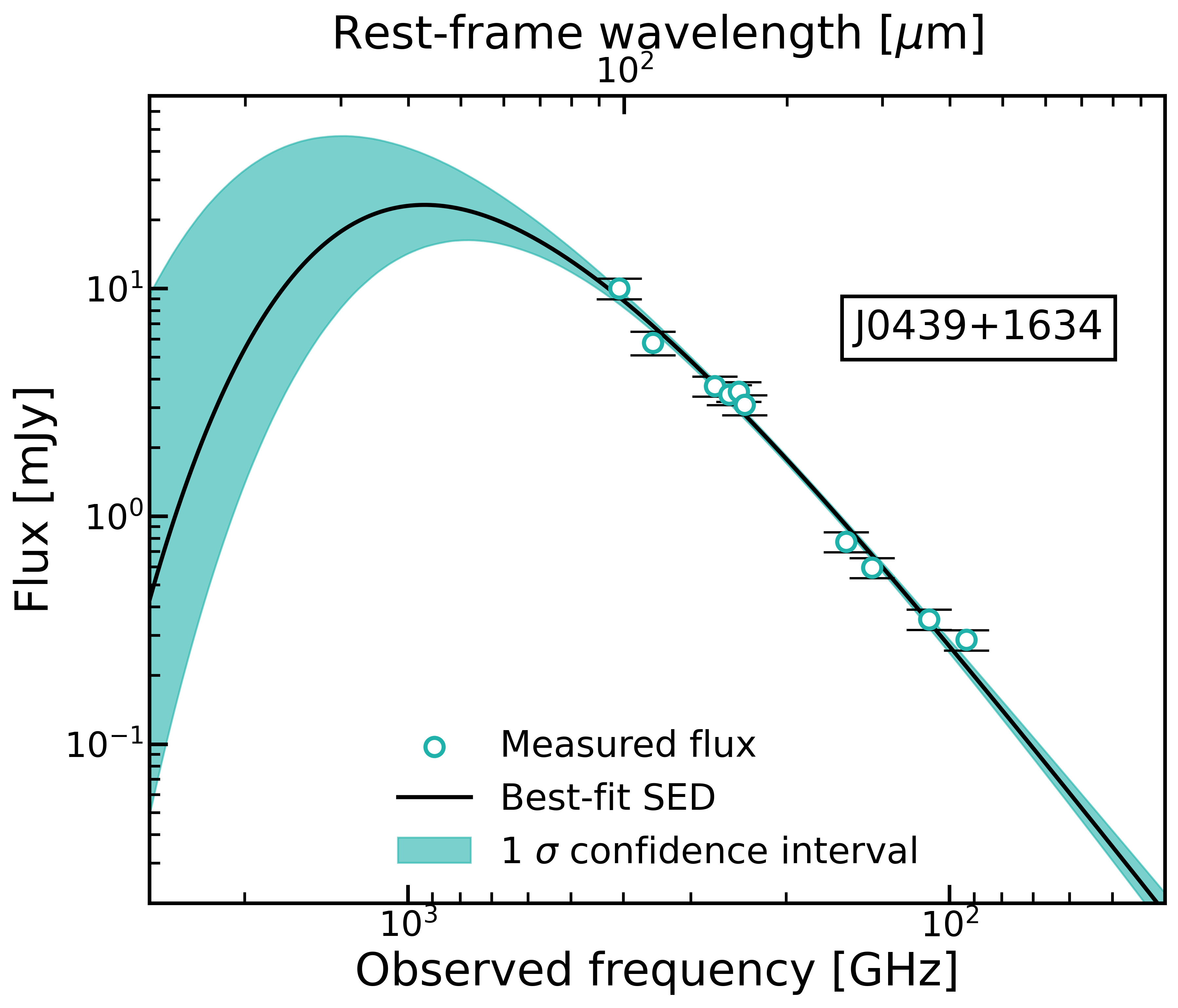}}
    \end{subfigure}
    \hspace{1.cm}
    \begin{subfigure}{\includegraphics[width=7.4cm]{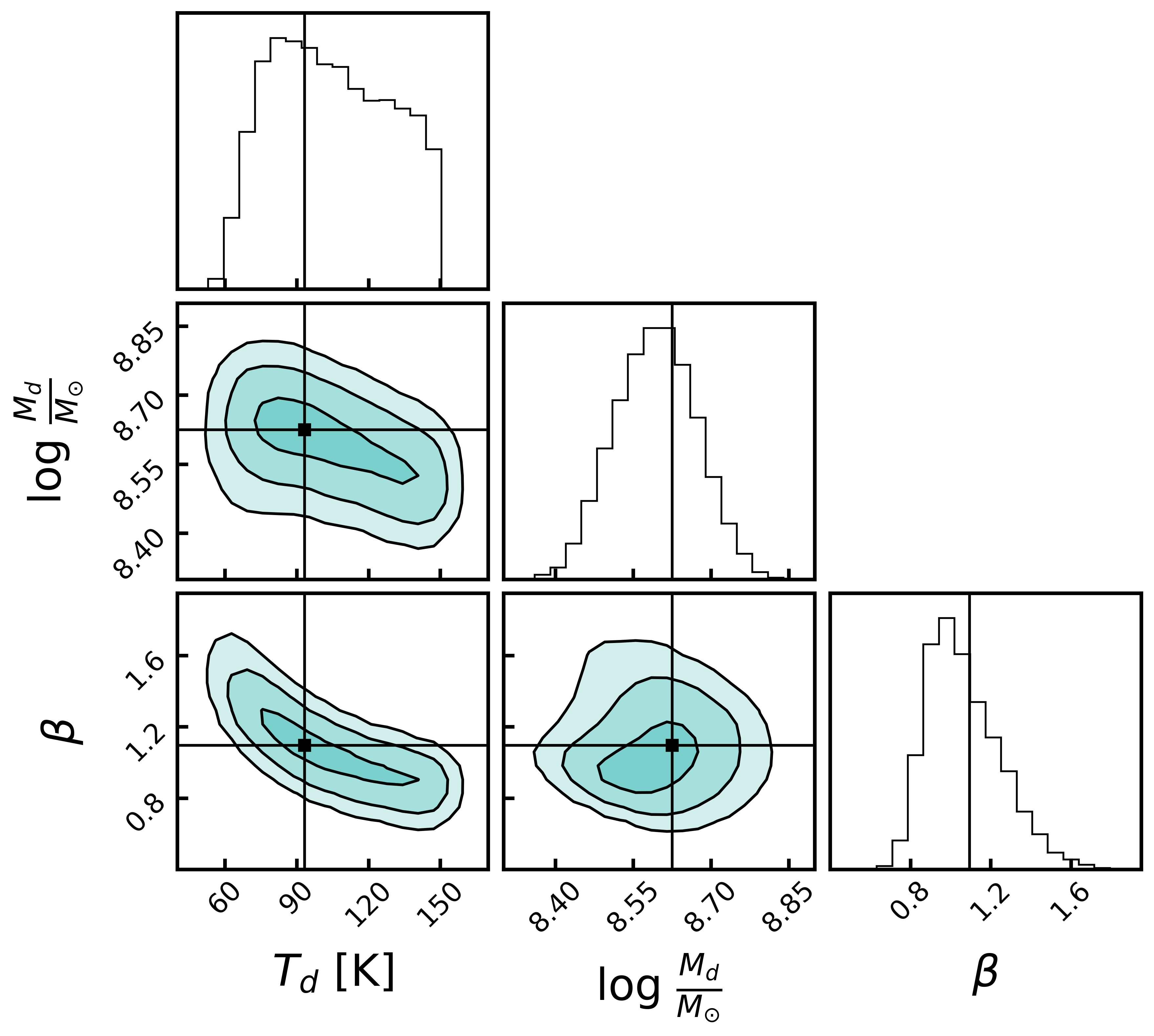}}
    \end{subfigure}
    
    \centering
    \begin{subfigure}{\includegraphics[width=8.5cm]{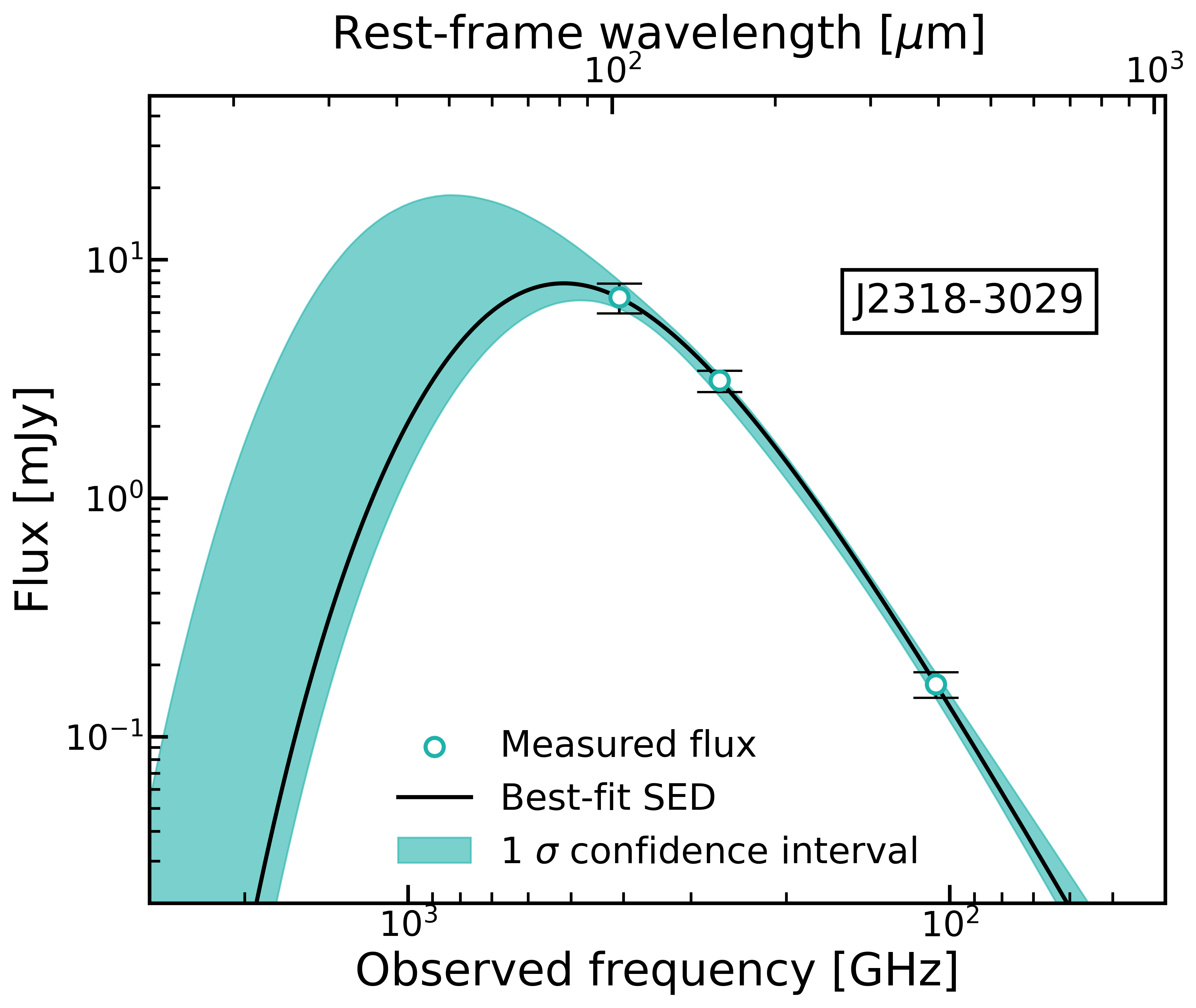}}
    \end{subfigure}
    \hspace{1.cm}
    \begin{subfigure}{\includegraphics[width=7.4cm]{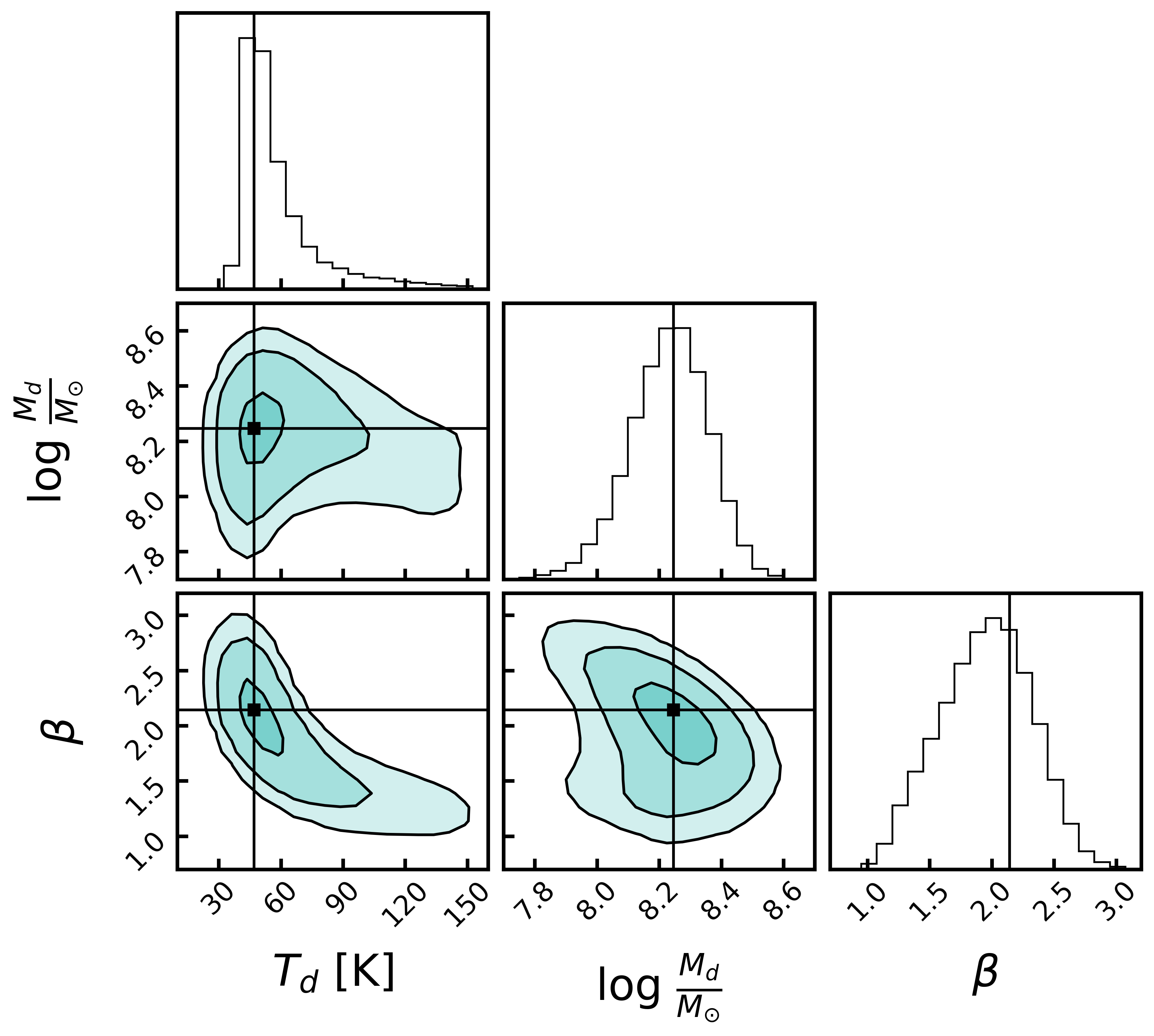}}
    \end{subfigure}
    
    \caption{Left panels: fits of the IR-SED of the QSOs in this sample. Right panels: corner plots of the parameters' posterior probability distribution. We only report the sources for which the algorithm converged}
    \label{sedimages}
\end{figure*}
\vfill
\setcounter{figure}{0}
\vfill
\begin{figure*}[h!]

    \centering
    \begin{subfigure}{\includegraphics[width=8.5cm]{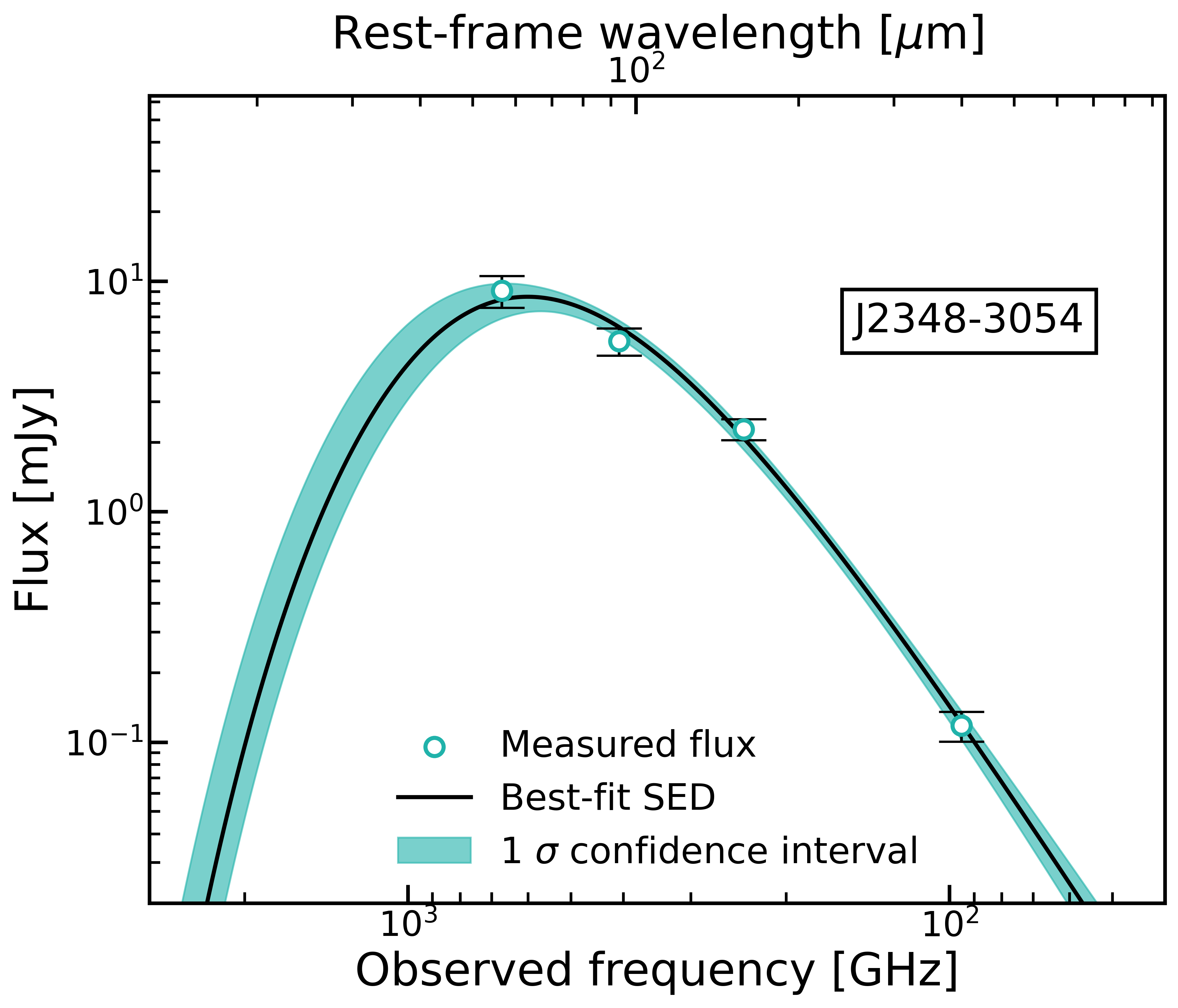}}
    \end{subfigure}
    \hspace{1.cm}
    \begin{subfigure}{\includegraphics[width=7.4cm]{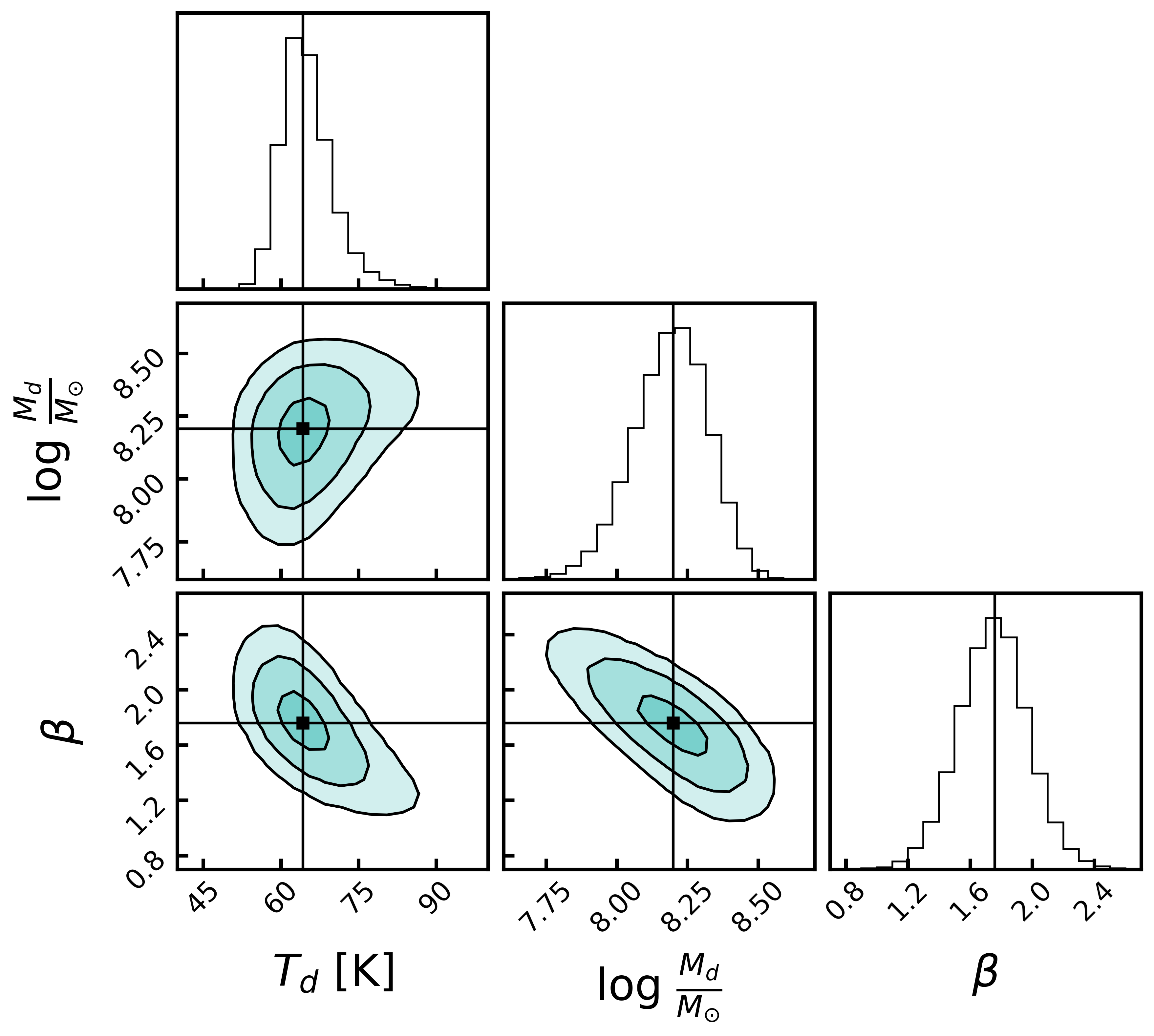}}
    \end{subfigure}

    \centering
    \begin{subfigure}{\includegraphics[width=8.5cm]{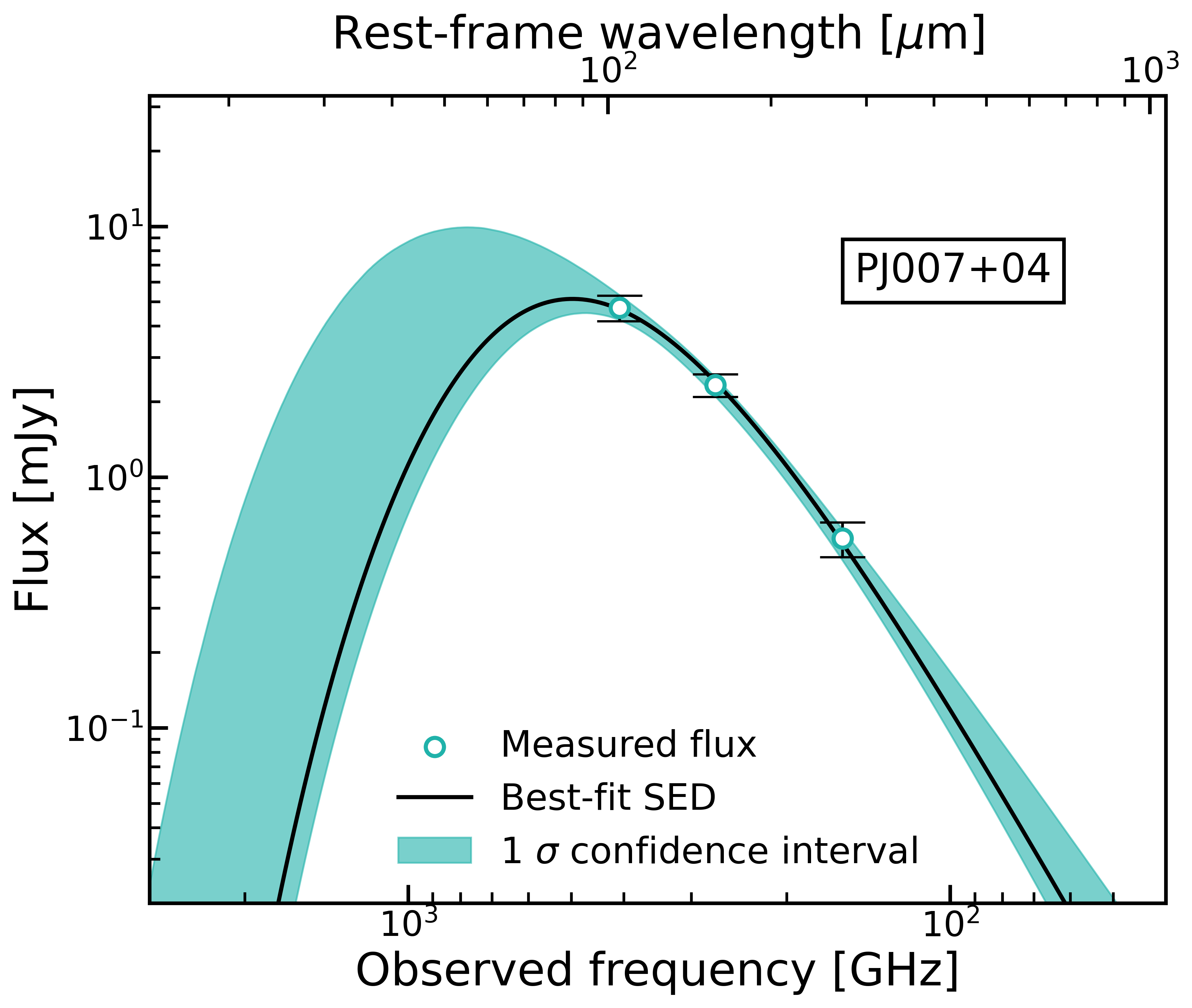}}
    \end{subfigure}
    \hspace{1.cm}
    \begin{subfigure}{\includegraphics[width=7.4cm]{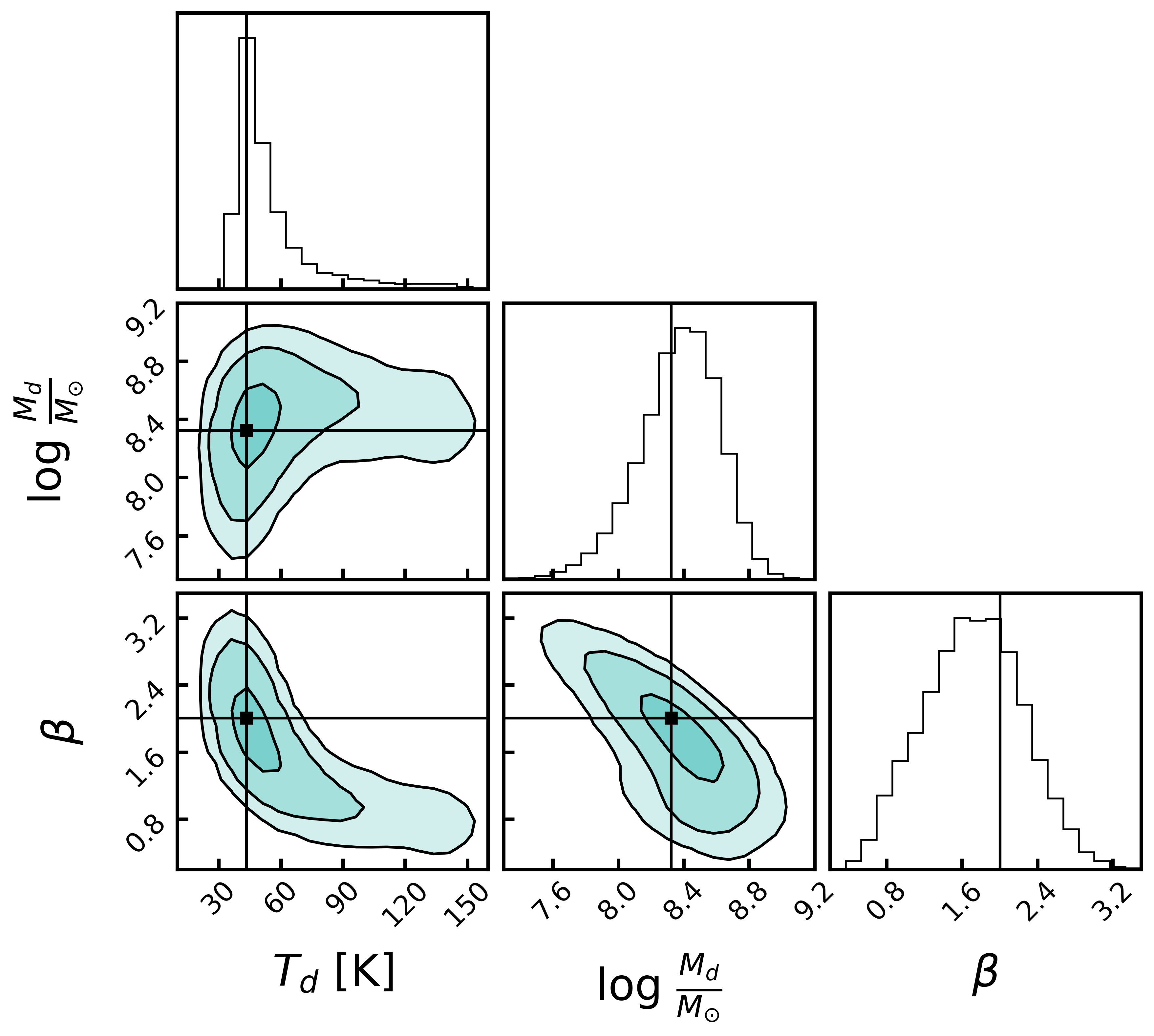}}
    \end{subfigure}
    
    \centering
    \begin{subfigure}{\includegraphics[width=8.5cm]{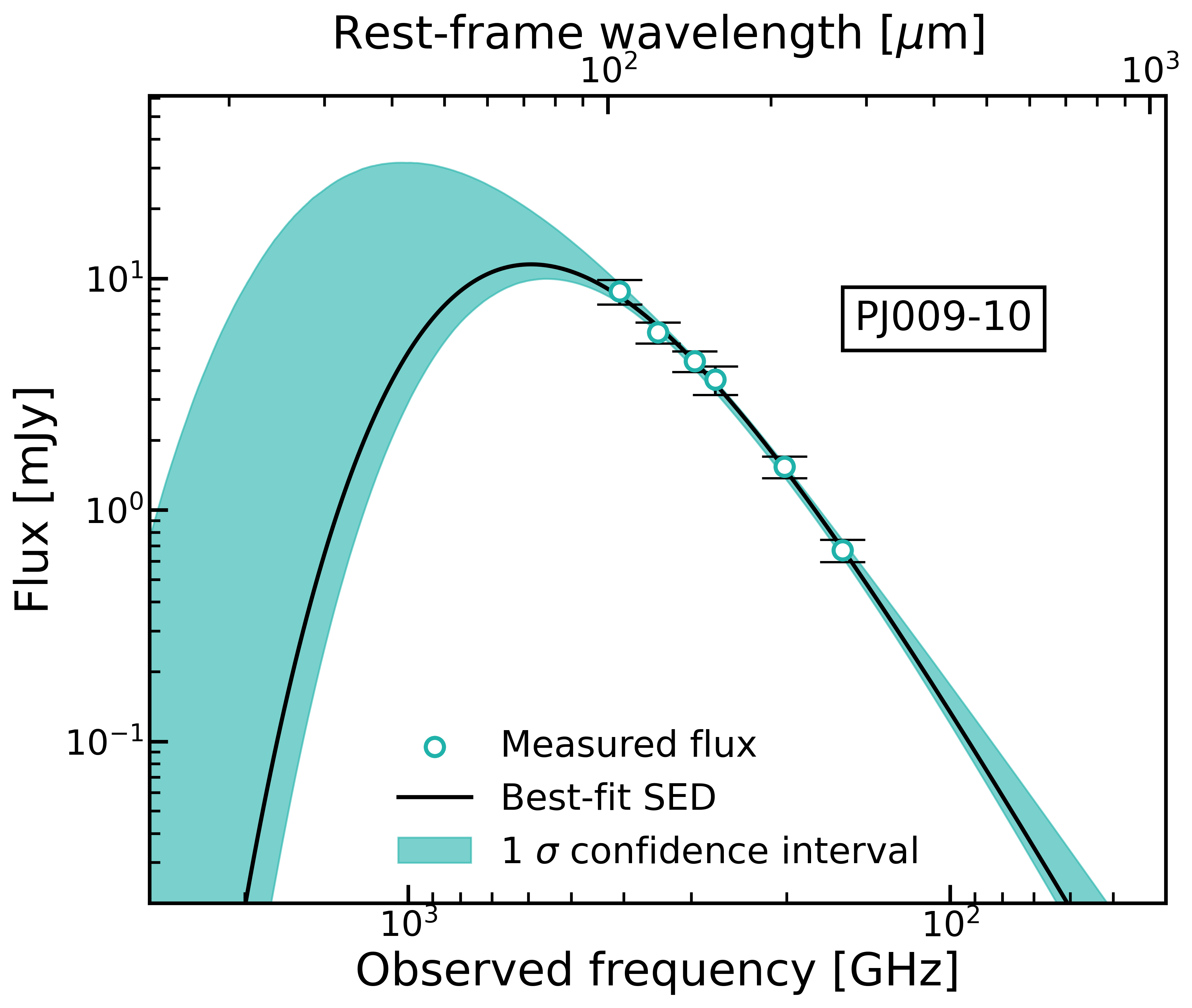}}
    \end{subfigure}
    \hspace{1.25cm}
    \begin{subfigure}{\includegraphics[width=7.4cm]{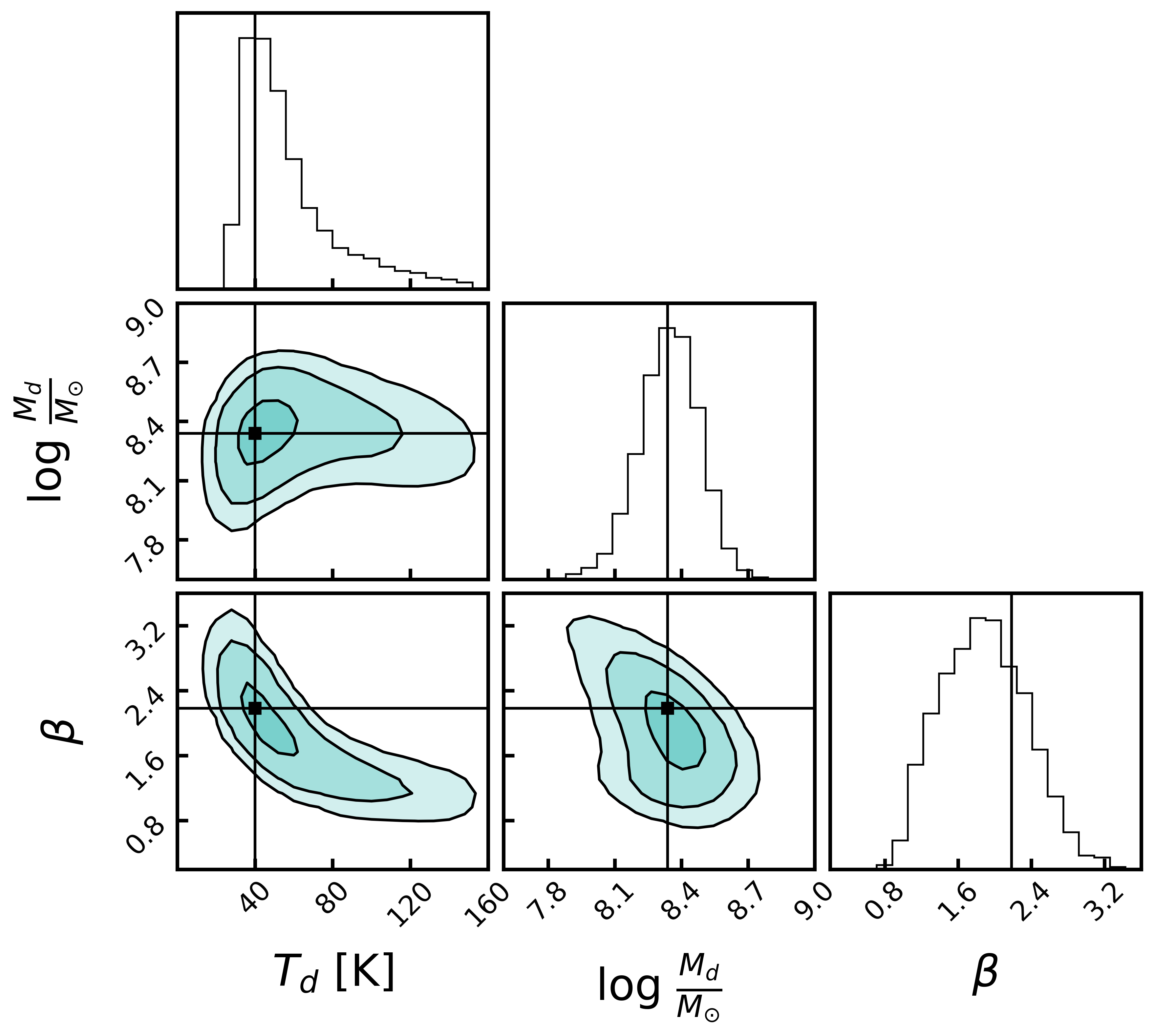}}
    \end{subfigure}
    
    \caption{continued.}
\end{figure*}
\vfill
\setcounter{figure}{0}
\vfill
\begin{figure*}[h!]

    \centering
    \begin{subfigure}{\includegraphics[width=8.5cm]{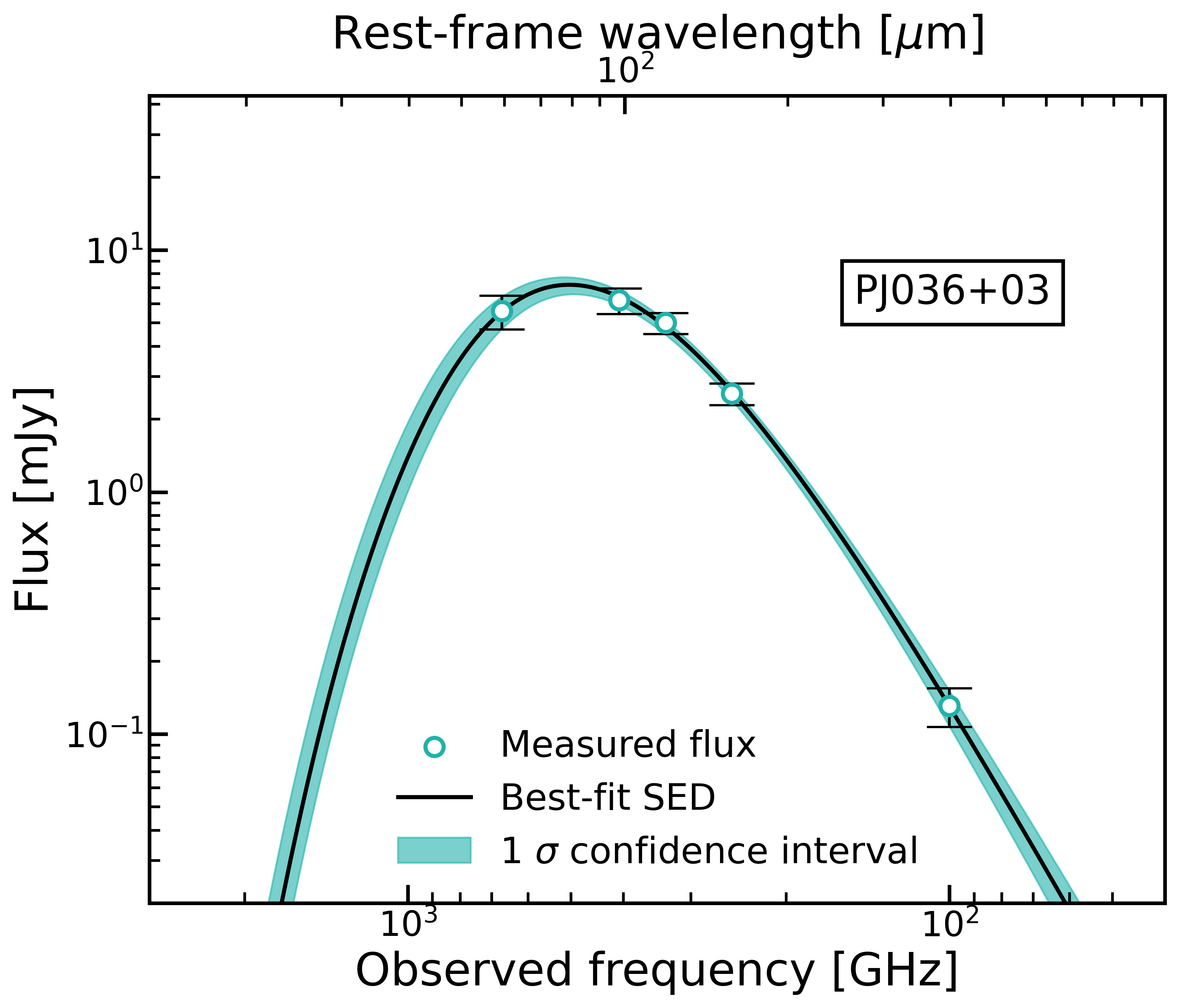}}
    \end{subfigure}
    \hspace{1.cm}
    \begin{subfigure}{\includegraphics[width=7.4cm]{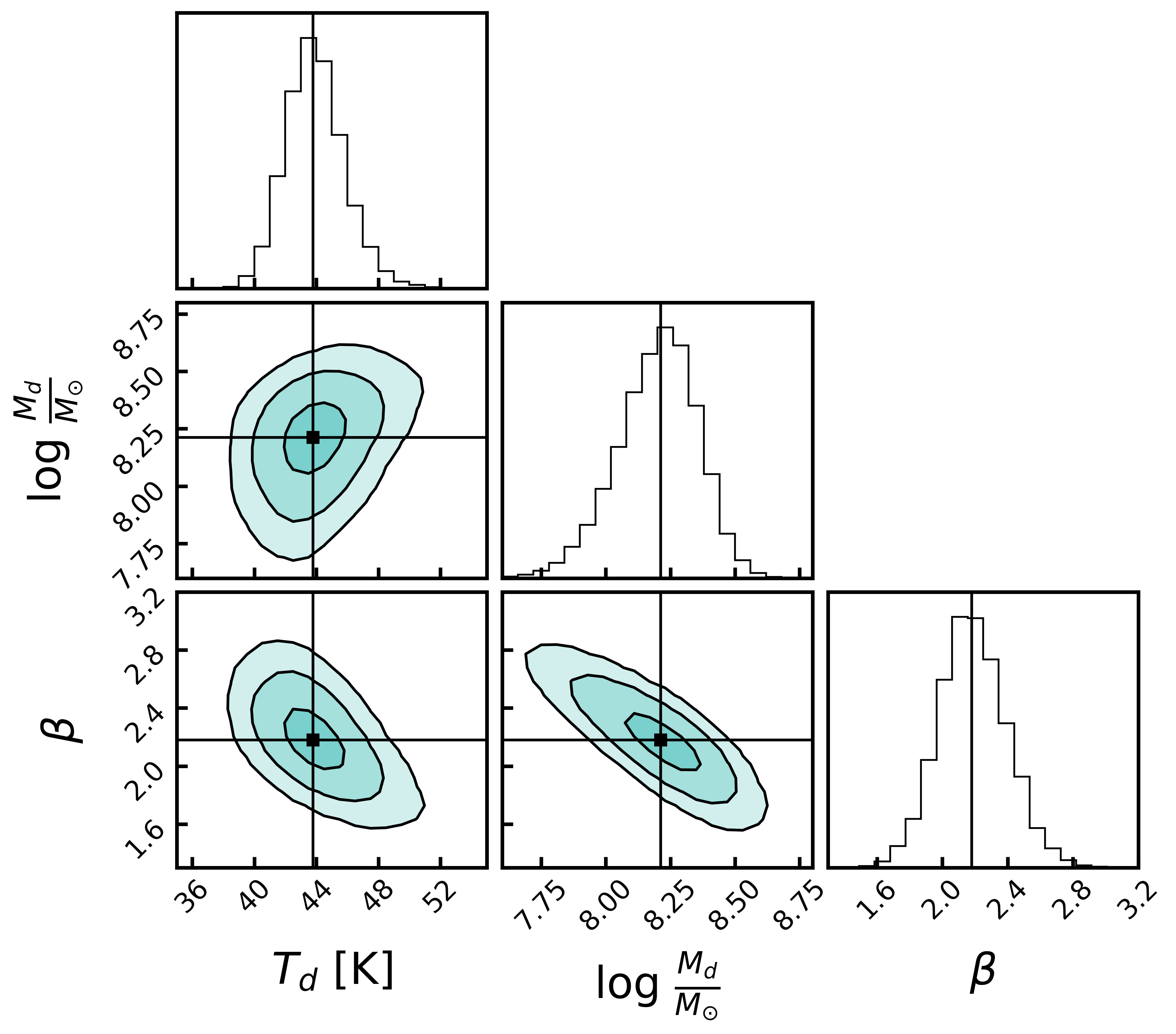}}
    \end{subfigure}

    \centering
    \begin{subfigure}{\includegraphics[width=8.5cm]{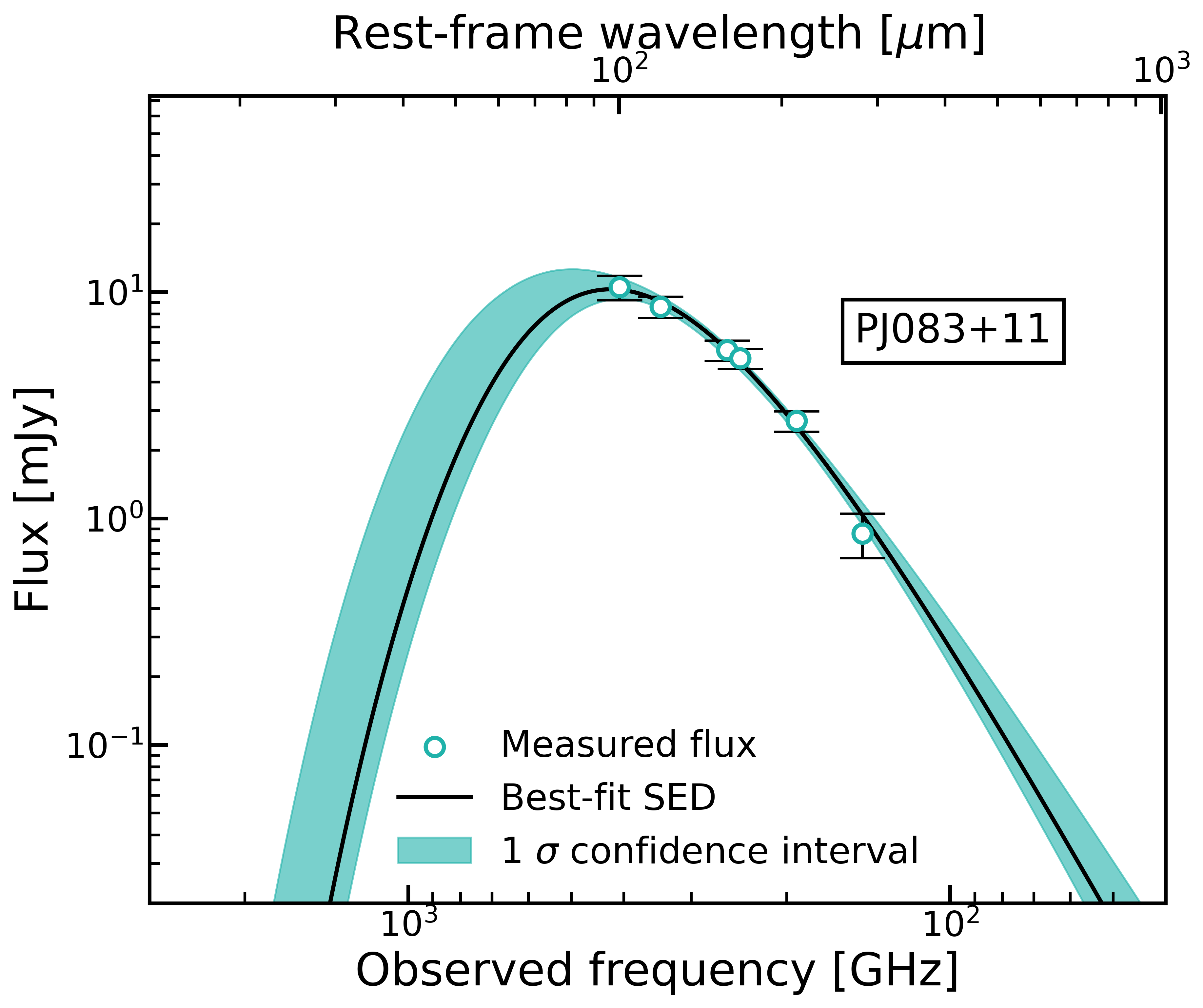}}
    \end{subfigure}
    \hspace{1.cm}
    \begin{subfigure}{\includegraphics[width=7.4cm]{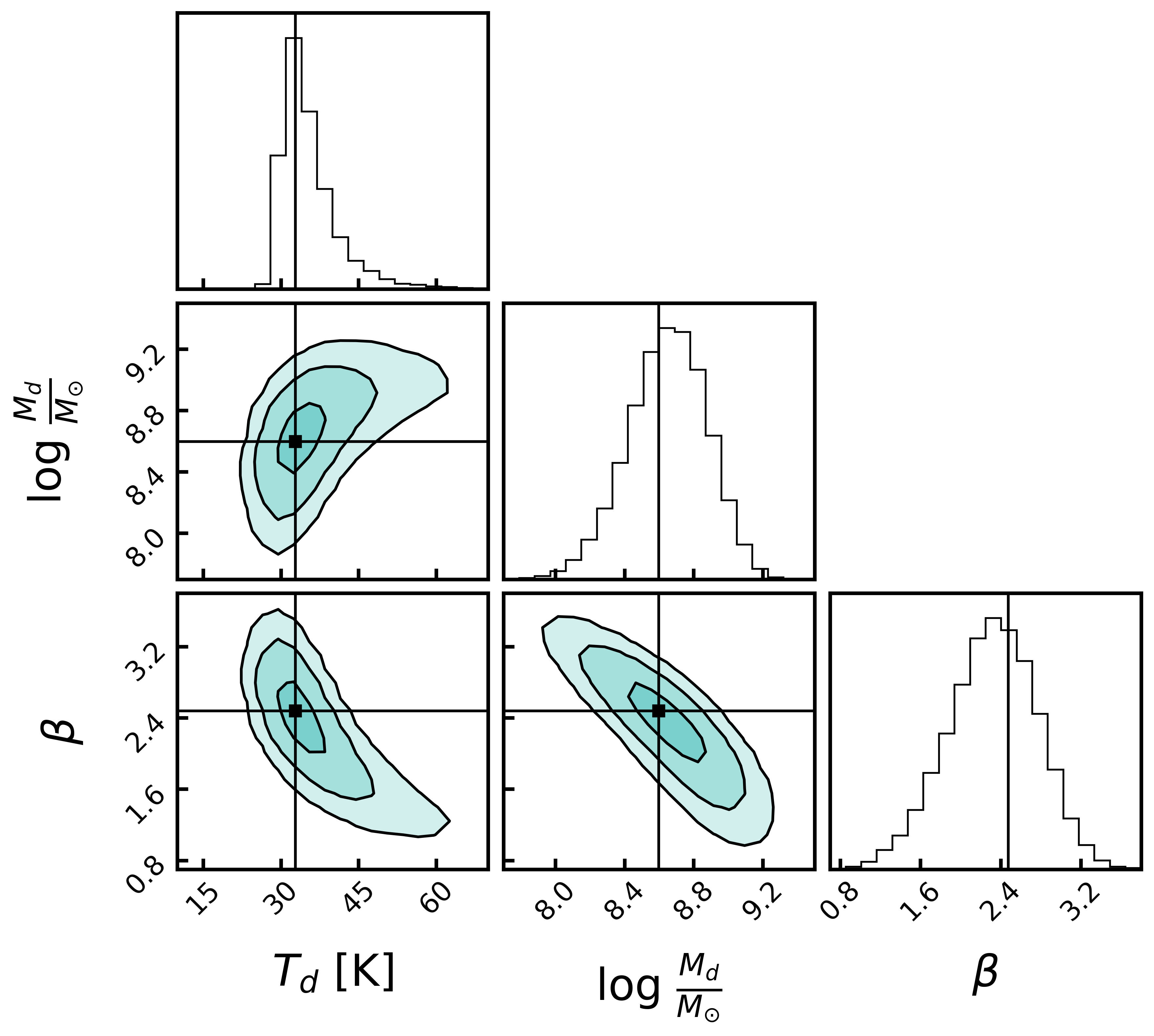}}
    \end{subfigure}
    
    \centering
    \begin{subfigure}{\includegraphics[width=8.5cm]{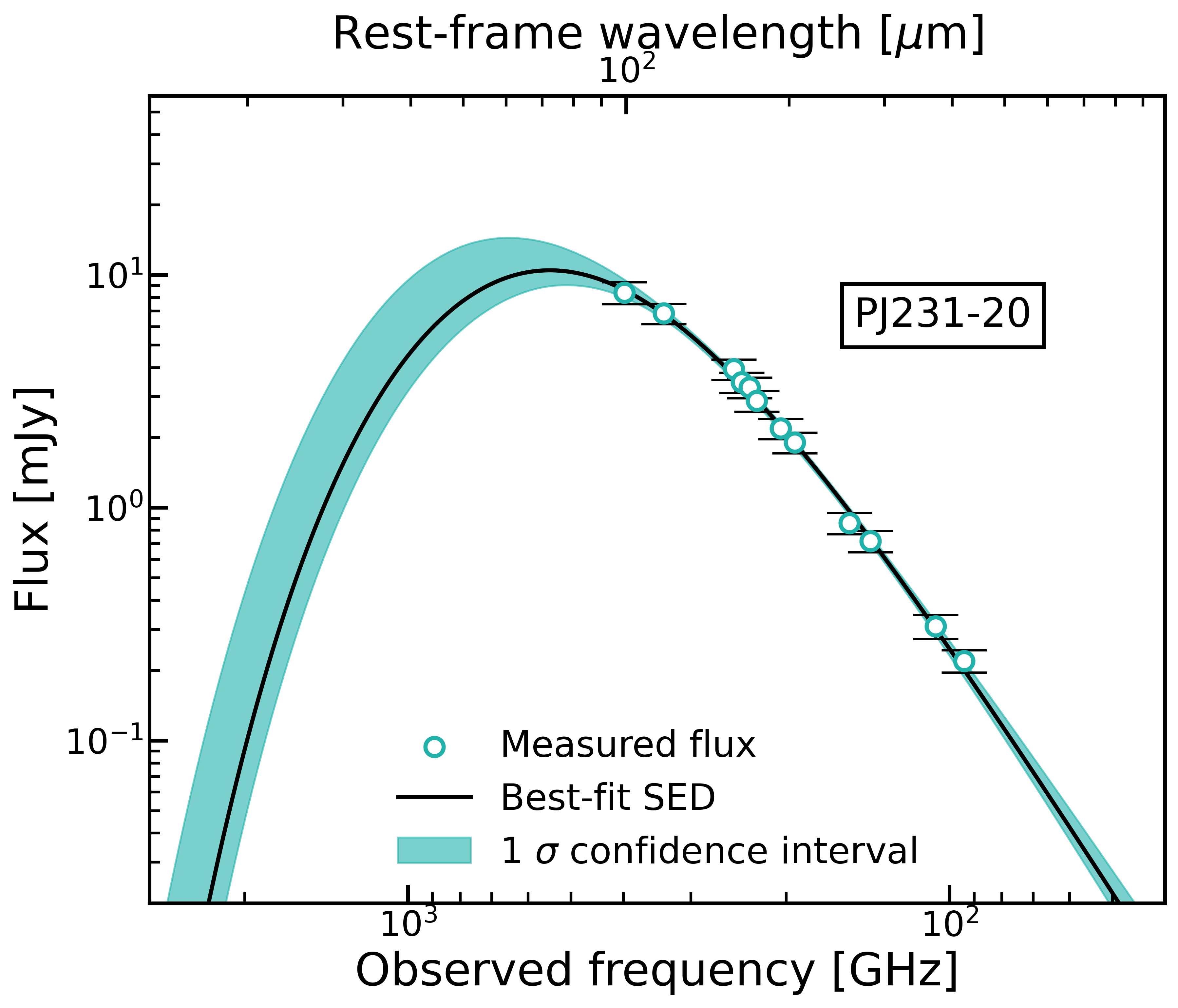}}
    \end{subfigure}
    \hspace{1.cm}
    \begin{subfigure}{\includegraphics[width=7.4cm]{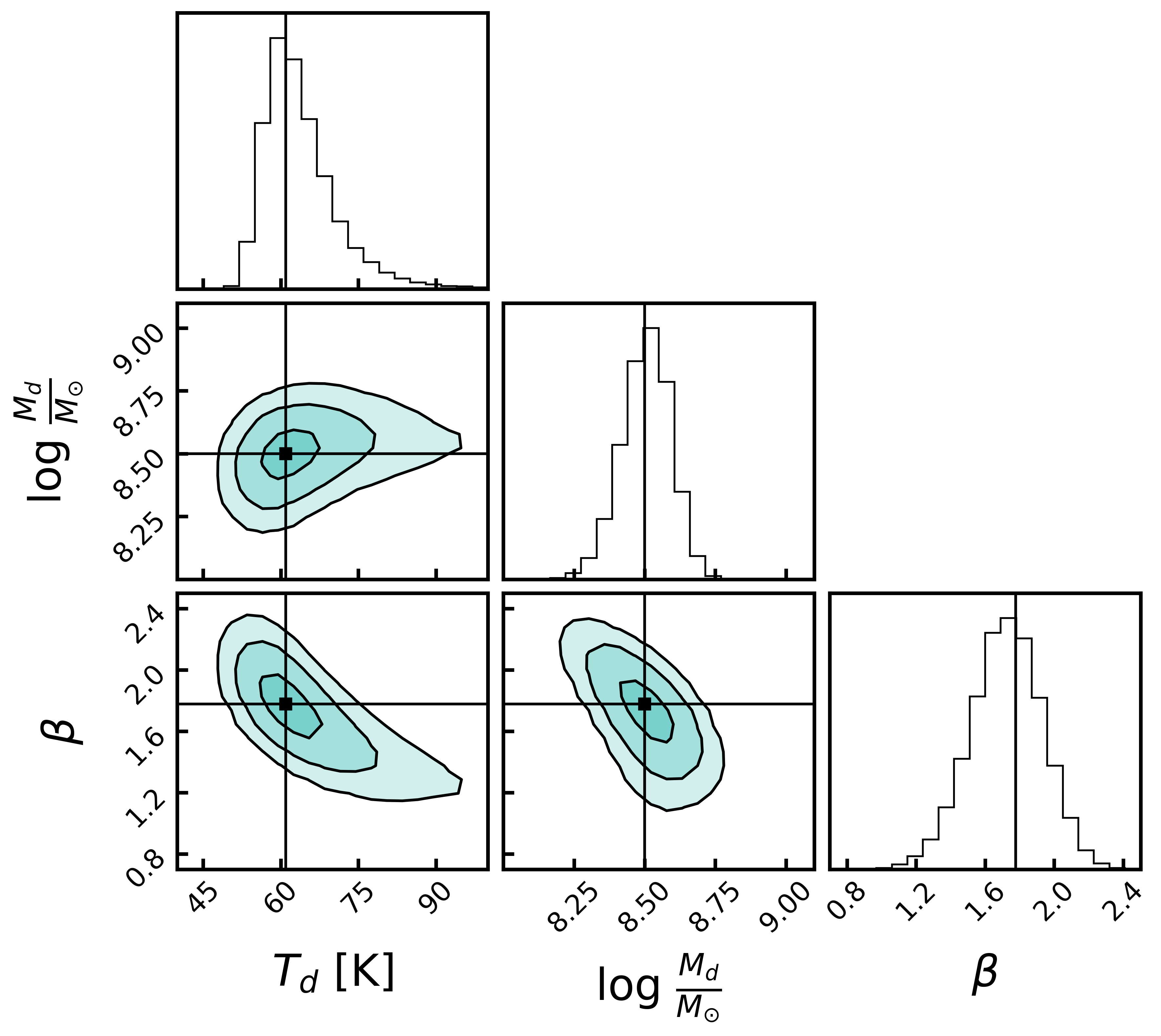}}
    \end{subfigure}
    
    \caption{continued.}
\end{figure*}

\end{document}